\documentclass[
aip,
chaos,
reprint,
amsmath,amssymb,
floatfix
]{revtex4-2}

\usepackage{graphicx}
\usepackage{dcolumn}
\usepackage{bm}

\usepackage[utf8]{inputenc}
\usepackage[T1]{fontenc}
\usepackage{mathptmx}
\usepackage{etoolbox}

\usepackage{lipsum}
\usepackage{amsmath}
\usepackage{amssymb}
\usepackage{graphicx}
\usepackage{bm}
\usepackage{refstyle}
\usepackage{multirow,array}
\usepackage{makecell}
\usepackage{mathtools}
\usepackage{empheq}
\usepackage{enumerate}
\usepackage{pifont}
\usepackage{longtable} 
\usepackage{booktabs}
\usepackage[normalem]{ulem}

\usepackage[
citecolor=blue,
colorlinks,
linkcolor=blue,
urlcolor=blue,
]{hyperref}
\usepackage[dvipsnames]{xcolor}
\usepackage{dcolumn}
\usepackage{bm}
\usepackage{comment} 
\usepackage{enumitem}
\usepackage{float}

\newcolumntype{C}[1]{>{\centering\arraybackslash}p{#1}}

\definecolor{deepgray}{RGB}{64,64,64}

\makeatother
\begin{document}

\preprint{AIP/123-QED}

\title{Adaptive dynamics of eco-evolutionary repeated games: Effect of reward and punishment}

\author{Prosanta Mandal}
\email{prosantam21@iitk.ac.in}
\affiliation{
 Department of Physics, Indian Institute of Technology Kanpur, Kanpur 208016, India
}

\author{Suman Chakraborty}
\email{schakraborty1001@gmail.com (Corresponding author)}
\affiliation{
	Department of Physics, Indian Institute of Technology Madras, Chennai 600036, India
}
\author{Vaibhav Madhok}
\email{madhok@physics.iitm.ac.in}
\affiliation{
	Department of Physics, Indian Institute of Technology Madras, Chennai 600036, India
}

\author{Sagar Chakraborty}
\email{sagarc@iitk.ac.in}
\affiliation{
 Department of Physics, Indian Institute of Technology Kanpur, Kanpur 208016, India
}

\begin{abstract}
Long-term evolutionary processes can strongly influence common-pool resource conservation by generating new traits or behaviours that modify the feedback between population strategies and the resource state. Here we develop an eco-evolutionary framework in which individuals repeatedly interact with the same opponent and follow direct reciprocity through reactive strategies. The strategic dynamics is coupled to a renewable common resource and analyzed using adaptive dynamics. After our exhaustive non-linear dynamical analysis of $2\times2$ strategic games, we focus on comparative and combined usefulness of institutional incentives in the form of rewards and punishments in preventing the tragedy of the commons even when defection dominates in the replete resource state. We also report possibility of robust stable oscillations---emerging via Hopf bifurcation---in resource state and population strategies.
\end{abstract}


\maketitle

\textbf{\color{black}In any population, individuals compete with one another while exploiting a common resource whose state, in turn, dictates how those individuals collectively and strategically interact. This intrinsic game-resource feedback loop is a central focus of eco-evolutionary dynamics. However, its existing frameworks rarely consider the strategic shifts that occur when individuals repeatedly interact with the same opponents. While the theory of repeated games has been instrumental over the last half-century in explaining the evolution of cooperation, its integration with dynamic resource feedback remains largely unexplored. In this work, we utilize the paradigms of evolutionary game theory and nonlinear dynamics to mathematically model the eco-evolutionary dynamics of simple yet non-trivial repeated games. We establish the mathematical conditions under which a shared resource is sustained and identify regimes where the resource exhibits periodic oscillations. Ultimately, our study maps these macroscopic resource dynamics directly to microscopic behavioural mechanisms, specifically the systems of punishment and reward adopted by the individuals.}

\section{Introduction}
The Tragedy of the Commons (ToC)~\cite{hardin1968, Rankin2007, Frischmann2019, Janssen2019, Ostrom2008, bromley1989wbp} refers to a social dilemma in which individuals, acting in their self-interest, overuse and ultimately deplete shared resources, resulting in collective harm. This phenomenon arises across diverse domains, including population growth~\cite{hardin1968}, environmental degradation~\cite{ZeaCabrera2006, Lal2009, Shiklomanov2000, Fraser2003}, property and communal rights~\cite{Ostrom1999ARPS}, wildlife crimes~\cite{Pires2011EJCPR, pandit2003ec}, and even plant competition~\cite{Falster2003}. The pivotal concern lies in designing mechanisms that manage common resources effectively, balancing individual incentives with long-term collective sustainability.

While one would naturally guess that cooperation in a society would prevent the ToC, there remains a fundamental puzzle in evolutionary biology and game theory about how cooperation can emerge and persist among selfish individuals. The Donation Game (DG)---a variant of the Prisoner’s Dilemma (PD)---offers a minimal yet powerful framework for studying this question~\cite{Nowak1998, LHN_2023_PLOS}. In this game, one individual pays a cost to confer a benefit to another. Mutual cooperation results in a socially optimal outcome, yet individuals are often tempted to defect and enjoy the benefits without incurring the cost. The tension between individually rational strategies and collectively desirable outcomes encapsulates a deep evolutionary paradox: if natural selection favours selfishness, why is cooperation so prevalent in nature and society? Numerous works have been done in this direction and `five rules of cooperation'~\cite{nowak2006science} (which includes direct reciprocity~\cite{Trivers1971, Axelrod1981, Hilbe2018, Garca2018}---focus of this paper) notably summarizes many of them within the paradigm of evolutionary game theory~\cite{Smith1982, NOWAK2006, SMITH1973, Hofbauer1998}.

Environmental variability plays a crucial role in shaping cooperation and social behaviour. Organisms constantly adapt to fluctuations in climate, resources, and habitat conditions~\cite{Melbinger2015, Sih2011, Levins1968}. A classic example is the peppered moth during the Industrial Revolution, where changes in tree coloration shifted selective advantages between morphs~\cite{Kettlewell1955}. Seasonal scarcity can intensify competition, alter social hierarchies, or trigger new cooperative strategies~\cite{Trivers1971}. Predator-prey dynamics, for instance, evolve with shifting environmental conditions, as prey adopt new defences and predators adjust tactics~\cite{Brown2004}. These adaptations ripple through ecological networks, reshaping interactions and enabling the emergence of symbioses~\cite{Muscatine1977}, behavioral innovation~\cite{Avila2023}, and even speciation events~\cite{Doebeli2011, Lenski2017, Meyer2016, Robertson1991, Post2009, Killingback2010, PassagemSantos2018, McKinnon2002}.

Empirical examples of eco-evolutionary feedbacks include algae–rotifer chemostats~\cite{Yoshida2003}, alewife–zooplankton dynamics in North American lakes~\cite{Brooks1965, Palkovacs2008EcoevolutionaryIB, Post2008}, guppies in Trinidadian streams~\cite{Reznick1997, Palkovacs2009}, Darwin’s finches in the Galápagos~\cite{Grant1986, Hairston2005, Grant2006}, and poplar trees in North America~\cite{Whitham2006}. These systems highlight how behavioural strategies and environmental context co-evolve, forming dynamic feedback loops.

Evidently, in an ecosystem, biotic interactions—such as competition, predation, and cooperation—are not isolated from the abiotic environment, including climate, temperature, and resource availability. These biotic and abiotic components are inherently coupled, meaning changes in one can influence the dynamics of the other. For instance, climate change (an abiotic shift) can alter species interactions, migration patterns, or reproductive strategies. In contrast, widespread changes in population behaviour can, in turn, affect nutrient cycling, atmospheric composition, or habitat structure.

To study such coupled dynamics, several eco-evolutionary game models have explicitly incorporated feedback between strategic behaviour and the state of a shared environment. In an original framework, Weitz et al.~\cite{weitz2016pnas} coupled the frequency dynamics of cooperators and defectors with a continuously changing environmental state, showing that feedback between strategic behaviour and environmental quality can generate oscillatory dynamics. Tilman et al.~\cite{TPA_2020_NAT_COM} subsequently generalized this framework by allowing the environment to possess intrinsic dynamics in addition to strategy-induced changes, thereby accommodating renewable, decaying, and tipping-point environments. More recently, Du et al.~\cite{DLZ_2024_CHAOS} extended game-environment feedback to an $n$-player public-goods game and incorporated mutation through replicator-mutator dynamics. Chen et al.~\cite{CGW_2025_PRE}, on the other hand, considered localized environmental feedback in a spatial Prisoner's Dilemma, thereby incorporating the interplay between spatial reciprocity and local game-environment feedback. Several other studies have further explored eco-evolutionary feedback under different payoff structures and evolutionary settings~\cite{Lin2019, BMC_2021_PRE, Mondal2022JoPC, BMC_2023_JOPC, MRC_2024_CHAOS, MSC_2025_PRSA}. These studies primarily describe evolutionary change through population level frequency dynamics of one-shot strategies. This naturally raises the question of how environmental feedback shapes evolutionary dynamics when individuals can instead adapt their behaviour based on their previous interactions.

Previous studies by Hilbe~\emph{et al.}~\cite{hilbe2018nature} have investigated the interplay between memory-based strategies~\cite{Nowak1993, Sigmund_2010_PUP, Hauert1997} and environmental feedback in repeated stochastic games with discrete environmental states. However, to the best of our knowledge, the eco-evolutionary dynamics of reactive strategies---characterized by continuous traits that condition behaviour on an opponent's previous action---have not been studied in settings where the environmental state itself evolves dynamically as a continuous variable. Our framework fills this lacuna by directly coupling the evolution of reactive strategies to continuous environmental feedback.

Also, the evolutionary dynamics studied has usually been short-term replicator dynamics~\cite{Hofbauer1998, Taylor1978, Schuster1983, Schuster1985, PAGE2002, cressman2003book} while ignoring the long-term evolutionary implications of game-environment feedback. In this paper, we focus on direct reciprocity---tendency of individuals strategize in reaction to their opponents' actions---which fosters cooperative behaviour deciding a resource's fate; and investigate how the feedback of this resource state on the strategies further impacts the ToC. To this end, we adopt the formalism of repeated games~\cite{Axelrod1981, VCS_2012_PLOS, Fundenberg1990} among agents endowed with reactive strategies~\cite{NOWAK2006, Nowak2004, VCS_2012_PLOS}; and the long-term evolutionary dynamics is modelled using the adaptive dynamics~\cite{Geritz1997, Geritz1998, Hofbauer1998, Dercole_2008, Kisdi2009, Avila2023}.

Additionally, after having presented the general framework of `adaptive dynamics of eco-evolutionary repeated games', we study the effects of institutional incentives---like rewards and punishments---on this game-environment feedback or eco-evolutionary dynamics.

It is well-known that across the natural world, from microbial symbioses to complex human societies, cooperation is often sustained through mechanisms of reward and punishment. For example, in the cleaner fish mutualism~\cite{Bshary2005}, the cleaner fish \textit{Labroides dimidiatus} benefits by eating ectoparasites off client reef fish. When it cheats by feeding on protective mucus instead, clients punish it by fleeing or switching partners. In the legume--rhizobium mutualism~\cite{Kiers2003}, plants sanction ineffective rhizobia by limiting oxygen supply and reducing nodule size. Similarly, in mycorrhizal associations~\cite{Kiers2011}, plants preferentially reward fungal hyphae that deliver more nutrients with greater carbon allocation. In human societies also, prosocial behaviour is promoted through reward and punishment~\cite{Fehr2002, Boyd2003, bowles2011cooperative}. Individuals may even incur personal costs to punish defectors or free-riders, sustaining cooperation through altruistic or norm-enforcing behavior. Such mechanisms are central to understanding the evolution and maintenance of cooperation.

In evolutionary public goods games (PGG), both reward and punishment can foster cooperation~\cite{Kiyonari2008, Hauert2010, Stoop2018, Banerjee2019}. Punishment enhances cooperation via altruistic enforcement, reciprocity, or exclusion~\cite{CluttonBrock1995, Fehr2002, Boyd2003, Sasaki2013, NagChowdhury2021, Chowdhury2021}. Targeted punishment, in particular, can transform a defector-dominated population into one of global cooperation through feedback loops~\cite{Johnson2015, Greenwood2016, Chen2018}. However, punishment alone may be insufficient or inefficient in maintaining cooperative equilibria, especially under environmental or strategic uncertainty.

A growing body of work has explored how the interplay of reward and punishment shapes social outcomes. Studies have examined their relative importance~\cite{Sigmund2001, Mondal2022JoPC}, roles in second-order punishment and indirect reciprocity~\cite{Lange_2014_BOOK_OUP}, and potential for addressing large-scale coordination challenges like climate change~\cite{Gis2019}. In engineered systems, reward-punishment mechanisms have been proposed to regulate unwanted network traffic by incentivizing cooperation~\cite{Liu2019}. Other work has modelled third-party interventions, showing that the combined use of reward and punishment achieves faster and more robust cooperation than either strategy alone~\cite{Fang2019}. Moreover, in peer-based PGGs, reward was found to facilitate mutual benefit in localized groups, though its role in large-scale cooperation was more limited~\cite{Ozono2020}.

Despite this growing understanding, in the context of this paper an obvious lacuna gap remains: what is the interplay of reward and punishment in promoting cooperation under long-term adaptive dynamics of eco-evolutionary  game dynamics? To the best of our knowledge, the literature lacks a clear demarcation of the maximal effectiveness of reward and punishment in reactive strategy spaces. To this end, in Sec.~\ref{model}, we introduce the mathematical framework, and present the generic results in Sec.~\ref{sec:Linear Stability Analysis}. Subsequently, in the light of those results, we understand the effects of reward and punishment in Sec.~\ref{Donation Game with Institutional Incentives} before concluding in Sec.~\ref{conclusion}.

\section{The Model}\label{model}
Consider a large (mathematically, infinite) unstructured population with access to a common resource. The individuals are randomly paired to allow them play a two-action--two-player game repeated infinitely using reactive (also called memory-half) strategies where each player consider only her opponent's most recent action. The reactive strategies are specified as a 3-tuple~\cite{Hofbauer1998, NOWAK2006, Nowak1990, Nowak1989, Nowak2004}, $E=(y,p,q)$, where the player with the strategy $E$ starts the repeated game with the action C (to cooperate) with probability $y$ or with action D (to defect) with probability $1-y$; furthermore, in every subsequent round of play, the player reciprocates with action C with probability $p$ (or action D with probability $1-p$) if her opponent cooperated in the immediately preceding round  and reciprocates with action C with probability $q$ (or action D with probability $1-q$) if her opponent defected in the immediately preceding round. In this paper, we confine our interest to the cases where $y,$ $p,$ and $q$ lie between zero to one. This renders the strategy space to be effectively an open unit square where each point is a strategy denoted by the coordinate, $(p,q)$. It is well-known that when a player with strategy $(p,q)$ plays with another player with strategy $(p',q')$, the $m^{th}$-level cooperation levels $C_m$  and $C'_{m}$ of the two player, respectively, asymptotically converges to 
\begin{eqnarray}
	&&C=\frac{q+rq'}{1-rr'}\quad{\rm and}\quad C'=\frac{q'+r'q}{1-rr'},\label{eq:cooperation_level}
\end{eqnarray}
where $r\equiv p-q$ and $r'\equiv p'-q'$.

As is customary in the literature of eco-evolutionary games where game-environment feedback is taken into account, the effect of state of the resource is incorporated in the payoff matrix as follows~\cite{weitz2016pnas}
\begin{equation}
	{\sf {\Pi}}
		\equiv
	\begin{bmatrix}
		R       & S \\
		T       & P \\
	\end{bmatrix}
	\equiv
	(1-n)
	\begin{bmatrix}
		R_0       & S_0 \\
		T_0       & P_0 \\
	\end{bmatrix}
	+n
	\begin{bmatrix}
		R_1       & S_1 \\
		T_1       & P_1 \\
	\end{bmatrix},
	\label{eq:matrix}
\end{equation}
where the state of the resource, $n$, is normalized to range from zero to one. Essentially, one is accounting for the fact the payoff matrix, and hence the equilibrium action of the players, can change with the resource-state. Here, the first row and the second row correspond to the focal player's action C and D, respectively; while the first column and the second column correspond to the opponent player's action C and D, respectively.  With this payoff  matrix in mind, after infinitely many repeated interactions, the average payoff of the focal player can easily be shown to be~\cite{Nowak1990, Hofbauer1998},
 \begin{equation}
 	A(E',E)=(\beta - \gamma)C'C - \beta C' + (\alpha + \gamma)C+P, \label{eq:payoff}
 \end{equation}
where, $\gamma\equiv T-R$, $\beta\equiv P - S$ and $\alpha\equiv R-P$.

Our focus is on the long-term evolutionary dynamics which occurs in the mutation-selection dominated regime which is conveniently studied through the theoretical framework of adaptive dynamics~\cite{Hofbauer1998, Nowak2004}. It considers the continuous evolution of strategies in response to small mutations and the resulting fitness landscape. Specifically, one considers strategy-wise homogeneous population where only a single mutant arises at any given instant, and it is either driven to extinction or fixed in the population before the appearance of the next mutant. If the mutant strategy is nearly identical to that of the resident, then the discrete-step substitution sequence can be approximated by the continuous trajectory of adaptive evolution in trait space~\cite{Hofbauer1990, Nowak1990, Dieckmann1996JMB} (i.e., $p$-$q$ space in our context).

\begin{figure*}[htbp]
	\centering	
	\includegraphics[scale=0.835]{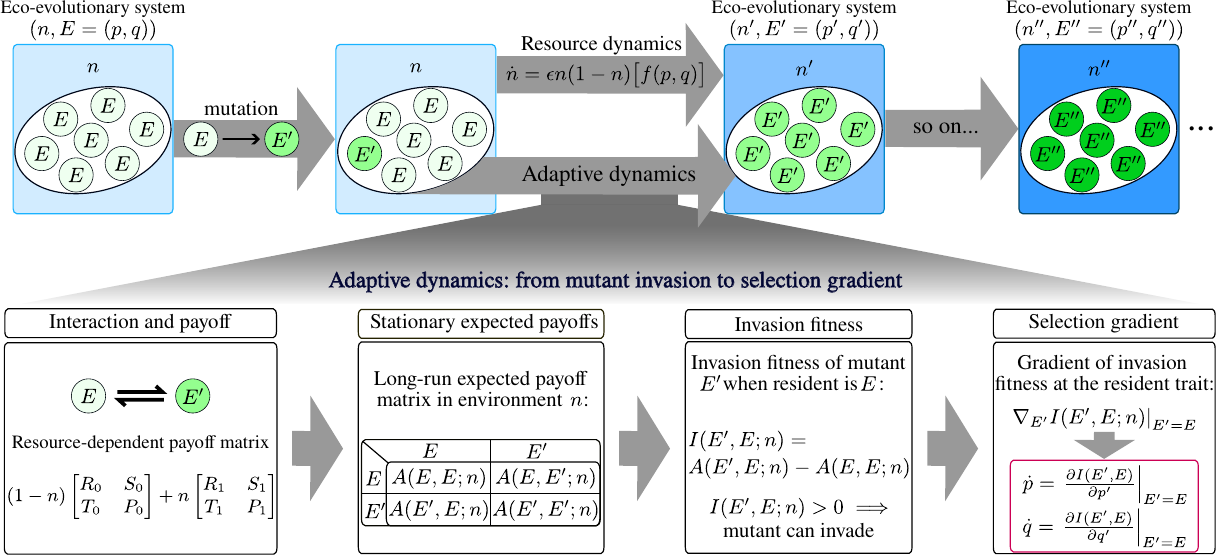}
	\caption{Schematic representation of the eco-evolutionary repeated-game framework. Individuals are characterized by reactive strategies $E=(p,q)$. For a given resource state $n$, repeated interactions determine the stationary cooperation probabilities and the corresponding long-run expected payoffs. The interaction payoffs depend on the resource state through the effective payoff matrix $(1 -n)\left[\begin{smallmatrix}R_0&S_0\\T_0&P_0\end{smallmatrix}\right] +n\left[\begin{smallmatrix}R_1&S_1\\T_1&P_1\end{smallmatrix}\right]$, thereby coupling strategic interactions to the environment.	Under the adaptive-dynamics framework, a rare nearby mutant $E'=(p',q')$ arises in the monomorphic resident population $E=(p,q)$. Its evolutionary success is determined by the invasion fitness obtained from the difference between the mutant--resident and the resident--resident long-run payoffs obtained via repeated games. The selection gradient of the invasion fitness with respect to the mutant traits, evaluated at $E'=E$, determines the adaptive evolution of the resident traits $p$ and $q$. Simultaneously, cooperative behaviour feeds back on the renewable resource through the resource dynamics, $\dot n=\epsilon n(1-n)f(p,q)$, yielding the coupled eco-evolutionary dynamics of $(p,q,n)$. Successive rare mutations and selection therefore generate long-term changes in the resident strategy---yielding a sequence of monomorphic populations---together with the associated resource state.}
	\label{fig:model_schematic}
\end{figure*}

To formalize the framework, one defines the relative fitness $I(E',E)$, also known as invasion fitness~\cite{Metz1992tee} of a mutant strategy $E'$ against the resident strategy $E$. In this paper, the strategy $E$ is basically the pair $(p,q)$. During evolution, these quantities change and therefore constitute the dynamical variables of adaptive dynamics. To determine their direction of evolution, we consider a rare mutant strategy $E'=(p',q')$ that differs only slightly from the resident strategy.
	
The invasion fitness of the mutant is defined as the difference between its payoff against the resident and the payoff obtained by the resident:	
\begin{eqnarray}
	&&I(E',E)\equiv A(E',E)-A(E,E).
	\label{eq:Invasion}
\end{eqnarray}
A positive value of $I(E',E)$, therefore indicates that the nearby mutant has an advantage over the resident.
At a given resource state, use of Eq.~(\ref{eq:matrix}) and Eq.~(\ref{eq:payoff}) transforms the definition of $I(E',E)$ to
\begin{eqnarray}
I(E',E)=[(1-n)(\beta_{0} - \gamma_{0})+n(\beta_{1} - \gamma_{1})] r \left(C' -\frac{1}{1-r}\right)^2.\,\,
	\label{eq:Invasion1}
\end{eqnarray}

In the theory of adaptive dynamics~\cite{Nowak1990,Hofbauer1998}, rare mutations appear in the neighbourhood of resident traits, and the local direction of selection is obtained by differentiating the invasion fitness with respect to the corresponding mutant trait and then evaluating the derivative at $E'=E$. This direction corresponds to what is best from the individual mutant's
point of view. Accordingly, the evolutionary dynamics of the resident traits are written as 
\begin{subequations}
	\begin{align}
		\dot p &=
		\left.
		\frac{\partial I(E',E)}{\partial p'}
		\right|_{E'=E}=\frac{q}{(1-r^2)(1-r)^2}~F(p,q,n),
		\label{eq:AD_p}\\
		\dot q &=
		\left.
		\frac{\partial I(E',E)}{\partial q'}
		\right|_{E'=E}=\frac{(1-p)}{(1-r^2)(1-r)^2}~F(p,q,n),
		\label{eq:AD_q}
	\end{align}
\end{subequations}
where the second equalities in each of the equations come from using Eq.~(\ref{eq:Invasion1}).
\noindent Here, $F(p,q,n)$ collects the common payoff-dependent contribution to the two selection gradients. It is not introduced as an independent or prescribed function; rather, it follows directly from the differentiation of the invasion fitness. It is given by  
\begin{eqnarray}
&&F(p,q,n)\equiv(1-n)\big[(\beta_0-\gamma_0)q(1+r)+(\alpha_0 + \gamma_0)r(1-r)\nonumber\\
&&\phantom{xxx}-\beta_0 (1-r)\big]+n\big[(\beta_1-\gamma_1)q(1+r)+(\alpha_1 + \gamma_1)r(1-r)\nonumber\\
&&\phantom{xxx}-\beta_1 (1-r)\big], \label{eq:F()}
\end{eqnarray}
where $\gamma_{0} \equiv T_0 -R_0$, $\beta_{0}\equiv P_0 - S_0$, $\alpha_0 \equiv R _0 - P_0$ , $\gamma_{1} \equiv T_1 -R_1$, $\beta_{1}\equiv P_1 - S_1$, and $\alpha_1 \equiv R _1 - P_1$.

Thus, $p$ and $q$ are evolutionary variables, whereas the parameters $\alpha_{0}$, $\alpha_{1}$, $\beta_{0}$, $\beta_{1}$,  $\gamma_{0}$, and $\gamma_{1}$ specify the payoff structure at the two limiting resource states. The function $F(p,q,n)$ links these payoffs and the current resource state to the direction of selection. In particular, its sign determines the local direction in which both reactive traits are selected to change, while the prefactors $q$ and $1-p$ determine their respective rates of response.

We pause a bit to clarify some basic aspects of the model. First, since $p$ and $q$ are conditional probabilities, individual actions are therefore stochastic, but we are not interested in explicitly tracking individual random action sequences; instead, the dynamics of our eco-evolutionary system, which uses the asymptotic cooperation probabilities and the corresponding expected long-run payoffs, deterministically tracks the evolution of traits $p$ and $q$. 

Second, it might appear we are dealing with two infinite timescales---one in the repeated game and the other in the adaptive dynamics where a mutant takes an infinite time to fixate in the limit of infinite population size. However, the model requires and strictly assumes that the timescale of repeated games to be much shorter than that of the adaptive dynamics. In this context, we would like to emphasize that the need of an infinite time horizon in repeated games is purely mathematical---in the infinite limit, unique analytical forms of cooperation probabilities and expected long-run payoffs exist. Then the question is how justified it is to use these forms in realistic situations where neither the population size would be infinite, nor the interactions would be repeated for an infinite number of rounds. The justification of the applicability of evolutionary game theoretical results pertaining random matching events in finite population has been long studied in detail~\cite{fudenberg1998theory} and the conclusion---especially in the light of experiments~\cite{F96EJ} on Prisoner's Dilemma game---is that when the population size is approximately six or more, the `small group effects' start becoming ignorable; thus, when population size is in merely few tens (especially in the context of $2\times 2$ games as are being used in this paper), the mathematical results of infinite population limit are very good theoretical estimates. As far as the reality of a finite number of interactions in a repeated game is concerned, the justification for choosing an infinite number of rounds in our paper is rather straightforward: the cooperation probabilities and the payoffs reach 95\% of their asymptotic values in less than ten interactions. Naturally, working with an infinite number of rounds is not only a mathematical convenience but also a valid approximation widely used in the literature. 

Finally, adaptive-dynamics models generally involve a separation between fast and slow evolutionary processes~\cite{Nowak2004}. On the fast timescale, a mutant competes with the resident population, and its growth or decline determines whether it can invade. During this process, the appearance of additional mutants is neglected. On the slow timescale, new mutants arise sufficiently infrequently that each mutant--resident competition is effectively completed before the next relevant mutation appears. In the reduced adaptive-dynamics formulation, the fast competition is not represented explicitly; its effect is incorporated through the mutant's invasion fitness, derived from the underlying microscopic model~\cite{Dercole_2008}. In the present model, invasion fitness is calculated by comparing the expected payoff of a rare mutant with that of the resident under the repeated-game dynamics. We approximate these payoffs using their asymptotic values, justified by the rapid convergence discussed above. The timescale associated with the appearance of new mutants is assumed to be much longer than both the repeated-game dynamics and the ensuing mutant--resident competition. Therefore, the conclusions of the model are conditional on this timescale-separation assumption, as is standard in adaptive-dynamics theory~\cite{Dercole_2008}.

It is easily realized that the invasion of a mutant strategy can influence the state of common resource, which in turn can modulate the fitness landscape of the population. If we consider that the resource undergoes a logistic growth such that its growth is sustained by the propensity of cooperation in the population (while the defectors only degrade the resource), then its dynamics as the form
\begin{equation}
	\dot{n}=\epsilon n(1-n)f(p,q).
	\label{eq:resource_gen}
\end{equation}
Here, $\epsilon$ (a positive real number) characterized the relative speed of the resource-state dynamics compared to that of the traits' and $f(p,q)$ may be modelled in two different ways: In one approach, we consider the state of the resource to be dependent on the players' trait values (i.e., their conditional probabilities of cooperating or defecting), while the second approach links the state of the resource to the asymptotic cooperation level. 

Thus, in the first approach, one may define a simple form $f(p,q)={a_r}[\theta \cdot p-(1-p)]+{a_g}[\theta \cdot q-(1-q)]$\label{subeq: recipro_generosity} so that the feedback of reciprocity $(p)$ and generosity $(q)$ traits on the resource is rather direct. The effects of reciprocity and generosity~\cite{Sadhukhan2022jop, Kurokawa2019, Kurokawa2019, Kurokawa2010, Park2017} are scaled by the positive parameters $a_r$ and $a_g$, respectively. For simplicity, and to reduce number of parameters, we assume that these effects are symmetric, i.e., $a_r=a_g$. Investigating the consequences of asymmetric effects $a_r\neq a_g$ represents an interesting direction for future research. The positive parameter~\cite{weitz2016pnas} $\theta$ is the ratio of the enhancement rate to degradation rate due to adoption of actions C and D. Alternatively, as the other approach, one could define $f(p,q)=\theta \cdot C-(1-C)$ where growth rate is effected by the asymptotic cooperation level (a robust measure of propensity of cooperation in the population) and the decay rate is effected by the asymptotic defection level. Note that here $C$ is a measure in monomorphic population; hence, using Eq.~(\ref{eq:cooperation_level}) with $p=p'$ and $q=q'$, we get $C=\frac{q}{1-r}$. The overall eco-evolutionary feedback mechanism underlying the model is illustrated schematically in Fig.~\ref{fig:model_schematic}.

Interestingly, when $\theta=1$, i.e., the cooperators and the defectors impact the resource symmetrically, the two forms of $f(p,q)$ are same except for a positive multiplicative factor, $\frac{1}{2(1-r)}$. This leads to unchanged conclusions about the existence of fixed points and their nature of stability in $p$-$q$-$n$ dynamics. 

Moreover, the dynamical outcomes of the strategy-resource remains qualitatively unchanged for all values of $\epsilon$. Thus, henceforth we set $\theta=1$ and $\epsilon=1/2$ to arrive at the following simple form of the resource dynamics:
\begin{eqnarray}
	\dot{n}={n(1-n)}\left[(p+q)-1\right].
	\label{eq:dynamics_n}
\end{eqnarray}
In Appendix~\ref{Appendix_F}, we have tested the model's sensitivity with respect to the parameter $\theta$.
\section{Linear Stability Analysis}{\label{sec:Linear Stability Analysis}}
The $p$-$q$-$n$ dynamics is, thus, governed by Eqs.~(\ref{eq:AD_p}), (\ref{eq:AD_q})~and~(\ref{eq:dynamics_n}). We now perform its stability analysis.
\subsection{Fixed points and saturated boundary points}
\begin{figure*}[htbp]
	\centering	
	\includegraphics[width=165mm, height=190mm]{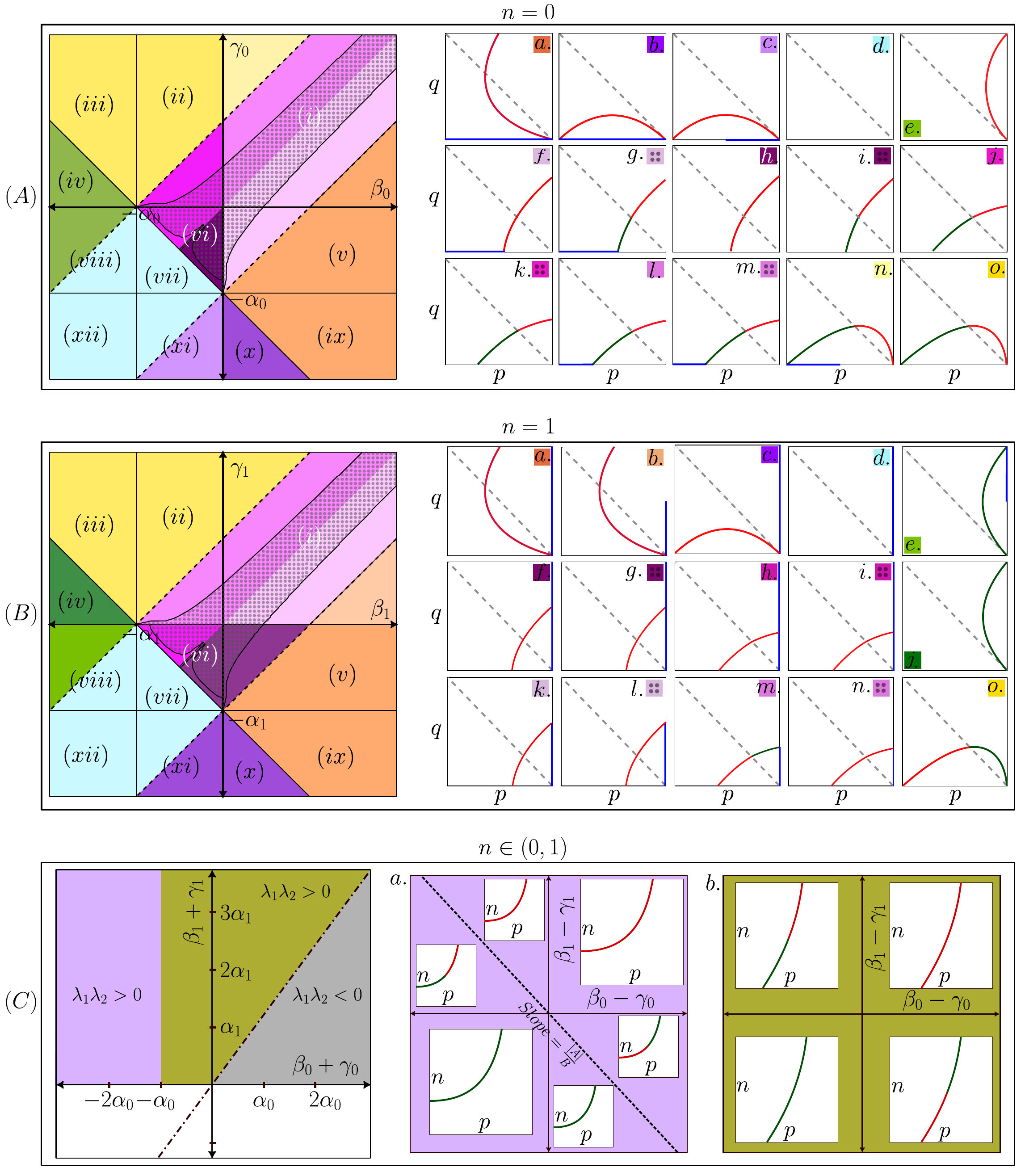}
	\caption{
		An exhaustive depiction of parameter space detailing the stability of fixed-points. Panels $(A)$ and $(B)$ correspond to fixed-point lines arising from the intersections of the surfaces $n=0$ and $F(p,q,n)=0$, and $n=1$ and $F(p,q,n)=0$, respectively. 
		The right-hand subpanels (a)–(o) illustrate the nature of stability of internal fixed-point structures corresponding to coloured parameter regions. Stable fixed-point curves appear in green, unstable fixed-point curves in red, and saturated fixed-point lines in blue.
		The dotted parameter region exhibits a distinctive feature, where the fixed-point continuum contains both the stable and the unstable segments. Panel $(C)$ shows the stability behaviour of the internal fixed-point line corresponding to the intersection of the surfaces $p+q=1$ and $F(p,q,n)=0$. Here $\lambda_1\lambda_2$ is product of to non-zero eigenvalues (see Eq.~(\ref{eq:conjugate})). The dot-dashed line corresponds to \(\beta_{1}+\gamma_{1}=\frac{\alpha_1}{\alpha_0}(\beta_{0}+\gamma_{0})\) and in sub-panel $a$, the slope \(=\frac{|A|}{B}=-\frac{|\alpha_0 + \beta_{0} +\gamma_{0}|}{\alpha_1 + \beta_{1} +\gamma_{1}}\). Twelve games different classes of distinct games, corresponding to different parameter regimes, are depicted in panels $(A)$ and $(B)$. They are labeled by $(i)$ to $(xii)$: (i) Prisoner’s dilemma, (ii) Chicken games, (iii) Leader games, (iv) Battle of sexes, (v) Stag-hunt, (vi) Harmony I, (vii) Harmony II, (viii) Deadlock II, (ix) Coordination I, (x) Coordination II, (xi) Harmony III, and (xii) Deadlock I.}
	\label{fig:boat4}
\end{figure*}

First we note that for this system with free parameters $\alpha_0, \alpha_1$, $\beta_0, \beta_1, \gamma_0$ and $\gamma_1$, there are six different types of sets of phase points of interest. Three of them are sets of non-isolated fixed points:
\begin{enumerate}[label=(\roman*)]
	\item non-isolated fixed points at the intersection of surfaces---$n=0$  and $F(p,q,n)=0$~(see Fig.~\ref{fig:boat4}($A$)) and Appendix \ref{Appendix_A} for the stability property of these fixed points),
	\item non-isolated fixed points at the intersection of surfaces---$n=1$ and $F(p,q,n)=0$~(see  Fig.~\ref{fig:boat4}($B$)) and Appendix \ref{Appendix_B} to get the stability property of these fixed points),
	\item \label{fp:internal}non-isolated fixed points at the intersection of surfaces---$p+q=1$ and $F(p,q,n)=0$~(see  Fig.~\ref{fig:boat4}($C$)) and Appendix \ref{Appendix_C} to get the stability property of these fixed points).
\end{enumerate}
Remaining two are saturated points~\cite{LHN_2023_PLOS}:
\begin{enumerate}
	\item[(iv)] (strictly) saturated boundary points---$(p,q,n)=(p,0,0)$,
	\item[(v)] (strictly) saturated boundary points---$(p,q,n)=(1,q,1)$.
\end{enumerate}
Note that nature of these points requires that the closure of the open reactive strategy space be considered in order to investigate the saturation behaviour in the limiting region approaching the boundary: The saturated boundary points lies on the boundary of the closure of phase space. In this context we recall that on the boundary of $[0,1]^3$ (i.e., set of all points $(p,q,n) \in [0,1]^3$---the unit phase cube---such that exactly one coordinate equals to $0$ or $1$) a point is (strictly) saturated if the phase velocity at that point tends to push it out of the unit phase cube; in other words, a strictly saturated boundary point is the $\omega$-limit of some phase trajectory emanating in the interior. 

First consider the possibility of saturated boundary points at $n=0$ plane. The condition to reach the $n=0$ plane is $\dot{n}<0$ which implies the following constraints on the possible reactive strategies $p+q<1$ (see Eq.~(\ref{eq:dynamics_n})). Therefore, at $n=0$ plane, the trajectories cannot reach $p=1$ and $q=1$. A glance at Eqs.~(\ref{eq:AD_p}) and ~(\ref{eq:AD_q}) further suggests at $q=0$ with $F(p,q,n=0)<0$, $\dot{q}<0$ (and $p$ does not change). Thus, points $(p,q,n)=(p,0,0)$ tend to get pushed out away from the phase space; and thus, quality as saturated boundary points. The condition for those to exist is: if $\alpha_0+\gamma_{0}>0$, then $p<\frac{\beta_{0}}{\alpha_0+\gamma_{0}}$; if $\alpha_0+\gamma_{0}<0$, then $p>\frac{\beta_{0}}{\alpha_0+\gamma_{0}}$; and \textcolor{black}{if $\alpha_0+\gamma_{0}=0$, it follows that $\beta_{0} > 0$ for any $p$}~(see Eq.~(\ref{eq:F()})).

Similarly, saturated boundary points can also appear at $n=1$ plane. In this case the condition for a trajectory to reach the $n=1$ plane is $p+q>1$  (see Eq.~(\ref{eq:dynamics_n})); the trajectories cannot reach $p=0$ and $q=0$. Again inspection of Eqs.~(\ref{eq:AD_p}) and ~(\ref{eq:AD_q}) suggests at $p=1$ with $F(p,q,n=0)>0$, $\dot{p}>0$ (and $q$ does not change). Thus, $(p,q,n)=(1,q,1)$ are saturated boundary points which tend to get pushed out away from the phase space. The condition for them to exist is: $\frac{\alpha_1+\beta_{1}-\gamma_{1}}{(\alpha_1+\beta_{1})}>q$ if $\alpha_1+\beta_{1}>0$; $\frac{\alpha_1+\beta_{1}-\gamma_{1}}{(\alpha_1+\beta_{1})}<q$ if $\alpha_1+\beta_{1}>0$; and \textcolor{black}{if $\alpha_1+\beta_{1}=0$, it follows that $\gamma_{1} > 0$ for any $q$} (see Eq.~(\ref{eq:F()})). 

\begin{figure}[htbp]
	\centering	
	\includegraphics[width=0.48\textwidth]{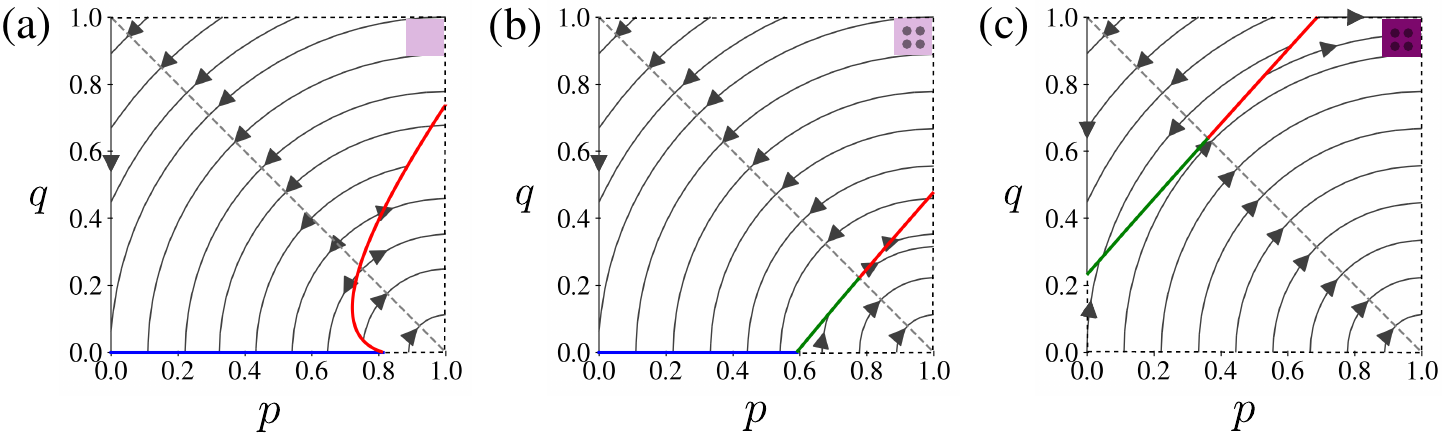}
	\caption{Representative phase portraits illustrating the stability of the continuum of non-isolated fixed points at the depleted resource state, $n=0$, for different parameter regimes: The shaded square in the upper-right corner indicates the corresponding  parameter region as listed in Fig.~\ref{fig:boat4}. The grey curves with arrows indicate the direction of the flow in the $(p,q)$ plane, while the coloured curves---green means stable, red means unstable and blue means saturated points---denote the corresponding fixed-point branches and saturated boundary points.}
	\label{fig:phase_dia}
\end{figure}

To furnish clearer picture of the dynamics around the stable fixed points, we present a few representative phase portraits (projected on $p$-$q$ subspace) in Fig.~\ref{fig:phase_dia} around the fixed points at $n=0$ corresponding to  Fig.~\ref{fig:boat4}(A)f, Fig.~\ref{fig:boat4}(A)g and Fig.~\ref{fig:boat4}(A)i. Nearby trajectories are directed toward the locally stable portions of the fixed-point set, whereas they are repelled from the unstable portions.

\subsection{Limit cycle via Hopf bifurcation}\label{para: hopf_bifurcation}
Besides the equilibrium evolutionary outcomes, there may be a possibility of periodically oscillating evolutionary outcomes which would correspond to existence of limit cycles. Emergence of the Hopf bifurcation~\cite{Marsden1976}, in particular, marks a critical transition where a fixed point changes its nature of stability leading to limit cycle oscillations as system parameters vary. 
In our framework, a Hopf bifurcation can arise only around the internal fixed points of type~\ref{fp:internal}. One of the three eigenvalues~(see Appendix~\ref{Appendix_C}) associated with the internal non-isolated fixed points is always zero, reflecting the neutral direction along the fixed points. The remaining two eigenvalue ($\lambda_1$ and $\lambda_2$) determine the local stability in the transverse directions of the fixed points. They are
\begin{equation}\label{eigen}
	\lambda_{1}=\frac{L - \sqrt{L^2-4\eta}}{2}\,\,\, {\rm and}\,\,\, \lambda_{2}=\frac{L + \sqrt{L^2-4\eta}}{2},
\end{equation}
where,
\begin{subequations}
	\begin{align}
		\eta &\equiv
		\frac{n(1-n)}{4pq}
		\frac{(\beta_{1}+\gamma_{1})\alpha_0 - (\beta_{0}+\gamma_{0})\alpha_1}
		{(1-n)(\beta_{0}+\gamma_{0}+2\alpha_0)+n(\beta_{1}+\gamma_{1}+2\alpha_1)},
		\label{eq:hopf_eta}
		\\[6pt]
		L &\equiv 
		\frac{1}{4q}\left[(\beta_0-\gamma_0)(1-n) + (\beta_1-\gamma_1)n\right].
		\label{eq:hopf_L}
	\end{align}
\end{subequations}
When $\eta=\lambda_1\lambda_2>0$, the condition $L=0$ defines a Hopf-bifurcation threshold. 

From the condition $L=0$ (see Eq.~(\ref{eq:hopf_L})), we obtain:
\begin{equation}\label{eq:n_cap}
	n=\hat{n}\equiv
	\frac{\beta_0-\gamma_0}
	{(\beta_0-\gamma_0)-(\beta_1-\gamma_1)}\in(0,1).
\end{equation}
From this expression, it is clear that the Hopf-bifurcation may happen about an internal fixed point occurring at the intersection of $p+q=1$ and $F(p,q,n)=0$ as listed in Sec.~\ref{sec:Linear Stability Analysis}A.

We note in Fig.~\ref{fig:boat4}C whose construction details are in Appendix~\ref{Appendix_C}1, the olive and the violet regions have $\lambda_{1}\lambda_{2}>0$, i.e., $\eta>0$. Therefore, the Hopf-bifurcation may arise in these two coloured regions. For $\lambda_1\lambda_2>0$, the transverse stability is determined by the sign of $\lambda_1+\lambda_2=L$. Specifically, $L<0$ corresponds to transverse stability, whereas $L>0$ corresponds to transverse instability. From Eq.~(\ref{eq:hopf_L}), we see that the condition $L=0$ depends explicitly on $(\beta_0-\gamma_0)$ and $(\beta_1-\gamma_1)$. It is therefore useful to represent the parameter space in the $(\beta_0-\gamma_0)$-$(\beta_1-\gamma_1)$ plane as done separately for the two regions in Fig.~\ref{fig:boat4}C(a) and Fig.~\ref{fig:boat4}C(b).

Three types of transverse stability structures of the non-isolated fixed-point continuum are observed: $(i)$ a completely stable continuum, shown in green; $(ii)$ a completely unstable continuum, shown in red; and $(iii)$ a continuum containing both stable and unstable segments, shown in green and red, respectively. Thus, the transverse stability structure of the fixed-point continuum depends strongly on the parameters, particularly on $(\beta_0-\gamma_0)$ and $(\beta_1-\gamma_1)$. It is obvious that whenever we change the parameters in such a fashion that a stable fixed point on the continuum becomes unstable, a Hopf-bifurcation should be witnessed. The details of the appropriate parameter change leading to the bifurcation are presented in Appendix~\ref{Appendix_C}2.

\begin{figure}[htbp]
	\centering	
	\includegraphics[scale=0.45]{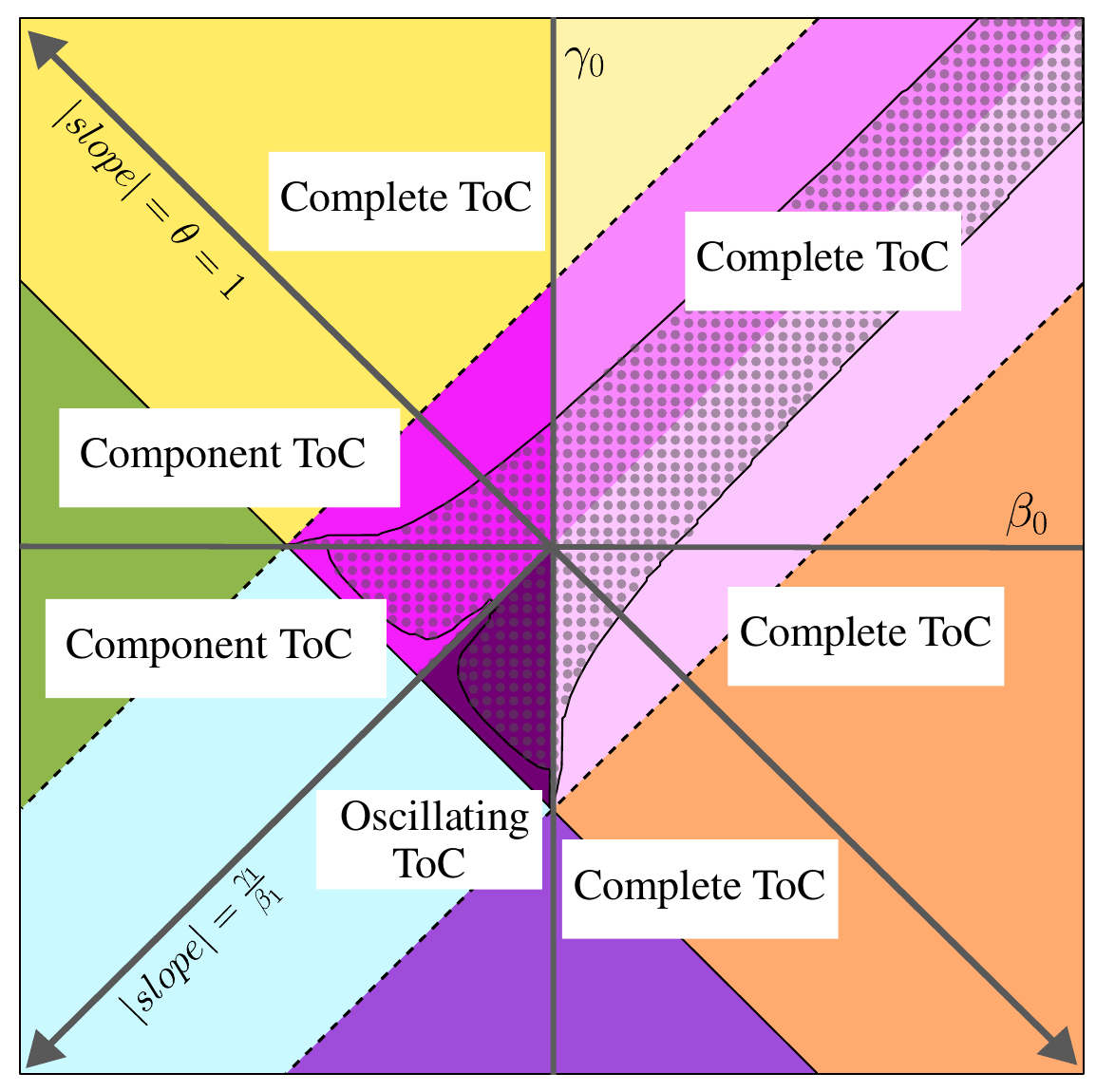}
	\caption{A unified comparison of short- and long-term evolutionary dynamics across the $\beta_{0}$--$\gamma_{0}$ parameter space. Seven regions of the parameter space are demarcated by the grey solid axes and the arrowed lines with absolute slopes $|\mathrm{slope}|=\theta$ and $|\mathrm{slope}|=\gamma_{1}/\beta_{1}$. These regions---classified here as Complete ToC, Component ToC, and Oscillating ToC---illustrate the possible outcomes of short-term eco-evolutionary dynamics mediated by replicator dynamics. These regions should be compared with Figure~\ref{fig:boat4} in its entirety; hence, for the sake of illustration, we have put the backdrop of Fig.~\ref{fig:boat4}(A) so that one can do a comparison with the long-term adaptive dynamics with $n=0$ fixed point in focus.
}
	\label{fig:Comparison}
\end{figure}

\subsection{Comparison Between Long-Term and Short-Term Evolutionary Dynamics}
From an evolutionary perspective, our analysis focuses on the long-term evolutionary process which is characterized by adaptive dynamics with additional feedback emanating from evolving resource state. We note that the reactive strategy space always contains a set of strategies in the vicinity of the \textit{tit-for-tat} (TFT) strategy ($y=1$, $p=1$, $q=0$) which can avert the ToC even for a defection-dominant strategy at both the replenished ($n=1$) and scarce ($n=0$) resource states. This result holds across multiple game classes, as shown in Appendix~\ref{appendix_E} and Fig.~\ref{fig:Bistability_across_parameters}. More specifically, the resource persists in the long run when the population evolves toward Generous Tit-for-Tat (GTFT)-like strategies~\cite{NS1993} (which are, unlike TFT, occasionally cooperating even after the opponent has defected), largely independent of the initial resource state. 
Furthermore, for sufficiently high initial resource states, resource persistence can also occur for a broader class of non-generous strategies $(p,~q=0)$, as illustrated in Appendix~\ref{appendix_E} and Fig.~\ref{fig:Bistability_across_parameters}. Moreover, the nonlinear feedback dynamics of strategy and resource can lead to a limit cycle kind of oscillation where the system ultimately settles down to an attractor where the strategy as well as resource state oscillates periodically due to the repeated invasion of the resident by mutants. 

To motivate the significance of our results let us compare with the findings of a recent paper by Weitz \textit{et al.}~\cite{weitz2016pnas} where they considered eco-evolutionary feedback dynamics involving one-shot game (instead of repeated interaction) and studied how the frequency  ($x$) of the cooperators and resource state co-evolve. The evolution of cooperator fraction followed replicator dynamics---a comparatively short-term dynamics. Fig.~\ref{fig:Comparison} captures the difference in evolutionary outcome between the short-term and the long-term dynamics. Short-term dynamics with one-shot game lead to three possible states: complete ToC ($n=0$), component ToC ($0<n<1$), and oscillatory ToC (periodically changing $n$); there is no possibility of complete aversion of ToC ($n=1$). In the long-term dynamics with repeated game, complete aversion of ToC is always possible for some initial choice of reactive strategy; however, for other strategies, the dynamics can lead to either complete, component ToC.
   
At first glance, this result may seem counterintuitive, since one might expect the outcomes in the short-term dynamics to be amplified in the long-term dynamics, given that the latter represents the cumulative effect of many short-term evolutionary steps. However, this contradiction is easily resolved \textcolor{black}{once we consider the long-term eco-evolutionary feedback dynamics of (history-independent) reactive strategies, corresponding to $p=q=x$ in our setting. In this case, the long-term dynamics exactly mirrors the evolutionary dynamics of the strategy frequency $x$ in the model of Weitz \textit{et al.}, which describes short-term evolution in a population with a fraction $x$ of cooperators and $1-x$ of defectors under repeated mutant invasion. To see it explicitly, we have to substitute $p=q=x$ and $p'=q'=x'$ in Eq.~(\ref{eq:cooperation_level}), which gives $C=x$ and $C'=x'$. Furthermore, the payoff Eq.~(\ref{eq:payoff}) of the focal player after infinitely repeated interaction simplifies to
	\begin{equation}
	A=(\beta - \gamma)x'x - \beta x' + (\alpha + \gamma)x+P.
	\end{equation}
	Therefore, the trait substitution sequence with eco--evolutionary feedback, incorporating resource dynamics given by Eq.~(\ref{eq:resource_gen}) under both resource-evolution approaches, leads to the following dynamical equations in the $x$--$n$ space:
	\begin{subequations}
		\begin{eqnarray}\label{eq:p=q}
		&&\dot{x}=\frac{\partial S}{\partial x'}\bigg|_{\{x'=x\}}=r(x)x(1-x)\left[r_1(x,n)-r_2(x,n)\right]~~\\&&
		\dot{n}=\epsilon n(1-n)\left[\theta x-(1-x)\right],
		\end{eqnarray}	
	\end{subequations}
where, $r(x)=1/x(1-x)$, $r_1(x,n)=[R_0(1-n)+R_1n]x+[S_0(1-n)+S_1n](1-x)$, and $r_2(x,n)=[T_0(1-n)+T_1n]x-[P_0(1-n)+P_1n](1-x)$. Note that the form of the above equation is similar to the short-term eco-evolutionary feedback dynamics considered by Weitz \textit{et al.}, except for a  positive multiplicative factor $r(x)$ appearing in front of Eq.~(\ref{eq:p=q}), which is finite since $x=0$ or $1$ is not allowed in the reactive strategy space.	
One can easily verify that the stability of all fixed points is the same as in the short-term dynamics, except for the four boundary fixed points $(0,0)$, $(0,1)$, $(1,0)$, and $(1,1)$, which are not admissible in the reactive strategy space without boundaries as adopted in this paper. However, these boundary points act as saturation boundaries, and the conditions for reaching saturation points are exactly the same as the corresponding stability conditions in the short-term dynamics. Therefore, the qualitative conditions for ToC prevention remain unchanged relative to the short-term dynamics, with the only minor difference that oscillatory ToC~\cite{weitz2016pnas} is replaced by stable limit cycle oscillations.} Main crux, however, is that allowing for the full reactive strategy space---which is infinitely larger than the binary choice of cooperating or defecting---introduces additional degrees of freedom that facilitate complete prevention of the ToC.

Although the proposed eco-evolutionary framework is sufficiently general to accommodate any symmetric $2\times 2$ game, it is too dense to render clearly understand the physical conditions under which ToC can be averted. To this end, the specific consideration of the Donation Game (DG)~\cite {Sigmund_2010_PUP}---a canonical yet straightforward representation of social dilemmas in which defection is individually optimal but socially costly---can simplify the route of our understanding. Moreover, thanks to the general framework, the comparative role of reward and punishment in averting ToC, can be closely studied in the context of DG.

\begin{figure*}[ht]
	\centering	
	\includegraphics[scale=0.8]{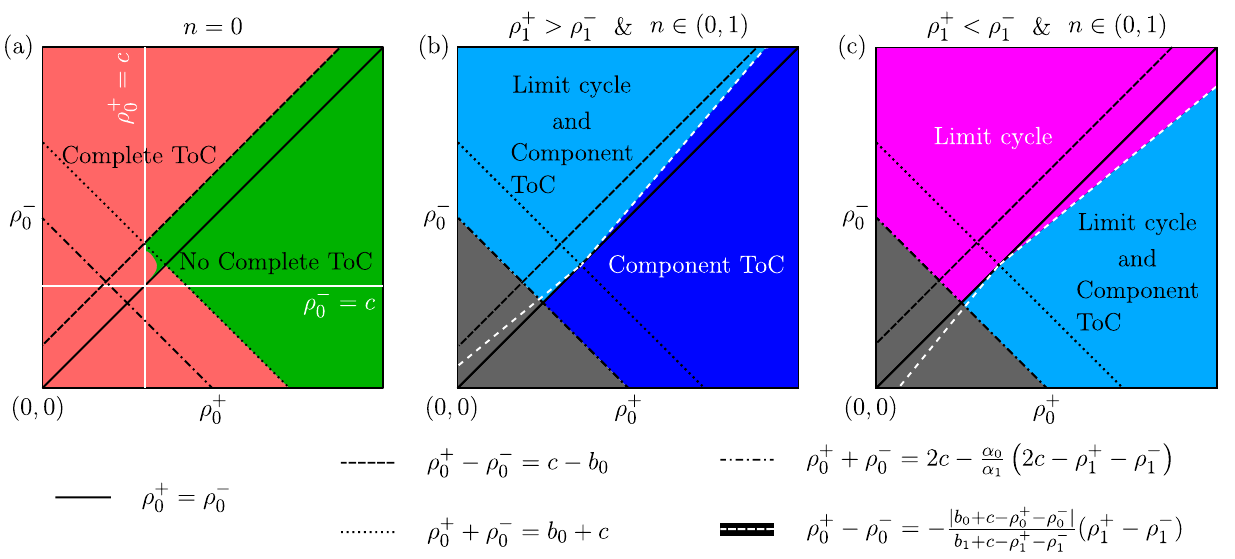}
	\caption{ToC and its prevention in the reward $(\rho_0^+)$--punishment $(\rho_0^-)$ parameter space. Subplot~(a) shows the stability properties of the fixed points on the $n=0$ plane. The green region corresponds to parameter values for which the fixed points are either completely unstable or do not exist; consequently, for any initial choice of reactive strategy, complete ToC is avoided in this region. In contrast, this is not possible in the light-red shaded region, where a stable fixed point exists on the $n=0$ plane.
	Subplots~(b) and~(c) show the stability properties of internal fixed points over the admissible range of reward and punishment. The light-blue region represents internal fixed points comprising both stable and unstable segments, while the dark-blue region corresponds to completely stable internal fixed points. The grey and magenta regions denote saddle and completely unstable internal fixed points, respectively. The corresponding dynamical outcomes in these regions are also indicated.}
	\label{fig:reward_punishment_n0n1}
\end{figure*}

\begin{figure}[ht!]
	\centering		
	\includegraphics[width=0.46\textwidth]{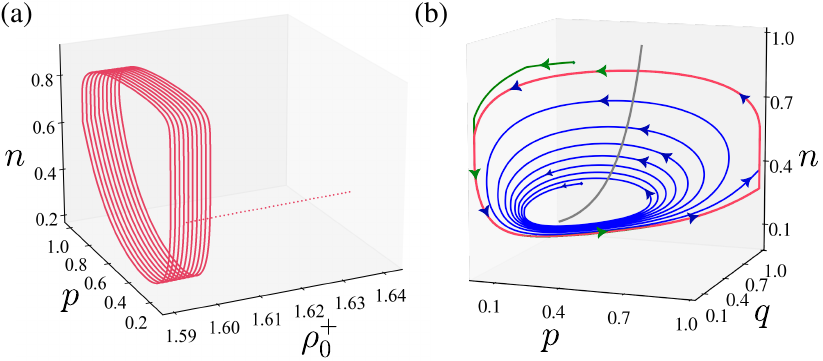}
	\caption{Emergence of oscillatory dynamics through a Hopf bifurcation in the presence of reward and punishment. Subplot~(a) shows the bifurcation scenario obtained by varying the reward parameter $\rho_0^+$ while keeping the punishment parameter $\rho_0^-=1.6$ fixed at the critical. For $\rho_0^+>1.6$, the system approaches the fixed-point branch, whereas decreasing $\rho_0^+$ through the critical value leads to sustained oscillatory dynamics for $\rho_0^+<1.6$. The three-dimensional representation shows with visual clarity the oscillatory trajectories in projected $(p,n)$ subspace.	 Subplot~(b) illustrates a limit cycle, for the representative value $\rho_0^+=1.5$. The dynamics are shown in the $(p,q,n)$ phase space for two different initial conditions, $(p=0.19,\,q=0.81,\, n=0.28)$ and $(p=0.15,\,q=0.80,\,n=0.85)$, whose respective blue and green trajectories approach the limit cycle shown in red, which surrounds the grey unstable fixed-point continuum. The remaining parameters used in these plots are $\rho^-_{0}=1.6$, $\rho^+_{1}=\rho^-_{1}=0.5$, $b_1=2.2$, $b_0=2.0$, and $c=1$.}
	\label{fig:hopf_bifuraction_reward_punishment}
\end{figure}  

\section{An Application: Donation Game with Institutional Incentives}\label{Donation Game with Institutional Incentives}
The DG---a simplified formulation of the PD game---offers one of the most transparent and analytically tractable frameworks for studying non-cooperative interaction systems observed in nature, where individuals face incentives to exploit shared resources or social partners~\cite{Nowak2005, McAvoy2017, LHN_2023_PLOS}. Its explicit cost-benefit structure makes it particularly well-suited for investigating the conditions under which cooperation can avert the ToC. In this game, a player incurs a cost $c$ whenever perform a cooperative act (and is then called a cooperator). In contrast, defectors obtain the benefit $b$ ($b > c > 0$) from interaction with cooperators without paying any cost.
While mutual cooperation yields the socially optimal outcome, the temptation to defect and obtain benefits without incurring costs drives the population toward behaviour that is individually advantageous yet collectively suboptimal. The payoff matrix of the DG is, thus, represented as follows:
\vskip0.5cm
	\begin{tabular}{c||c|c}
		&C (Cooperate) & D (Defect) \\
		\hline\hline
		C (Cooperate) & $b - c,\ b - c$ & $-c,\ b$ \\
		\hline
		D (Defect) & $b,\ -c$ & $0,\ 0$ \\
		\hline
	\end{tabular}
	\vskip0.5cm
\noindent Following our general framework in~Eq.~(\ref{eq:matrix}), the environment-coupled payoff matrix becomes: 
\begin{equation}
	{\sf {\Pi}}
	\equiv
	(1-n)
	\begin{bmatrix}
		b_0 - c   & - c \\
		b_0       & 0 \\
	\end{bmatrix}
	+n
	\begin{bmatrix}
		b_1 - c   & -c \\
		b_1       & 0 \\
	\end{bmatrix},
	\label{eq:bmatrix}
\end{equation}
where $b_1$ and $b_0$ (with $b_1>b_0>c>0$) are benefits at replete and deplete resource state, respectively. 

\subsection{Implementation of state dependent reward and punishment}
In this simple yet nontrivial setting, our one objective is to investigate the comparative roles of reward and punishment in averting the ToC. Hardin's seminal formulation~\cite{hardin1968} of the ToC describes a situation in which a shared resource is subject to unrestricted access, such that no individual can exclude others from use and no effective mechanism exists to regulate exploitation, ultimately leading to overuse and resource depletion. Subsequent studies, in particular, the ones by Ostrom and others demonstrated~\cite{Ostrom2015, Cole2010} that many communities successfully avoid overexploitation through locally evolved institutions, monitoring mechanisms, sanctions, and collective governance arrangements.
From this perspective, institutional incentive mechanisms in the form of rewards and punishments play a crucial role in aligning individual incentives with collective welfare. The costs of rewards are borne collectively through the management institution, whereas punishments correspond to fines or sanctions imposed on individuals who overexploit the resource.
By altering the costs and benefits associated with behavioural choices, these mechanisms can discourage selfish exploitation while promoting cooperative actions that sustain shared resources. Motivated by this idea, we investigate how rewards and punishments~\cite{Mondal2022JoPC, Fang2019, Ozono2020, MSC_2025_PRSA} influence ToC in eco-evolutionary feedback dynamics. Specifically, individuals engaging in selfish or exploitative behaviour may incur external punishments that reduce their benefits and, consequently, their payoffs or fitnesses. Conversely, cooperative or prosocial behaviour may be reinforced through external rewards, thereby increasing individual benefits.

To this end, we modify the payoff matrix so that a cooperative act against defection receive rewards and a defective act against cooperation incur punishment.  Specifically, at deplete resource state~($n=0$), a cooperator (or donor) receives a reward~$\rho^+_0$ when interacting with a defector, whereas a defector incurs a punishment~$\rho^-_0$ for defecting against a cooperator. At the replete state~($n=1$), the corresponding modifiers are~$\rho^+_1$ and $\rho^-_1$, respectively. These extensions yield the following coupled payoff matrix:
\begin{equation}
	\Pi = (1 - n)
	\begin{bmatrix}
		b_0 - c & -c + \rho^+_0 \\
		b_0 - \rho^-_0& 0
	\end{bmatrix}
	+ n
	\begin{bmatrix}
		b_1 - c & -c + \rho^+_1 \\
		b_1 - \rho^-_1 & 0
	\end{bmatrix}.
	\label{eq:q_matrix_rp}
\end{equation}
Thus, rewards and punishments are imposed externally through an institution and are not provided by individual players. Therefore, the present model does not explicitly include the second-order free-rider problem~\cite{SST_2010_NAT, P_2012_Sc_R}, where individuals may benefit from the incentive mechanism without contributing to its implementation cost. Extending the model such that the individuals can actively choose whether to contribute to the reward or punishment mechanism would be an interesting direction for future work.

Under such incentive structures, we expect that cooperation is more likely to be sustained; however, the interplay between reward and punishment is not clear {\emph a priori}.

In terms of the parameters $\alpha$, $\beta$, and $\gamma$, we have $\gamma_0= c - \rho^-_0$, $\gamma_1=c - \rho^-_1$, $\beta_0 = c - \rho^+_0$, $\beta_1=c - \rho^+_1$, $\alpha_0 = b_0 - c$ and $\alpha_1 = b_1 - c$. Since $b_1> b_0$, it follows directly that $\alpha_1 > \alpha_0$. Thus, we now translate the general relations on $\beta, \gamma$, and $\alpha$ in terms of reward ($\rho^+$), punishment ($\rho^-$), benefit ($b$), and cost ($c$) to physically understand the conditions for ToC aversion and appearance of limit cycle in DG.

Firstly, following  Fig.~\ref{fig:boat4}(A), one can directly talk about the conditions on reward ($\rho^+_0$) and punishment ($\rho^-_0$) for aversion complete ToC just by mapped from the parameter space~\(\beta_{0}\text{--}\gamma_{0}\) to the reward--punishment~\(\rho^+\text{--}\rho^-\) (see Fig.~\ref{fig:reward_punishment_n0n1}(a)). The axis $\beta_{0}=0$ and $\gamma_{0}=0$  corresponds to $\rho^+_0=c$ and $\rho^-_0=c$, and conditions \(\beta_{0}+\gamma_{0}=-\alpha_0~\text{and}~\beta_{0}-\gamma_{0}=\alpha_0\) correspond to \(\rho^+_0 +\rho^-_0=b_0+c\) and \(\rho^+_0 -\rho^-_0=c-b_0\), respectively. 

Similarly, following Fig.~\ref{fig:boat4}(C), the region corresponding to partial or component ToC at the intermediate resource state $n\in(0,1)$ is mapped from the parameter space~\(\beta_{0}\text{--}\gamma_{0}\) to reward--punishment~\(\rho^+_0\text{--}\rho^-_0\), under the conditions  \(\rho^+_1>\rho^-_1\) and \(\rho^+_1<\rho^-_1\) (see Fig.~\ref{fig:reward_punishment_n0n1}(b) and Fig.~\ref{fig:reward_punishment_n0n1}(c)). The detailed calculations are shown in Appendix~\ref{appendix_D}. 

Finally, the conditions for Hopf bifurcation in reward-punishment parameter space, which is archived just by transforming the conditions of Hopf bifurcation calculated in terms of $\beta$, $\gamma$ in Section~\ref{para: hopf_bifurcation} in terms of $\rho^+$, $\rho^-$ as follows: The condition \(\rho^+_0(b_1+c-2\rho^-_1)-\rho^-_0(b_1+c-2\rho^+_1)=(\rho^+_1-\rho^-_1)(b_0+c)\) corresponds to $\alpha_1(\beta_0-\gamma_0)+\alpha_0(\gamma_1-\beta_1)=2(\beta_1\gamma_0-\gamma_1 \beta_0)$ and \(\rho^+_0+\rho^-_0= \frac{(\alpha_1-\alpha_0)2c}{\alpha_1}+\frac{\alpha_0}{\alpha_1}(\rho^+_1+\rho^-_1)\) correspond to constraint \((\beta_{1} + \gamma_{1})\alpha_0 = (\beta_{0} + \gamma_{0})\alpha_1\). 

To illustrate the Hopf bifurcation more clearly, Fig.~\ref{fig:hopf_bifuraction_reward_punishment} shows the dynamics as the reward parameter $\rho_0^+$ is varied while keeping $\rho_0^-=1.6$ fixed. The critical value occurs at $\rho_{0,c}^+=\rho_0^-=1.6$. For $\rho_0^+>1.6$, the trajectories approach the corresponding fixed-point branch. As $\rho_0^+$ is decreased through the critical value, the fixed point loses stability and sustained oscillations emerge for $\rho_0^+<1.6$. This transition is displayed in Fig.~\ref{fig:hopf_bifuraction_reward_punishment}(a), which provides a three-dimensional bifurcation representation in the $(\rho_0^+,p,n)$ space.

To examine the oscillatory regime in greater detail, we consider the representative value $\rho_0^+=1.5$, which lies below the critical value. Figure~\ref{fig:hopf_bifuraction_reward_punishment}(b) further illustrates the dynamics in the $(p,q,n)$ phase space for two representative initial conditions. In both cases, the trajectories approach the closed oscillatory orbit---shown in red colour---surrounding the unstable fixed-point continuum. Together, these representations make the emergence and phase-space structure of the sustained oscillatory dynamics explicit.

\subsection{Results}
\begin{enumerate}
	\item[(i)] {It is evident from Fig.~\ref{fig:boat4}(B) that, for any choice of parameter values, either stable fixed points or saturation points exist. Thus, for any values of reward and punishment, there always exists a reactive strategy close to TFT such that choosing it allows one to avert the ToC and attain the maximum resource state.}
	\item[(ii)] {The necessary condition to avert complete ToC for all possible initial choices of reactive strategies is $\rho_0^+ > c$ (see Fig.~\ref{fig:reward_punishment_n0n1}(a)).} 
	\item[(iii)] Fig.~\ref{fig:reward_punishment_n0n1}(a) shows that sufficiently high reward in the depleted resource state leads to the aversion of the complete ToC irrespective of the amount of punishment. In contrast, when punishment exceeds reward, the effectiveness of averting complete ToC depends on how much the benefit $b_0$ is more that the cost $c$---larger differences resulting in higher possibility of ToC mitigation by rewarding and punishing.
	\item[(iv)] One more fact can be read from Fig.~\ref{fig:reward_punishment_n0n1}(a): Complete ToC is precluded for all initial reactive strategies when the reward--punishment parameters at the depleted resource state satisfy
	$
	\rho^+_0+\rho^-_0>c+b_0
	\quad \text{and} \quad
	\rho^+_0-\rho^-_0>c-b_0 .
	$
	These conditions can be recast as
	$
	\rho^+_0-c>b_0-\rho^-_0
	\quad \text{and} \quad
	(\rho^+_0-c)+(b_0-\rho^-_0)>0,
	$ respectively. In game-theoretic terms, this corresponds to a Harmony-type interaction in the depleted state, where cooperation is a strictly dominant strategy. In particular, when one player cooperates while the opponent defects, the cooperator receives a higher payoff than the defector, and the sum of the two players' payoffs exceeds that obtained when both defect.
	
	Moreover, even when the condition $\rho^+_0-c>b_0-\rho^-_0$ is violated, there exist region with punishment exceeding reward for which complete ToC can still be averted for all initial strategy choices. This region increases with the difference ($b_0-c$) between benefit and cost in the deplete resource state. 
	 
	\item[(v)] {As per Fig.~\ref{fig:reward_punishment_n0n1}(b)-(c), below the dot-dashed (grey region), i.e., when
	$
	\frac{\alpha_0}{\alpha_1}\bigl(2c-\rho^+_1-\rho^-_1\bigr)
	<
	2c-\rho^+_0-\rho^-_0,
	$} intermediate resource states---either through limit cycles or component ToC---are impossible.
	In this region, the dynamical outcome is bistable: depending on the initial strategy choice, the system evolves either to complete ToC or to aversion of ToC. Consequently, the long-term outcome becomes highly sensitive to the initial strategy choice, and some initial strategies may lead to severe resource depletion.
	This region corresponds to a defection-dominant payoff structure in the depleted state, given that defection is already the dominant strategy in the replete state. Moreover, the extent of this region depends on the ratio $(\alpha_0/\alpha_1)$ of the benefit--cost differences between the depleted and replete states. {When the benefit--cost difference in the depleted state is smaller than that in the replete state, this region is more likely to occur.}
	Interestingly, when the magnitudes of benefit and cost are equal in the replete state, this region is never reached, and intermediate resource levels are always observed, even under defection-dominated payoffs. 
	\item[(vi)] {In the region above the dotted line, bounded by the white dashed line and the black dashed line, in Fig.~\ref{fig:reward_punishment_n0n1}(c), the dynamics either lead to limit cycle oscillations or to the maximum resource state [note also~Fig.~\ref{fig:reward_punishment_n0n1}(a)]. Interestingly, our numerical results show that when the punishment in the replete state is equal to the cost, and for any allowed value of the reward, the long-term eco-evolutionary feedback dynamics always converge to the maximum resource state for all possible initial choices of reactive strategies.}

	\item[(vii)]{Our numerical analysis illustrates the occurrence of a Hopf bifurcation for fixed parameter values $\rho_0^- = 1.6$ and $\rho_1^+ = \rho_1^- = 0.5$ (see Fig.~\ref{fig:hopf_bifuraction_reward_punishment}(a)). In this regime, the reward--punishment mechanism can give rise to periodically oscillatory evolutionary outcomes, such as stable limit cycles, in which both the reactive strategies and the resource state oscillate indefinitely.} 

\end{enumerate}

\section{Discussion and Conclusions}\label{conclusion}
A key result of our comprehensive analysis is the qualitative contrast between short-term and long-term evolutionary outcomes in $2\times2$ games in shaping the ToC. Short-term dynamics---driven by one-shot interaction and population-level frequency updates---exhibit either complete ToC or component ToC; complete aversion of ToC is impossible in this regime.
In contrast, long-term mutation--selection dynamics---driven by repeated interactions and trait substitution sequences in the reactive strategy space---can give rise to complete aversion of ToC for all parameter values when the population adopts reactive strategies close to TFT.

The introduction of reward and punishment mechanisms fundamentally reshapes the effective payoff landscape experienced by individuals, thereby influencing their cooperative or defective tendencies. Through modifying the relative costs and benefits associated with resource use, these mechanisms stabilize cooperative behaviour and suppress exploitative strategies that would otherwise drive the system toward the ToC. Importantly, this influence persists even over long evolutionary timescales, where transient advantages are filtered out and only dynamically stable outcomes remain. Thus, reward and punishment do not merely produce short-term compliance but instead contribute to the emergence of sustained, system-level regulation of shared resources.

A significant amount of work has been conducted on the basis of rewards and punishments to explore the emergence of cooperation, which is directly intertwined with ToC. The punishment leads more effectively has a significant importance to avert ToC in the work \cite{Mondal2022JoPC, kareva2013bmb, Johnson2015, Greenwood2016, Chen2018}; specifically, Kerava~\cite{kareva2013bmb} studied both adaptive dynamics and replicator dynamics averting ToC for a one-shot game with reward and punishment, and has concluded that the punishment has more importance than rewards in those scenarios. 

Our results further demonstrate that rewards and punishments applied in the depleted resource state have a subtle impact on averting ToC. This effect is constrained by the cost–benefit parameters and is particularly evident in regimes dominated by punishment. In contrast, in the intermediate resource regime—where punishment at the replete state exceeds reward---the depleted-state punishment exerts a robust influence on partial ToC aversion through the emergence of limit cycles, whereas depleted-state rewards tend to promote component ToC accompanied by limit cycle dynamics. We also observe that even when reward dominate punishment at high values, averting ToC can still occur, either via component ToC or through limit cycle behaviour. The periodically oscillatory evolutionary outcomes, via a limit cycle, where both reactive strategies and resource state oscillate indefinitely is indeed an interesting feature of our model. All these findings imply that institutional rewards alone may be insufficient to sustain cooperation in resource-sharing systems.

The model describes a class of systems in which individuals repeatedly interact over a shared, depletable resource, and where the resource state feeds back into future interaction opportunities and payoffs. Model predictions can be tested in repeated common-pool-resource laboratory experiments or field studies. Researchers can independently manipulate the resource-regeneration rate, initial stock, probability of future access, and the severity of sanctions. Observable outcomes would include individual extraction levels, resource-stock trajectories, interaction duration, and the frequency of strategy switching, allowing for direct validation of the predicted coupling between behaviour and ecological dynamics. We outline two empirically tractable illustrations of this structure: one drawn from human social-ecological systems and one from microbial ecology.

In human social-ecological systems, the environmental state corresponds to a physical resource stock---for example, the biomass of a fishery, the standing volume of a forest, or the water level in an irrigation system. Each round of the repeated game represents a harvesting period. Cooperative strategies correspond to restrained extraction that keeps the stock above a productive threshold, whereas defection corresponds to overexploitation. Because current extraction determines the stock available in the next round, users face a shadow-of-the-future trade-off: defecting yields an immediate gain but reduces future yields for all. This feedback selects for reactive strategies such as monitoring neighbours' extraction, imposing graduated sanctions on overharvesters, and forgiving repentant violators---behaviours extensively documented in long-enduring common-pool resource institutions \cite{Ostrom2015, Ostrom2008}. The model predicts that cooperation is most stable when (i) the resource regeneration rate is high enough that restraint is visibly rewarded, (ii) the coupling between individual actions and the resource trajectory is sufficiently strong that defection carries a detectable future cost, and (iii) sanctions are moderate rather than punitive, since excessively severe punishment can trigger partner withdrawal, destabilising the cooperative equilibrium.

A structurally analogous feedback loop operates in microbial populations where cells engage in repeated rounds of public-goods production within spatially structured environments such as biofilms or colonies \cite{west2006social, nadell2016spatial}. Here, the environmental state is the local concentration of a shared molecular resource---for instance, siderophores that scavenge iron or extracellular enzymes that digest complex substrates. Cooperators pay a metabolic cost to produce the public good; defectors withhold production while still benefiting from the shared pool. The resource accumulates through cooperative production and decays through degradation, diffusion, and consumption. Each growth phase constitutes a round: the public-good concentration at the start of a round determines the growth payoff available to all cells, and the resulting population composition determines the environmental state for the next round. This repeated interaction generates selection for conditional cooperation: cooperators can be favoured when spatial structure ensures that defectors disproportionately experience the consequences of local resource depletion, and lineages employing retaliatory exclusion of non-producing cheaters can outcompete unconditional strategies \cite{raina2018microbial, smith2019evolution}. The model predicts that cooperation persistence depends on the same three factors identified for human systems---regeneration (production) rate, coupling strength between individual action and resource trajectory, and the intensity of exclusion or punishment mechanisms---making microbial populations a tractable experimental platform for testing the model's core predictions under controlled laboratory conditions \cite{harrison2008microbial, drescher2014biofilm}.

An immediate extention of the present work would be to incorporate asymmetric role of non-reciprocity and generosity in promoting cooperation~\cite{Sadhukhan2022jop}. Another promising direction is to extend the present framework to multiplayer public goods interactions with continuous cooperative investments. Integrating adaptive dynamics with ecological feedback in public goods games may provide deeper insights into how collective investments and environmental conditions jointly shape the long-term evolution of cooperation and resource persistence~\cite{Luo2025}. While the present model considers a globally shared renewable resource, another interesting extension would be to incorporate spatially explicit resource dynamics with local regeneration and diffusion. This would allow the investigation of eco-evolutionary processes in heterogeneous environments, where individuals experience different local resource conditions and ecological feedbacks~\cite{Wakano2007}. Finally, before we conclude, we would like to recall that in the paradigm of adaptive dynamics, a central question concerns the fate of trait evolution---whether the population remains monomorphic or undergoes evolutionary branching~\cite{Geritz1997}: Understanding how evolutionary branching is affected by the co-evolving environmental resources constitutes an important open direction for exploration.

\acknowledgements
Abhishek Yogi is thanked for his inputs during the initial stages of this work. Financial assistance from the University Grants Commission (UGC), New Delhi, in the form of a Senior Research Fellowship is gratefully acknowledged by the first author.
\appendix
\section{Non-isolated fixed points at the intersection of surfaces---$n=0$  and $F(p,q,n)=0$}\label{Appendix_A}
The eigenvalues of the Jacobian matrix at the deplete state ($n=0$) are given by:
\begin{enumerate}[label=(\roman*)]\label{eq:eigen_dpeleted}
	\item $\lambda_1=(p+q-1)$,
	\item $\lambda_2=qA_0+(1-p)B_0$,
	\item $\lambda_3=0$,
\end{enumerate}
 where, $A_0=q(\beta_0-\gamma_0)+(1-2r)(\alpha_0+\gamma_0)+\beta_0$ and $B_0=(1+r)(\beta_0-\gamma_0)-A_0.$
 The zero eigenvalue $\lambda_3=0$ indicates neutral stability along the corresponding eigenvector, meaning that small perturbations in this direction does not time-evolve. Naturally, the fixed points and their stability at $n=0$ depend solely on the game played at the depleted state and are independent of the game played at $n=1$.  The existence and stability properties of the non-isolated fixed points and boundary points at the depleted state are analysed in detail below with a view to showing how Subplot~(A) of Fig.~(\ref{fig:boat4}) is constructed to exhibit the distinct stability regions associated with different underlying games.
\subsection{Construction of  Fig.~\ref{fig:boat4}$(A)$}
To fully determine the stability of the continuous line of fixed points, we must know the signs of both eigenvalues $\lambda_1$ and $\lambda_2$. The sign of $\lambda_1$ can be readily inferred from the structure of the fixed points: $\lambda_1>0$ for fixed points lying above the line $p+q=1$---shown in grey dashed line in subplots, and $\lambda_1<0$ for those below it. In contrast, predicting the sign of $\lambda_2$ along the continuous line of fixed points is more challenging because of its complex functional dependence. However, we identify a simpler approach that allows us to determine the sign of $\lambda_2$ for all fixed points and parameter combinations.
\subsubsection{Predicting the Sign of $\lambda_2$}
{The eigenvector corresponding to $\lambda_2$ is orthogonal to the resource direction and lies entirely within the $p$--$q$ plane. Hence, the sign of $\lambda_2$ is determined exclusively by the dynamics of the reactive strategies Eq.~(\ref{eq:AD_p}) and Eq.~(\ref{eq:AD_q})  in the vicinity of $n=0$.
}

In the resulting two-dimensional setting at $n=0$, the behaviour is straightforward: if $F(p,q,0)>0$, then $\dot{p}>0$ and $\dot{q}>0$; if $F(p,q,0)<0$, then $\dot{p}<0$ and $\dot{q}<0$. A continuous line of fixed points arises precisely when $F(p,q,0)=0$. Thus, by determining the sign of $F(p,q,0)$ above and below this line, we can directly infer the stability of the fixed points---or equivalently, the sign of $\lambda_2$.

First, we rewrite the function $F(p,q,0)$ in a more convenient form:
\begin{subequations}\label{eq:F(pq0)}
	\begin{eqnarray}
		\nonumber&&F(p,q,0)= f(c,r)  \equiv \frac{(\beta_0-\gamma_0)}{2}c(1+r)-\frac{(\beta_0-\gamma_0)}{2}r(1+r)\\&&+(\alpha_0+\gamma_0)r(1-r)-\beta_0 (1-r),
	\end{eqnarray}
\end{subequations}
where $c=p+q$. Note that the curves $c=a_1$ and $r=p-q=a_2$ (with $a_1$ and $a_2$ being arbitrary constants) are orthogonal in the $p$--$q$ coordinate plane; i.e., $c$ and $r$ form two independent coordinates. Thus, we can vary the value of $c$ while keeping $r$ fixed. 

Consider an arbitrary point with $c=c^*$ and $r=r^*$ on the continuous line of fixed points. If we increase the value of $c$ from $c^*$ while keeping $r$ unchanged, and if $\beta_0>\gamma_0$, then $f(c,r)>0$, which implies $\dot{p}>0$ and $\dot{q}>0$ (see Eqs.~(\ref{eq:AD_p}) and (\ref{eq:AD_q})). Conversely, if we decrease the value of $c$ from $c^*$, then for
$\beta_0>\gamma_0$ we obtain $f(c,r)<0$, yielding $\dot{p}<0$ and $\dot{q}<0$. A similar argument holds for $\beta_0<\gamma_0$, if $c$ is increased (decreased) from $c^*$, then $f(c,r)<0$~($f(c,r)>0$), implying $\dot{p}<0$, $\dot{q}<0$~($\dot{p}>0$, $\dot{q}>0$). Therefore, for $\beta_0>\gamma_0$, the eigenvalue $\lambda_2$ at the fixed point corresponding to $(c^*,r^*)$ is positive, while for $\beta_0<\gamma_0$, it is negative. Since the argument holds for any arbitrary fixed point on the continuous line, the conclusion applies to the entire line of fixed points. 

\begin{figure}
	\centering	
	\includegraphics[scale=0.57]{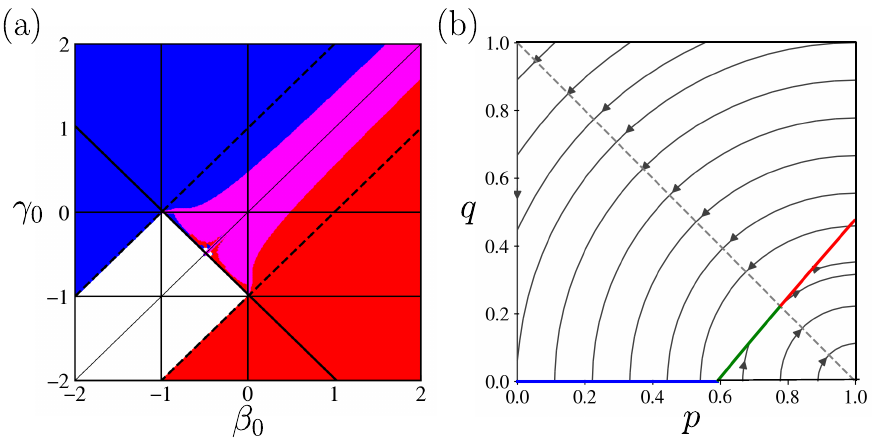}
	\caption{Summary of the sign of the eigenvalue $\lambda_2$:
	Subplot~(a) summarizes the sign of the eigenvalue $\lambda_2$ in the $\beta_0$--$\gamma_0$ parameter plane for the fixed-point curve arising from the intersection of the surfaces $n=0$ and $F(p,q,n)=0$. In this plot, blue and red colors indicate regions where $\lambda_2$ is negative and positive, respectively. The magenta region denotes parameter values for which $\lambda_2$ is negative along one portion of the fixed-point curve and positive along another portion. The white region represents parameter values for which no fixed-point curve exists. Subplot~(b) corresponds to the magenta region and illustrates representative trajectories approaching or departing from the fixed-point curve, demonstrating that one portion of the curve is stable while the remaining portion is unstable.
	}
	\label{fig:boat5}
\end{figure}

Caution needs to be exercised in the magenta region shown in Fig.~\ref{fig:boat5}(a), where the sign of $\lambda_2$ may differ at different points along the same fixed-point line due to its special structure. As shown in Fig.~\ref{fig:boat5}(b), even though increasing (decreasing) $c$ from $c^*$ at fixed $r$ always leads to $f(c,r)<0$ ($f(c,r)>0$) for fixed values of $\beta_0$ and $\gamma_0$, certain segments of the fixed-point line are stable while others are unstable. The underlying reason is that the orbits of the evolutionary dynamics in the $p$--$q$ space intersect the fixed-point line twice. Consequently, although our method---based on the sign of $f(c,r)$ in Eq.~(\ref{eq:F(pq0)}) evaluated at $n=0$---correctly predicts the direction of flow throughout the phase space, the sign of $\lambda_2$ may still vary depending on the precise location along the fixed-point line. This behavior is exhibited by the fixed-point line corresponding to the dotted region of parameter space in Subplots (A) and (B) of Fig.~\ref{fig:boat4}.
\subsubsection{Predicting the sign of $\lambda_1$}
{Determining the sign of $\lambda_1$ requires knowledge of the structure of the fixed-point line for different parameter values. Specifically, if the fixed-point line (or a segment of it) lies above the line $p+q=1$, then $\lambda_1<0$ along that segment. Conversely, if the fixed-point line (or a segment of it) lies below the line $p+q=1$, then $\lambda_1>0$.}

The function $F(p,q,0)$ is always zero at $p=1$ and $q=0$; i.e., the continuous line of fixed points always passes through point $(1,0)$. By examining the slope of this fixed-point line at ($1,0$) point and determining the parameter conditions under which the fixed-point curve intersects the line $p+q=1$, we can divide the parameter space into six different regions, each exhibiting a different fixed-point structure. These six regions are shown in Subplot~$(A)$ of Fig.~(\ref{fig:boat4}), represented by six different colours, viz., orange, cyan, green, yellow, violet, and magenta (different shades of same colours are also used).

\subsubsection{Demarcation of six regions in the parameter space}
At the $n=0$ plane, the continuous line of fixed points satisfies $F(p,q,0)=0$, which in explicit form can be written as
\begin{eqnarray}\label{eq:Fpq0_sim}
	&&-(\alpha_0+\gamma_0)p^2-(\alpha_0+\beta_0)q^2+(2\alpha_0+\beta_0+\gamma_0)pq\nonumber\\&&+(\alpha_0+\beta_0+\gamma_0)p -(\alpha_0+2\gamma_0)q-\beta_0=0.
\end{eqnarray}
This equation defines a hyperbola, and it is straightforward to verify that the point \((1,0)\) is always a solution for all admissible parameter values. The slope of the curve at the point \((1,0)\) relative to the \(q = 0\) axis is given by 
\begin{equation}\label{eq:slope}
	\frac{dq}{dp}=-\frac{(\beta_0-\gamma_0)-\alpha_0}{(\beta_0-\gamma_0)+\alpha_0}.
\end{equation}
To determine whether the continuous line of fixed points intersects the line \(p + q = 1\) at a point other than \((1,0)\), we substitute \(p = 1 - q\) into Eq.~(\ref{eq:Fpq0_sim}). This yields a quadratic equation in \(q\), whose solutions are $q=0$ and $q=\alpha_0/(2\alpha_0+\beta_0+\gamma_0)$.
The solution $q=\alpha_0/(2\alpha_0+\beta_0+\gamma_0)$ provides us with new information, where the intersection point of the continuous line of fixed points and the $p+q=1$ line is physical when $\alpha_0/(2\alpha_0+\beta_0+\gamma_0) \in (0,1]$, which implies: 
\begin{subequations}
	\begin{eqnarray}\label{eq:parameter1}
		&&\frac{\alpha_0}{2\alpha_0+\beta_0+\gamma_0}> 0 \implies (\beta_0+\gamma_0) > -2\alpha_0,\\\label{eq:parameter2}&&
		\frac{\alpha_0}{2\alpha_0+\beta_0+\gamma_0}\le 1 \implies (\beta_0+\gamma_0) \ge -\alpha_0.
	\end{eqnarray}
\end{subequations}

It is evident that the condition in Eq.~(\ref{eq:parameter2}) is stronger and therefore automatically implies condition~(\ref{eq:parameter1}). Hence, we conclude that the continuous line of fixed points intersects the line $p+q=1$ at a second physical point (in addition to \((1,0)\)) whenever $(\beta_0+\gamma_0) \geq -\alpha_0$.

We next characterize the possible configurations of the continuous line of fixed points. There are two cases in which the tangent angle of this curve at $(1,0)$, measured relative to the $q=0$ axis, lies in the interval [$\pi/2,\pi$].

\begin{enumerate}
	\item $|(\beta_0-\gamma_0)-\alpha_0|>|(\beta_0-\gamma_0)+\alpha_0|$\\
	This case occurs only when $(\beta_0-\gamma_0) \le -\alpha_0$.
	\begin{enumerate}[label=(\roman*)]
		\item When $(\beta_0-\gamma_0) \le -\alpha_0$ and $(\beta_0+\gamma_0) \ge -\alpha_0$, the continuous line of fixed points has a tangent angle $\phi$ at $(1,0)$ that is in the range $3\pi/4<\phi<\pi/2$ and intersects the line $p+q=1$ within the physical region.  
		{Therefore, in this parameter region, $\lambda_1<0$ on some segments of the fixed-point line and $\lambda_1>0$ on others, as indicated by the yellow region in Fig.~\ref{fig:boat4}(A).}
		\item When $(\beta_0-\gamma_0) \le -\alpha_0$ but $(\beta_0+\gamma_0) < -\alpha_0$, the continuous line of fixed points
		again has a tangent angle $\phi$ at $(1,0)$ that is in the range $3\pi/4<\phi<\pi/2$ but does not intersect the line $p+q=1$. {Therefore, in this parameter region, $\lambda_1>0$ everywhere, and the fixed-point line is completely unstable, as indicated by the green region in Fig.~\ref{fig:boat4}(A).}
	\end{enumerate}
	
	\item $|(\beta_0-\gamma_0)-\alpha_0|<|(\beta_0-\gamma_0)+\alpha_0|$\\
	This case occurs only when $(\beta_0-\gamma_0) \ge \alpha_0$.
	\begin{enumerate}[label=(\roman*)]
		\item When $(\beta_0-\gamma_0) \ge \alpha_0$ and $(\beta_0+\gamma_0) \ge -\alpha_0$, the continuous line of fixed points has a tangent angle $\phi$ at $(1,0)$ that is in the range $\pi<\phi<3\pi/4$ and intersects the line $p+q=1$ within the physical region. {Therefore, in this parameter region, $\lambda_1<0$ on some segments of the fixed-point line and $\lambda_1>0$ on others. This corresponds to the orange region in Fig.~\ref{fig:boat4}(A). Nevertheless, the fixed-point line is completely unstable in this parameter range because $\lambda_2>0$ everywhere.} 
		
		\item When $(\beta_0-\gamma_0) \ge \alpha_0$ but $(\beta_0+\gamma_0) < -\alpha_0$, the continuous line of fixed points has a tangent angle $\phi$ at $(1,0$) that is in the range $\pi<\phi<3\pi/4$  but does not intersect the line $p+q=1$. {Therefore, in this parameter region, $\lambda_1<0$ everywhere. This corresponds to the violet region in Fig.~\ref{fig:boat4}. However, the fixed-point line is completely unstable in this parameter range because $\lambda_2>0$ throughout the region.} 
	\end{enumerate}
\end{enumerate}

When $-\alpha_0<(\beta_0-\gamma_0) < \alpha_0$, the branch of the hyperbola passing through the point $(1,0)$ lies outside the physical region. However, other portions of the same hyperbola remain within the physical region. In such cases, the fixed-point curve must intersect the boundary of the physical region at points other than $(1,0)$. The possible boundaries and the corresponding crossing points are given below:

\begin{widetext}
	\begin{subequations}\label{eq:boundary_points}
		\begin{eqnarray}
			&&(i)~ p=1 \implies q=1-\frac{\gamma_0}{\alpha_0+\beta_0},\\&& (ii)~
			p=0 \implies q_{\pm}=\frac{1}{2(\alpha_0+\beta_0)}\left[-(\alpha_0+2\gamma_0)\pm \sqrt{(\alpha_0+2\gamma_0)^2-4\beta_0(\alpha_0+\beta_0)}\right],\\&& (iii)~ q=0 \implies p=\frac{\beta_0}{(\alpha_0+\gamma_0)}, \\&&(iv)~
			q=1 \implies p_{\pm}=\frac{1}{2(\alpha_0+\gamma_0)} \left[(3+2\beta_0+2\gamma_0)\pm \sqrt{(3\alpha_0+2\beta_0+2\gamma_0)^2-8(\alpha_0+\gamma_0)(\alpha_0+\beta_0+\gamma_0)}\right].
		\end{eqnarray}
	\end{subequations}
\end{widetext}
Within the parameter region $-\alpha_0<(\beta_0-\gamma_0) < \alpha_0$ and $(\beta_0+\gamma_0) \ge -\alpha_0$, the fixed-point curve defined by Eq.~(\ref{eq:Fpq0_sim}) always intersects the line $p+q=1$ at a point other than $(1,0)$), and this lies within the physical region. Consequently, a branch of the fixed-point curve that does not pass through $(1,0)$ always exists whenever $(\beta_0+\gamma_0) \ge -\alpha_0$. {Moreover, since the fixed-point line crosses the line $p+q=1$, $\lambda_1>0$ on some portions of it and $\lambda_1<0$ on the remaining portions. This parameter regime corresponds to the pink region in Fig.~\ref{fig:boat4}(A).}

We next consider the case $(\beta_0+\gamma_0) < -\alpha_0$ together with $-\alpha_0<(\beta_0-\gamma_0) < \alpha_0$. In this region, both $\beta_0$ and $\gamma_0$ are negative as indicated by Fig.~(\ref{fig:boat4})(A). Since no branch of the fixed-point curve intersects the line $p+q=1$ in this regime, the only remaining question is whether more than one boundary fixed point--corresponding to the cases listed in (i.e., (\ref{eq:boundary_points}a), (\ref{eq:boundary_points}b), (\ref{eq:boundary_points}c), (\ref{eq:boundary_points}d)) can exist.\\

\begin{enumerate}[label=(\roman*)]
	\item  Since in our region of interest, $(\alpha_0+\beta_0)>\gamma_0$ always holds, and $\beta_0$ and $\gamma_0$ are always negative, we can directly conclude that the fixed point at $p=1$ and $q=1-\gamma_0/(\alpha_0+\beta_0)$ is not physical.
	\item \begin{enumerate}[label=(\alph*)]
		\item 
		When $(\alpha_0+\beta_0)<0$, the necessary condition for the fixed point~(\ref{eq:boundary_points}b) (both $q_{+}$ and $q_{-}$) to be physical is $(\alpha_0+2\gamma_0)>0$ and $|2(\alpha_0+\beta_0)|>|(\alpha_0+2\gamma_0)|$. However, in that case, the term inside the square root is negative, and the fixed points become imaginary.
		
		\item When $(\alpha_0+\beta_0)>0$, then the necessary condition for the fixed point to exist is $(\alpha_0+2\gamma_0)<0$. For both the fixed points, corresponding to $q_{+}$ and $q_{-}$, to be greater than or equal to zero, the necessary condition is the following:
		\begin{equation*}
			-(\alpha_0+2\gamma_0)> \sqrt{(\alpha_0+2\gamma_0)^2-4\beta_0(\alpha_0+\beta_0)}.
		\end{equation*}
		The above equation ultimately boils down to $0>-4\beta_0 (\alpha_0+\beta_0)$, which is impossible as $\beta_0<0$ in our case.
	\end{enumerate}
	\item In light of the domain of interest, $\beta_0<(\alpha_0+\gamma_0)$ always, and $\beta_0$ and $\gamma_0$ are always negative, we can directly conclude that the fixed point at $p=\beta_0/(\alpha_0+\gamma_0)$ and $q=0$ can't be physical.
	\item 
	\begin{enumerate}[label=(\alph*)]
		\item When $(\alpha_0+\gamma_0)> 0$, the following condition need to satisfy for $p_{\pm}\ge 0$:
		\begin{eqnarray*}
			&&(3\alpha_0+2\beta_0+2\gamma_0)\nonumber\\&&\pm \sqrt{(3\alpha_0+2\beta_0+2\gamma_0)^2-8(\alpha_0+\gamma_0)(\alpha_0+\beta_0+\gamma_0)}\ge 0.
		\end{eqnarray*}
		
		It ultimately implies, $-8(\alpha_0+\gamma_0)(\alpha_0+\beta_0+\gamma_0) \le 0$, which is impossible because $(\alpha_0+\beta_0+\gamma_0)<0$ for our case. So we can conclude that for $(\alpha_0+\gamma_0)> 0$, $p_{\pm}$ can not be non-negative.
		\item When $(\alpha_0+\gamma_0)<0$, sign of $(\alpha_0+\gamma_0)$ have to be same with the sign of $(3\alpha_0+2\beta_0+2\gamma_0)$, otherwise, one of the fixed point is negative. Now, if $2|\alpha_0+\gamma_0|\ge|3\alpha_0+2\beta_0+2\gamma_0|$, then  $2|\alpha_0+\gamma_0+\beta_0|>|3\alpha_0+2\beta_0+2\gamma_0|$; since $\beta_0$ and $\gamma_0$ are always negative, therefore, $|\alpha_0+\gamma_0+\beta_0|>|\alpha_0+\gamma_0|$. So the term inside the square root becomes negative, and the fixed points become imaginary. If, on the other hand, $2|\alpha_0+\gamma_0|<|3\alpha_0+2\beta_0+2\gamma_0|$ then the first term becomes greater than one and one of the fixed points should surely be outside the physical region. 
	\end{enumerate}
\end{enumerate}
So, we can conclude that in the region such that $-\alpha_0<(\beta_0-\gamma_0) < \alpha_0$ and $(\beta_0+\gamma_0)<-\alpha_0$, i.e., in the cyan region of Fig.~(\ref{fig:boat4}), no fixed point line can exist inside the physical region.

We next examine the boundary points and determine whether they are saturated in each parameter region. At $n=0$, the saturated boundary points can occur only along the boundary $q=0$ (as discussed in Section~\ref{sec:Linear Stability Analysis}). A boundary point on $q=0$ is saturated only when it satisfies the following relation:
\begin{equation}\label{eq:boundary_points2}
	(\alpha_0+\gamma_0)p<\beta_0, ~~~\forall p \in [0,1].
\end{equation}
{\textit{Orange zone:}}
Since in this region $\beta_0>\alpha_0+\gamma_0$ and $\beta_0>0$, the boundary $q=0$ consists entirely of saturated points for all $p \in [0,1]$.\\
{\textit{Violet zone:}}
In this region $\beta_0>\alpha_0+\gamma_0$ and $(\alpha_0+\gamma_0)<0$. The behaviour of boundary points depends on the sign of $\beta_0$. When $\beta_0>0$, all the boundary points are saturated (Subplot $b.$). In contrast, when $\beta_0<0$, only those boundary points near $(1,0)$ are saturated (Subplot $c.$), in accordance with Eq.~(\ref{eq:boundary_points2}).\\
{\textit{Yellow zone:}}
In this region $\beta_0<\alpha_0+\gamma_0$ and $\alpha_0+\gamma_0>0$. Again, the behaviour of boundary points depends on the sign of $\beta_0$. When $\beta_0>0$, some of the boundary points are saturated (Subplot $g.$). {In contrast, when $\beta_0<0$, no boundary point is saturated (Subplot $f.$)}, in accordance with Eq.~(\ref{eq:boundary_points2}).\\
{\textit{Green zone:}}
In this region $\beta_0<\alpha_0+\gamma_0$ and $\beta_0<0$, {the boundary $q=0$ has no saturated point.}\\
{\textit{Magenta zone:}}    
Although the condition $\beta_0<(\alpha_0+\gamma_0)$ holds throughout this region, the behaviour of the boundary points depends on the sign of $\beta_0$, as determined by the criterion in Eq.~(\ref{eq:boundary_points2}). Specifically:\\
$\bullet$ When $\beta_0>\gamma_0$ and $\beta_0<0$, all the boundary points are unsaturated (Subplot $h.$ and $i.$).\\
$\bullet$ When $\beta_0>\gamma_0$ and  $\beta_0>0$, only some of the boundary points are saturated (Subplot $f.$ and $g.$).\\
$\bullet$ When $\beta_0<\gamma_0$ and $\beta_0>0$, only some of the boundary points are saturated (Subplot $l.$ and $m.$).\\
$\bullet$ Finally, when $\beta_0<\gamma_0$ and $\beta_0<0$, no boundary point is saturated (Subplot $j.$ and $k.$).\\ 
{\textit{Cyan zone:}}     
In this region, $\beta_0<0$, while $\alpha_0+\gamma_0$, may have either sign. If $\alpha_0+\gamma_0>0$ then Eq.~(\ref{eq:boundary_points2}) requires $p<0$, which lies outside the physical domain $p\in [0,1]$. Hence, no saturated boundary point can exist in this case.
If instead $\alpha_0+\gamma_0<0$, then within the cyan region we have, $\beta_0<(\alpha_0+\gamma_0)$, and Eq.~(\ref{eq:boundary_points2}) again cannot be satisfied. Thus, no saturated boundary point exists in this case either.

\section{Non-isolated fixed points at the intersection of surfaces---$n=1$  and $F(p,q,n)=0$} \label{Appendix_B}
The eigenvalues of the Jacobian matrix at the replete state ($n=1$) are given by:
\begin{enumerate}[label=(\roman*)]\label{eq:eigen_repleted}
	\item $\lambda_1=(1-(p+q))$,
	\item $\lambda_2=qA_1+(1-p)B_1$,
	\item $\lambda_3=0$,
\end{enumerate}
where, $A_1=q(\beta_1-\gamma_1)+(1-2r)(\alpha_1+\gamma_1)+\beta_1$ and $B_1=(1+r)(\beta_1-\gamma_1)-A_1.$ The zero eigenvalue $\lambda_3=0$ indicates neutral stability along the corresponding eigenvector.

Subplot~$(B)$ of Fig.~(\ref{fig:boat4}) highlights the distinct stability regions associated with different underlying games. The existence and stability properties of the non-isolated fixed points and boundary points at the depleted state are analysed in detail in the following subsection. Importantly, the fixed points and their stability at $n=1$ depend solely on the game played at the replete state and are independent of the game played at $n=0$.

\subsection{Construction of  Fig.~\ref{fig:boat4}$(B)$}	
The expressions for $\lambda_2$ and for the fixed-point condition $F(p,q,1)=0$, at $n=1$ have exactly the same structure as their counterparts $\lambda_2$ and $F(p,q,0)=0$ at $n=0$, except that with the parameters ($\alpha_0$,  $\beta_0$, $\gamma_0$) replaced by ($\alpha_1$,  $\beta_1$, $\gamma_1$). Therefore, we conclude, from our earlier analysis, that whenever $\beta_1>\gamma_1$, the sign of the eigenvalue $\lambda_2$ is positive, and when $\beta_1<\gamma_1$, then the sign of the eigenvalue $\lambda_2$ is negative. The fixed-point curve retains the same qualitative structure as in the $n=0$ (see subplot $(A)$ of Fig.~\ref{fig:boat4}), with the sole exception that  $\lambda_1$ becomes negative whenever $p+q>1$.

However, the analysis of the saturated boundary points differs from the $n=0$ case. At $n=1$, the only possible saturated boundary points lie on the boundary  $p=1$ (see discussion in Section~\ref{sec:Linear Stability Analysis}). A saturated point on the boundary $p=1$  arises only when the generosity level $q$ satisfies,
\begin{equation}\label{eq:boundary_points3}
	(\alpha_1+\beta_1)(1-q)>\gamma_1,~~~~\forall q \in [0,1]. 
\end{equation}	
In the following discussion, we have analysed the existence of saturated points for all relevant parameter values:\\
{\textit{Orange zone:}}
In this region, $\alpha_1+\beta_1>\gamma_1$. Since $\gamma_1$ may be either positive or negative, the number of saturated points depends on the sign of $\gamma_1$ as indicated by Eq.~(\ref{eq:boundary_points3}). When $\gamma_1<0$, all the boundary points are saturated (see Subplot $b.$). In contrast, when $\gamma_1>0$, only a subset of the boundary points is saturated (see Subplot~$a.$).\\
{\textit{Violet zone:}}
In the violet region, $\alpha_1+\beta_1>\gamma_1$. Therefore, according to Eq.~(\ref{eq:boundary_points3}), the boundary $p=1$ is always saturated.\\
{\textit{Yellow zone:}}
In the yellow region, where $\alpha_1+\beta_1<\gamma_1$, and hence Eq.~(\ref{eq:boundary_points3}) implies that no boundary point is saturated.\\
{\textit{Green zone:}}
In the green region, even though $\alpha_1+\beta_1<\gamma_1$, the behavior of the boundary $p=1$ depends on the sign of $\gamma_1$. When $\gamma_1>0$, no boundary point is saturated (see Subplot $j.$). In contrast, when $\gamma_1<0$, a subset of the boundary points becomes saturated~(see Subplot $e.$).\\
{\textit{Pink zone:}}    
In the pink region, $\beta_1$ may be either greater or less than $\gamma_1$. Although this region satisfies $\alpha_1+\beta_1>\gamma_1$, the parameter $\gamma_1$ itself may take either positive or negative values. From Eq.~(\ref{eq:boundary_points3}), it follows that $\gamma_1<0$, all boundary points are saturated. In contrast, when $\gamma_1>0$, only a subset of the boundary points becomes saturated. \\
{\textit{Cyan zone:}}    
Since $\alpha_1+\beta_1>\gamma_1$ in this region, and both $\gamma_1$ and $\alpha_1+\beta_1$ are negative, Eq.~(\ref{eq:boundary_points3}) implies that all boundary points on $p=1$ are saturated. 

\section{Non-isolated fixed points at the intersection of surfaces---$p+q=1$  and $F(p,q,n)=0$} \label{Appendix_C}
The eigenvalues of the Jacobian matrix at the intermediate state of resource $n\in(0,1)$ are given by: 
\begin{enumerate}[label=(\roman*)]
	\item $\lambda_1+\lambda_2=\frac{2pq}{(1-r)^2 (1-r^2)}\left[(1-n)(\beta_0-\gamma_0)+n(\beta_1-\gamma_1)\right]$ \label{eq:eigen_sum_nm}
	\item \small $\lambda_1\lambda_2=\frac{n(1-n)}{4pq}\left[(\beta_1+\gamma_1+2\alpha_1-\beta_0-\gamma_0-2\alpha_0)q - (\alpha_1-\alpha_0)\right]$ \label{eq:eigen_prod_nm}
	\item $\lambda_3=0$
\end{enumerate}

Using Eq.~(\ref{eq:internal}) for the internal fixed points, we ultimately get,
\begin{equation}
	\lambda_1\lambda_2= \frac{n(1-n)}{4pq}\frac{(\beta_{1} + \gamma_{1})\alpha_0 - (\beta_{0} + \gamma_{0})\alpha_1}{(1-n)(\beta_{0}+\gamma_{0}+2\alpha_0)+n(\beta_{1}+\gamma_{1}+2\alpha_1)} \label{eq:conjugate}.
\end{equation}

For a fixed point to be stable, the sum of the relevant eigenvalues must be negative $\lambda_1+\lambda_2<0$ and their product must be positive $\lambda_1 \lambda_2>0$~\cite{Strogatz2018}. Thus, the stability of the internal fixed points depends on the games played at both $n=0$ and $n=1$.

Subplot~$(C)$ of Fig.~(\ref{fig:boat4}) illustrates the corresponding stability regions: where in parameter space of \(\beta_{1}+\gamma_{1}\) vs \(\beta_{0}+\gamma_{0}\) the fixed point stability structure has been shown in subsubplots $a.$ and $b.$ for violate region and olive region respectively.
\subsection{Construction of  Fig.~\ref{fig:boat4}$(C)$}
It illustrates the existence and stability properties of the internal fixed points, defined by the conditions $p + q = 1$ and $F(p, q, n) = 0$.  The fixed-point curve follows 
\begin{eqnarray}\label{eq:Fpn_internal}
	&& F(p,n)=(1-n)[(\beta_0-\gamma_0)p+(\alpha_0+\gamma_0)(2p-1)-\beta_0]\nonumber\\
	&&+n[(\beta_1-\gamma_1)p+(\alpha_1+\gamma_1)(2p-1)-\beta_1]=0,
\end{eqnarray}
which implies
\begin{eqnarray}\label{eq:internal}
	q = \frac{(1-n)\alpha_0 + n \alpha_1}{(1-n)(\beta_0+ \gamma_0+ 2\alpha_0)+ n(\beta_{1} + \gamma_{1} + 2 \alpha_1)}.
\end{eqnarray}
Note that for our system $\alpha_0, \alpha_1>0$ always, so the necessary condition for existence of the fixed point is $(1-n)(\beta_0+ \gamma_0+ 2\alpha_0)+ n(\beta_{1} + \gamma_{1} + 2 \alpha_1)>0$, which implies:
\begin{equation}
n>\frac{\beta_{0}+\gamma_{0}+2\alpha_0}{(\beta_{1}+\gamma_{1}+2\alpha_1)-(\beta_{0}+\gamma_{0}+2\alpha_0)}
\end{equation}
whenever, $\beta_{1}+\gamma_{1}+2\alpha_1>\beta_{0}+\gamma_{0}+2\alpha_0$.
From Eq.~(\ref{eq:Fpn_internal}), we observe that the fixed-point curve intersects the $n=0$ plane at 
\begin{equation}\label{eq:p_n_eq_0}
	p=\frac{(\alpha_0+\beta_0+\gamma_0)}{\alpha_0+(\alpha_0+\beta_0+\gamma_0)},
\end{equation}
Similarly, the fixed-point curve intersects the $n=1$ plane at
\begin{equation}\label{eq:p_n_eq_1}
	p=\frac{(\alpha_1+\beta_1+\gamma_1)}{\alpha_1+(\alpha_1+\beta_1+\gamma_1)}.
\end{equation}
In addition, the curve intersects the plane $p=0$ at
\begin{equation}\label{eq:intermediate_n}
	n_m=\frac{-(\alpha_0+\beta_0+\gamma_0)}{(\alpha_1 + \beta_1+\gamma_1)-(\alpha_0 + \beta_0+\gamma_0)},
\end{equation}
and does not intersect the plane $p=1$. 
Just from the eigenvalue expressions---\ref{eq:eigen_sum_nm}, \ref{eq:eigen_prod_nm}, we can directly draw the following conclusions: 
Whenever \((\beta_{1} + \gamma_{1})\alpha_0 < (\beta_{0} + \gamma_{0})\alpha_1\), then the product of the  eigenvalues $\lambda_1\lambda_2<0$ (see Eq.~(\ref{eq:conjugate})), since the denominator in Eq.~(\ref{eq:conjugate}) is always positive whenever the fixed point line Eq.~(\ref{eq:internal}) exists. Therefore, the corresponding fixed points are saddles. On the other hand, whenever \((\beta_{1} + \gamma_{1})\alpha_0 > (\beta_{0} + \gamma_{0})\alpha_1\) the eigenvalue product $\lambda_1\lambda_2>0$, which implies the fixed point can be stable or unstable depending on the sign of $\lambda_1+\lambda_2$ and hence the sign and relative values of $\beta_1-\gamma_1$ and $\beta_0-\gamma_0$.
We now examine these conditions in detail to clarify the existence and stability of internal fixed points across different parameter regimes.

\textit{Grey zone:}
In this parameter region \((\beta_{1}+\gamma_{1})\alpha_0<(\beta_{0}+\gamma_{0})\alpha_1\), implies that the eigenvalues $\lambda_1$ and $\lambda_2$ have opposite signs. Consequently, the system possesses saddle fixed points.

\textit{Olive zone:}
In this parameter regime, the condition $(\beta_1+\gamma_1)\alpha_0>(\beta_0+\gamma_0)\alpha_1$ esures that $\lambda_1 \lambda_2>0$, so the two eigenvalues share the same sign. Moreover, in this region both $(\beta_1+\gamma_1)>0$ ($0$, as our assumption, its corresponds to defections dominant game parameters) and $(\beta_0+\gamma_0)>-\alpha_0$, consequently from Eq.~(\ref{eq:p_n_eq_0}) and (\ref{eq:p_n_eq_1}), the fixed-point curve intersects both the $n=0$ and $n=1$ plane.
To determine the stability of the fixed points---that is, the sign of $\lambda_1+\lambda_2$---we must analyze the relative signs of $(\beta_0-\gamma_0)$ and $(\beta_1-\gamma_1)$. Since $\lambda_1+\lambda_2$ is a linear and monotonic function of $n$, the stability of the entire fixed-point curve can be inferred solely from its behaviour at the two endpoints, $n=0$ and $n=1$. Accordingly, in the first quadrant, the fixed points are always unstable, whereas in the third quadrant, they are always stable. In contrast, in the second and fourth quadrants, the stability behaviour differs between the two endpoints; thus, in these regions, part of the fixed-point curve is stable while the remaining part is unstable.\\
\textit{Violet zone:}
In this parameter regime $(\beta_0+\gamma_0)<-\alpha_0$ and $(\beta_1+\gamma_1)>0$, we have  $\lambda_1 \lambda_2>0$. Furthermore, from Eq.~(\ref{eq:p_n_eq_0}), (\ref{eq:p_n_eq_1}), and (\ref{eq:intermediate_n}), the fixed-point curve is seen to intersect both the $n=1$ plane and the $p=0$ plane.
To determine the stability of these fixed points---that is, the sign of $\lambda_1+\lambda_2$---we must examine the relative signs of $(\beta_0-\gamma_0)$ and $(\beta_1-\gamma_1)$, as summarized in subplot $a.$ of Fig.~(\ref{fig:boat4}). 
In the first quadrant of Subplot $a.$, where $(\beta_0-\gamma_0)> 0$ and $(\beta_1-\gamma_1)> 0$, we have $\lambda_1+\lambda_2>0$ and the fixed points are unstable. In contrast, in the third quadrant of Subplot $a.$, where $(\beta_0-\gamma_0)< 0$ and $(\beta_1-\gamma_1)< 0$, which obtain $\lambda_1+\lambda_2<0$, and the fixed points are stable. In the second quadrant of Subplot~$a.$, the condition $(\beta_1-\gamma_1)>0$, implies that the fixed point lying on the $n=1$ plane is always unstable. The stability of the fixed point on the $p=0$ plane, however, depends on the value of $n$ at that fixed point. Substituting $n=n_m$ into the expression for $\lambda_1+\lambda_2$, one finds that the sign of $\lambda_1+\lambda_2$ is determined by the sign of $(\alpha_1+\beta_1+\gamma_1)(\beta_0-\gamma_0)-(\alpha_0+\beta_0+\gamma_0)(\beta_1-\gamma_1)$. 
Hence, the fixed point is unstable whenever
\begin{equation}\label{eq:unstable_v_zone}
	(\beta_0-\gamma_0)>-K,
\end{equation}
and stable whenever
\begin{equation}\label{eq:stable_v_zone}
		(\beta_{0}-\gamma_{0}) 
		< -K, 
\end{equation}
{where } 
$K = {|A|\,(\beta_{1}-\gamma_{1})}/{|B|}$, $A=\alpha_0+\beta_0+\gamma_0$ and $B=\alpha_1+\beta_1+\gamma_1$.
We therefore conclude that in the second quadrant the fixed-point curve is entirely unstable when condition~(\ref{eq:unstable_v_zone}) is hold; otherwise, only part of the curve is unstable. Similarly, in the fourth quadrant, the fixed-point curve is completely stable whenever condition~(\ref{eq:stable_v_zone}) is satisfied; otherwise, only a portion of it is stable.

\subsection{Finding parameter ranges for Hopf-bifurcation} \label{app:hopf_parameter_conditions}
	As established in the main text, the necessary Hopf bifurcation at the internal fixed-point continuum can appear depending on the value of the resource state $n= \hat{n}$, where Eq.~(\ref{eq:n_cap}) shows that $\hat n\in(0,1)$ only under either of the following two mutually exclusive parameter configurations: $(i)~\beta_0>\gamma_0~\text{and}~\beta_1<\gamma_1$ or $(ii)~\beta_0<\gamma_0~\text{and}~\beta_1>\gamma_1$.
	Furthermore, for a Hopf bifurcation to occur, $\eta$ has to be positive; only then eigenvalues~\(\lambda_{1}~\text{and}~\lambda_{2}\) become purely imaginary. On inspecting the numerator and denominator of RHS in Eq.~(\ref{eq:hopf_eta}), this is found to occur when either
	\begin{subequations}
		\begin{align}
			(\beta_1+\gamma_1)\alpha_0 >(\beta_0+\gamma_0)\alpha_1,
			\label{eq:condition11}\\ \text{and~}
			(1-\hat{n})(\beta_{0}+\gamma_{0}+2\alpha_0)+\hat{n}(\beta_{1}+\gamma_{1}+2\alpha_1)>0 ,
			\label{eq:condition12}
		\end{align}
	\end{subequations}
	or 
	\begin{subequations}
		\begin{align}
			(\beta_1+\gamma_1)\alpha_0<(\beta_0+\gamma_0)\alpha_1, 
			\label{eq:condition21}\\ \text{and~}
			(1-\hat{n})(\beta_{0}+\gamma_{0}+2\alpha_0)+\hat{n}(\beta_{1}+\gamma_{1}+2\alpha_1)<0.
			\label{eq:condition22}
		\end{align}
	\end{subequations}
	One can note from Fig.~\ref{fig:boat4}(C), that a Hopf-bifurcation can arise in the olive and violet shaded regions only, where, depending on the parameter values, the internal fixed point can lose its stability. In contrast, in the grey region, the fixed point remains a saddle point irrespective of the parameter values. Therefore, we focus on the olive and violet regions to determine the parameter criteria for the occurrence of a limit cycle.
	\begin{enumerate}
		\item \textit{Olive region:} This region is demarcated by the one of the inequality $(\beta_1+\gamma_1)\alpha_0>(\beta_0+\gamma_0)\alpha_1 $ (rest are shown in the earlier section~1.), hence the necessary and sufficient condition for $\eta$ to be positive is 
		\begin{equation}
			(1-\hat{n})(\beta_0+\gamma_0+2\alpha_0)+\hat{n}(\beta_1+\gamma_1+2\alpha_1)>0
		\end{equation}
		This condition is automatically satisfied because both 
		\((\beta_0+\gamma_0+2\alpha_0)\) and \((\beta_1+\gamma_1+2\alpha_1)\) 
		are positive throughout the olive region. 
				
		Therefore, a Hopf bifurcation can occur in the olive regions in both the second and fourth quadrants of Fig.~\ref{fig:boat4}(C)b.  
		Here, each quadrant fixed point structure is shown on the $(n,p)$ subspace, where the resource level $n=\hat{n}$ (when $L=0$) demarcates the transversely stable and unstable segments of the non-isolated fixed-point continuum. Furthermore, the third and first quadrant has a complete stable and unstable fixed point structure, because there is no such $n=\hat{n}$, when $L=0$. 
		
		Further moving from the third to the first quadrants by simultaneously changing $(\beta_0-\gamma_0)$ and $(\beta_1-\gamma_1)$ from negative to positive produces a qualitative reorganization of the transverse stability, with the fixed-point continuum changing from completely stable to completely unstable.
		
		\item \textit{Violet region:} This region is demarcated by the conditions 
		\(\beta_1+\gamma_1>0\) and \(\beta_0+\gamma_0+\alpha_0<0\), 
		which always satisfies the inequality 
		\((\beta_1+\gamma_1)\alpha_0>(\beta_0+\gamma_0)\alpha_1\). 
		Hence, the necessary and sufficient condition for \(\eta\) to be positive, whenever $\hat{n} \in (0,1)$, is
		\begin{equation}
			(1-\hat{n})(\beta_0+\gamma_0+2\alpha_0)+\hat{n}(\beta_1+\gamma_1+2\alpha_1)>0,
		\end{equation} 
		which is always true given the intermediate fixed points with $q>0$ exists  (see Eq.~(\ref{eq:internal})). 
		However, the admissible range of the equilibrium resource state is more constrained, because it attains its minimum physical value at $n = n_m~(>0)$~(see Eq.~(\ref{eq:intermediate_n})). Since,  $n_m<\hat{n}<1$, this imposes stricter constraints on the parameter space:
		\begin{itemize}
			\item[(i)] \(\beta_0 > \gamma_0\) and \(\beta_1 < \gamma_1\), and $\alpha_1(\beta_0-\gamma_0)+\alpha_0(\gamma_1-\beta_1)>2(\beta_1\gamma_0-\gamma_1 \beta_0)$;
			\item[(ii)] \(\beta_0 < \gamma_0\) and \(\beta_1 > \gamma_1\), and $\alpha_1(\beta_0-\gamma_0)+\alpha_0(\gamma_1-\beta_1)<2(\beta_1\gamma_0-\gamma_1 \beta_0)$.
		\end{itemize}
		Cases~(i) and~(ii) correspond to the regions spanning the second and fourth quadrants of Fig.~\ref{fig:boat4}(C)a, where the internal equilibrium resource level $n=\hat{n}$ demarcates the transversely stable and unstable segments of the fixed-point continuum. Thus, in these two quadrants, a Hopf-bifurcation can occur. 
	\end{enumerate}

\section{Analysis for Constructing Subplots (b) and (c) in Fig.~\ref{fig:reward_punishment_n0n1}}\label{appendix_D} 
Based on the generic analysis presented in Appendix~\ref{Appendix_C}, we now examine its physical interpretation in terms of reward and punishment mechanisms under different resource states. Specifically, we consider reward $(\rho^{+}_{0})$ and punishment $(\rho^{-}_{0})$ at the depleted resource state, and reward $(\rho^{+}_{1})$ and punishment $(\rho^{-}_{1})$ at the replete resource state. These are analysed within the framework of donation games, characterized by benefits $b_{0}$ and $b_{1}$ at the depleted and replete resource states, respectively, and a cost $c$. The generic symbolic parameters introduced in the analytical treatment correspond to the following physical quantities:
\(\beta_{0}=c-\rho^+_0\), \(\beta_{1}=c-\rho^+_1\), \(\gamma_{0}=c-\rho^-_0\), \(\gamma_{1}=c-\rho^-_1\), \(\alpha_0=b_0-c\) and \(\alpha_1=b_1-c\).

We impose a predefined assumption that individual behaviour is defection-dominated when the resource state is replete. Consequently, both $\beta_{1} > 0$ and $\gamma_{1} > 0$ hold. Under these conditions, we construct the corresponding dynamical outcomes shown in Fig.~\ref{fig:reward_punishment_n0n1}.

In this figure, the grey region is bounded by the condition \(\rho^+_0+\rho^-_0<2c-\frac{\alpha_0}{\alpha_1}\left(2c-\rho^+_1-\rho^-_1\right)\); and represents the saddle fixed-point curve region of the $(\rho^{+}_{0},\,\rho^{-}_{0})$ parameter space, which is indicated by the chain-dotted line. The complementary regime, \(\rho^+_0+\rho^-_0>2c-\frac{\alpha_0}{\alpha_1}\left(2c-\rho^+_1-\rho^-_1\right)\), admits the possibility of either stable or unstable continuum of fixed points. Within this regime, we further distinguish two sub-cases depending on whether\(\rho^+_0+\rho^-_0 < b_0 + c\) or \(\rho^+_0+\rho^-_0 > b_0 + c\); which are separated by the dotted black line in the parameter space.

\begin{enumerate}[label=(\alph*)]
	\item \(\rho^+_0+\rho^-_0>2c-\frac{\alpha_0}{\alpha_1}\left(2c-\rho^+_1-\rho^-_1\right)\) and \(\rho^+_0+\rho^-_0 > b_0 + c\):
	By expressing the generic coefficients $(\beta_{0}, \gamma_{0}, \beta_{1}, \gamma_{1})$ in terms of reward and punishment parameters, Panel~(a) of subplot~(C) in Fig.~\ref{fig:boat4}, corresponding to the intermediate resource state $n \in (0,1)$, reveals the associated stability regions, which are shown in Fig.~\ref{fig:reward_punishment_n0n1}. Here, the stability of the fixed-point curve depends on the relative magnitudes and signs of $(\rho^{-}_{0} - \rho^{+}_{0})$ and $(\rho^{-}_{1} - \rho^{+}_{1})$, as discussed below.\\
	When $\rho^-_1-\rho^+_1>0$, for $(\rho^-_0-\rho^+_0)>-K$, the fixed-point curve is unstable, and for $(\rho^-_0-\rho^+_0)<-K$, part of the fixed-point curve is stable, while the remaining part is unstable, where \(K = -\frac{|b_0+c-\rho^+_0-\rho^-_0|}{b_1+c-\rho^+_1-\rho^-_1}(\rho^+_1-\rho^-_1)\). On the other hand,
	Whenever $(\rho^-_1-\rho^+_1)<0$, for $(\rho^-_1-\rho^+_1)<K$, the fixed-point curve is stable, and for $(\rho^-_0-\rho^+_0)>K$, part of the fixed-point curve is stable, while the remaining part is unstable.
	
	\item \(\rho^+_0+\rho^-_0>2c-\frac{\alpha_0}{\alpha_1}\left(2c-\rho^+_1-\rho^-_1\right)\) and \(\rho^+_0+\rho^-_0 < b_0 + c\):
	By expressing the generic coefficients $(\beta_{0}, \gamma_{0}, \beta_{1}, \gamma_{1})$ in terms of reward and punishment parameters, Panel~(b) of subplot~(C) in Fig.~\ref{fig:boat4}, corresponding to the intermediate resource state $n \in (0,1)$, reveals the associated stability regions, which are shown in Fig.~\ref{fig:reward_punishment_n0n1}. Here, the stability of the fixed-point curve depends on the relative magnitudes and signs of $(\rho^{-}_{0} - \rho^{+}_{0})$ and $(\rho^{-}_{1} - \rho^{+}_{1})$, as discussed below.\\
	When $(\rho^-_1-\rho^+_1)>0$ the corresponding fixed point at $n=1$ is unstable. Now if $(\rho^-_0-\rho^+_0)>0$, the $\lambda_1+\lambda_2<0$ for any allowed values of $n$ (see Appendix~\ref{Appendix_C}) and hence the whole fixed-point curve is unstable, and if $(\rho^-_0-\rho^+_0)<0$, the part of the fixed-point curve is stable and the remaining is unstable depending on the relative values of $(\rho^-_0-\rho^+_0)$ and $(\rho^-_1-\rho^+_1)$. 
	On the other hand,		
	when $(\rho^-_1-\rho^+_1)<0$ the corresponding fixed-point curve at $n=1$ is stable. Now, If $(\rho^-_0-\rho^+_0)<0$, the $\lambda_1+\lambda_2<0$ (see Appendix~\ref{Appendix_C}) for any allowed values of $n$, therefore, the whole fixed-point curve is stable. On the other hand, if $(\rho^-_0-\rho^+_0)>0$, the part of the fixed-point curve is stable and the remaining is unstable depending on the relative values of $(\rho^-_0-\rho^+_0)$ and $(\rho^-_1-\rho^+_1)$.
\end{enumerate}
\section{Eco-Evolutionary Outcomes in the Reciprocal $p$-$q$ Trait Space}\label{appendix_E}
\begin{figure}[htbp]
	\centering	
	\includegraphics[scale=0.41]{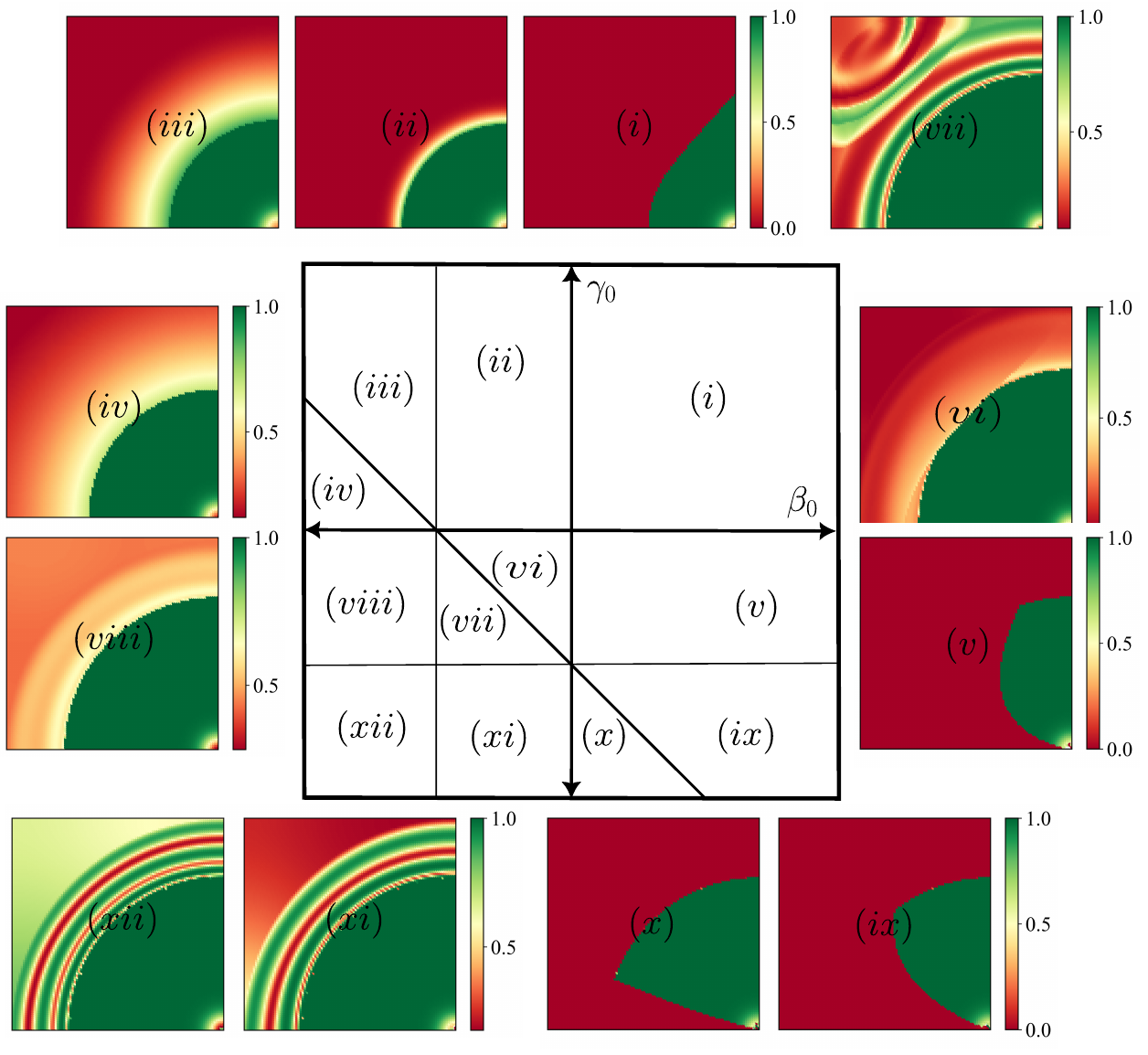}
	\caption{Phase diagram illustrating the emergent outcomes in a TOC scenario as a function of two key parameters (initial $p$ and initial $q$, each ranging from 0 to 1) for all possible games (i) Prisoner’s dilemma, (ii) Chicken games, (iii) Leader games, (iv) Battle of sexes, (v) Stag-hunt, (vi) Harmony I, (vii) Harmony II, (viii) Deadlock II, (ix) Coordination I, (x) Coordination II, (xi) Harmony III, and (xii) Deadlock I. The colour-bar indicates the equilibrium level resource availability, ranging from red (complete resource collapse) to green (resources sustained). The figure illustrates the nonlinear dependence of collective outcomes on institutional and ecological incentives, and demonstrates how their interplay can prevent the ToC. The results correspond to $\beta_{1}=0.6$, $\gamma_{1}=0.5$, $\alpha_0=1.0$, $\alpha_1=1.2$ and an initial resource level $n_0=0.2$.}
	\label{fig:Bistability_across_parameters}
\end{figure}

	We explored the system dynamics across the $\beta_{0}$--$\gamma_{0}$ parameter space while fixing $\beta_{1}=0.6,~\gamma_{1}=0.5$, corresponding to a scenario in which individuals engage in a Prisoner's Dilemma under resource-abundant conditions. It is worth emphasizing that this analysis is not restricted to the Donation Game. While the game at the replete resource state is fixed as a Prisoner's Dilemma, varying $(\beta_0,\gamma_0)$ spans several distinct $2\times2$ game classes at the depleted resource state, including the Chicken, Stag-Hunt, Harmony, Coordination, and other games, as classified in  Fig.~\ref{fig:boat4} of subplots $(A)$ and $(B)$ for depleted and replete resource state respectively. In particular, the Chicken game belongs to the same anti-coordination class as the Snowdrift/Hawk--Dove game.\\
	Corresponding dynamical long-term evolutionary outcomes are obtained in Fig.~\ref{fig:Bistability_across_parameters} by considering all initial conditions $p_0,q_0\in(0,1)$ with an initial resource level $n_0=0.2$. 
	
	Here, all these games are explicitly demarcated in the $(\beta_0,\gamma_0)$ parameter space, and all twelve possible game classes are identified in the caption of Fig.~\ref{fig:Bistability_across_parameters}, following the classifications of Pandit et al.~\cite{Pandit2018}, Hummert et. al. \cite{HBB_2014_MOLB}, and Sadhukhan et. al. \cite{SCR_2021_PRE} Hauert ch. \cite{H_2002_IJBC}. This representation makes it possible to directly relate the fixed-point and stability regions of our general eco-evolutionary framework to the corresponding underlying game structures. 
	
	Several distinct dynamical regimes emerge throughout the parameter space. In some regions, the long-term resource state exhibits two alternative outcomes, indicating bistability and a strong dependence on initial conditions. In other regions, the equilibrium resource level varies continuously with the model parameters, producing smooth gradients in the asymptotic resource state. Furthermore, complex and irregular patterns appear in certain parameter regions. A detailed dynamical analysis reveals that these patterns correspond to stable limit cycle behavior~(Fig.~\ref{fig:hopf_bifuraction_reward_punishment}), leading to persistent oscillations in both the strategic and environmental variables.
	
\begin{figure}[htbp]
	\centering	
	\includegraphics[scale=0.42]{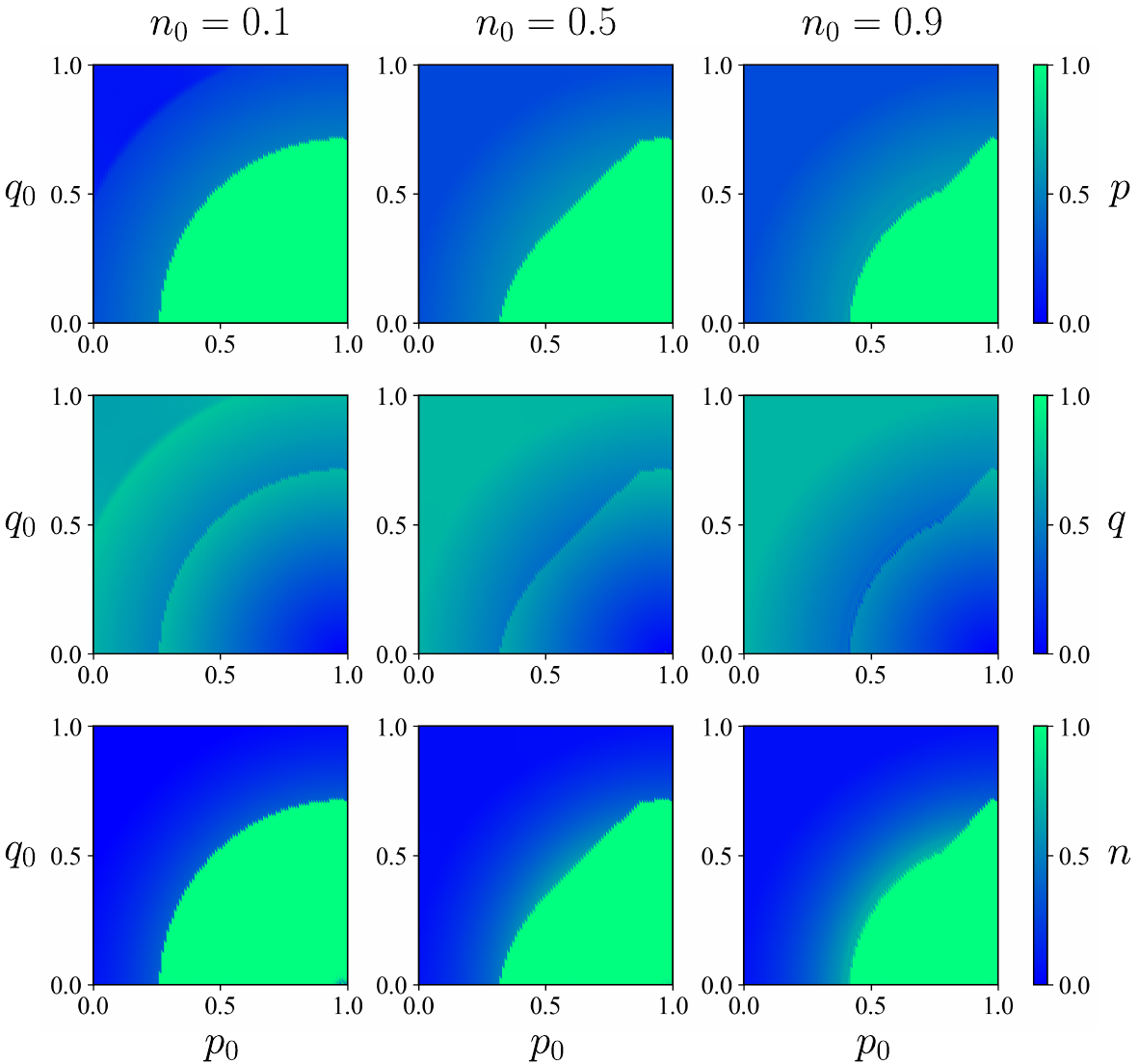}
	\caption{Long-term eco-evolutionary outcomes of the reactive traits $p$ and $q$ and the resource state $n$ over the full range of initial reciprocity $p_0$ and generosity $q_0$, for three different initial resource states, $n_0=0.1$, $0.5$, and $0.9$. The common parameter values are $\beta_0=-0.5$, $\gamma_0=-0.2$, $\beta_1=0.6$, $\gamma_1=0.5$, $\alpha_0=1.0$, and $\alpha_1=1.2$. The first, second, and third rows show the long-term values of $p$, $q$, and $n$, respectively.}
	\label{fig:sensitivity_analysis}
\end{figure}

	To further characterize the system dynamics, we examined the long-term behavior of the variables $p$, $q$, and $n$ for all initial conditions $p_0,q_0\in(0,1)$ and for three representative initial resource levels, $n_0\in{0.1,0.5,0.9}$. The corresponding results are presented in Fig.~\ref{fig:sensitivity_analysis} for $\beta_{0}=-0.5$, $\gamma_{0}=-0.2$, $\beta_{1}=0.6$,  $\gamma_{1}=0.5$, $\alpha_0=1.0$, $\alpha_1=1.2$.
	
	The figure demonstrates that the resource is sustained in the long run when the population evolves toward Generous Tit-for-Tat (GTFT)-like strategies, regardless of the initial resource level. Interestingly, when the initial resource abundance is sufficiently high, resource persistence can also occur for non-generous strategies. These findings highlight the joint influence of environmental conditions and evolutionary strategy dynamics on long-term resource sustainability. In particular, both the environment's initial state and the evolving behavioural strategies play an important role in determining whether the resource is ultimately maintained or depleted.


\section{For the $\theta$ dependence of the system’s eigenvalues}\label{Appendix_F}
So far, in Sec.~\ref{model}, we have considered two formulations of the resource dynamics: 1. \emph{reciprocity--generosity-driven resource feedback}: one in which the environmental feedback depends directly on the reciprocity $p$ and generosity $q$ of the resident population, and 2. \emph{cooperation-level-driven resource feedback}: one in which the environmental feedback depends on the resulting cooperation level. In the main analysis, the feedback strength was taken to be symmetric, corresponding to $\theta=1$. To examine the sensitivity of our results to this assumption, we now generalize both formulations to $\theta\neq1$ and analyse the corresponding Jacobian and stability conditions.

\subsection{\emph{Reciprocity--Generosity-driven resource feedback}}\label{reci_geno_res_feed}
In this feedback type, we should use in Eq.~(\ref{eq:resource_gen}),
\begin{equation}
	f(p,q)=(1+\theta)(p+q)-2,
\end{equation}
with $a_r=a_g=1$, yielding the corresponding resource dynamics as
\begin{equation}
	\dot n
	=
	\epsilon n(1-n)
	\left[(1+\theta)(p+q)-2\right].
	\label{eq:resource_gen1}
\end{equation}
The condition $\dot n=0$ consequently yields the two boundary resource states, $n=0$ and $n=1$, together with the internal resource nullcline
\begin{equation}
	p+q=\frac{2}{1+\theta}.
	\label{eq:resource_nullcline_rg}
\end{equation}
Combining these conditions with the evolutionary equilibrium condition $F(p,q,n)=0$ obtained from Eqs.~(\ref{eq:AD_p}) and (\ref{eq:AD_q}), the system admits three classes of non-isolated fixed points:
\begin{enumerate}[label=(\alph*)]
	\item non-isolated fixed points at the intersection of surfaces---$n=0$  and $F(p,q,n)=0$,
	\item non-isolated fixed points at the intersection of surfaces---$n=1$ and $F(p,q,n)=0$,
	\item non-isolated fixed points at the intersection of surfaces---$p+q=\frac{2}{1+\theta}$ and $F(p,q,n)=0$, with $0<n<1$.
\end{enumerate}
We next examine the local stability of each of these fixed-point continua from the eigenvalues of the corresponding Jacobian.

\subsubsection{Non-isolated fixed points at the intersection of surfaces---$n=0$  and $F(p,q,n)=0$}
Evaluating the Jacobian along the depleted-resource $(n=0)$ fixed-point continuum yields the three eigenvalues
\begin{enumerate}[label=(\roman*)]\label{eq:eigen_dpeleted_theta_rg}
	\item $\lambda_1=\epsilon (1+\theta)(p+q-\frac{2}{1+\theta})$,
	\item $\lambda_2=qA_0+(1-p)B_0$,
	\item $\lambda_3=0$,
\end{enumerate}
where, $A_0\equiv q(\beta_0-\gamma_0)+(1-2r)(\alpha_0+\gamma_0)+\beta_0$ and $B_0\equiv (1+r)(\beta_0-\gamma_0)-A_0.$
\subsubsection{Non-isolated fixed points at the intersection of surfaces---$n=1$  and $F(p,q,n)=0$}
The eigenvalues of the Jacobian matrix at the replete state ($n=1$) are given by:
\begin{enumerate}[label=(\roman*)]
	\item $\lambda_1=\epsilon (1+\theta)(\frac{2}{1+\theta}-(p+q))$,
	\item $\lambda_2=qA_1+(1-p)B_1$,
	\item $\lambda_3=0$,
\end{enumerate}
where, $A_1\equiv q(\beta_1-\gamma_1)+(1-2r)(\alpha_1+\gamma_1)+\beta_1$ and $B_1\equiv (1+r)(\beta_1-\gamma_1)-A_1.$ 
\subsubsection{Non-isolated fixed points at the intersection of surfaces---$p+q=\frac{2}{1+\theta}$  and $F(p,q,n)=0$}
The eigenvalues of the Jacobian matrix at the intermediate state of resource $n\in(0,1)$ for $p+q=\frac{2}{1+\theta}$ are given by: 
\begin{enumerate}[label=(\roman*)]
	\item $\displaystyle
	\begin{aligned}[t]
		\lambda_1+\lambda_2
		={}&
		\frac{q}{(1-r)^2(1-r^2)}\times\\[-2pt]
		&\hspace{-1.5cm}{\scriptsize\left[\left(1-\frac{2}{1+\theta}\right)G +(1+r)\left((1-n)(\beta_0-\gamma_0) +n(\beta_1-\gamma_1)\right)\right],}
	\end{aligned}
	$
	\item $\lambda_1\lambda_2=\frac{-\epsilon n(1-n)\left[q(1+\theta)+(1-\frac{2}{1+\theta})\theta\right]}{(1-r)^2 (1-r^2)}H,$
	\item $\lambda_3=0$
\end{enumerate}
where, 
\[
\begin{aligned}
	G\equiv{}&(1-n)\left[(2\alpha_0+\beta_0+\gamma_0)p-2(\alpha_0+\beta_0)q-\alpha_0-2\gamma_0\right]\\
	&+n\left[(2\alpha_1+\beta_1+\gamma_1)p-2(\alpha_1+\beta_1)q-\alpha_1-2\gamma_1\right],
\end{aligned}
\] and
\[
\begin{aligned}
	H\equiv{}&(\beta_1-\gamma_1-\beta_0-\gamma_0)q(1+r)+(\alpha_1+\gamma_1-\alpha_0-\gamma_0)r(1-r)\\
	&-(\beta_1-\beta_0)(1-r).
\end{aligned}
\]

In order to comprehend the sensitivity of the results on $\theta$, let us look at the schematic Fig.~\ref{fig:theta_sensitivity}(a). Here we have a representative black solid curve of non-isolated fixed points. Assume we are focussing on the $n=0$ plane. In general, the black curve can have one of the twelve possible stability structures (depending on the signs of real part of all the three eigenvalues) as shown in Fig.~\ref{fig:boat4}(A). We note that there is a grey dashed line $p+q=1$ in the figure---same as the one we have plotted in Fig.~\ref{fig:theta_sensitivity}(a) and marked with $\theta=1$; the fixed point lying on it has naturally $\lambda_1=0$, the fixed points to the left of it have $\lambda_1<0$, and  the fixed points to the right of it have $\lambda_1>0$. Now if $\theta>1$ (or $\theta<1$), then the straight dashed line demarcating the fixed points' values of $\lambda_1$ is parallel to the one with $\theta=1$ and to the left (or to the right). Thus, if the continuum of the fixed points is already completely unstable owing to positivity of $\lambda_2$, $\theta\ne1$ brings no change to the dynamics. In case the continuum of the fixed points is stable and unstable in parts, then due to $\theta\ne1$, the range of stable portion of the fixed point continuum can at worst reduce which just means that the measure of initial conditions asymptotically reaching a fixed point reduces---qualitatively, the results, thus, remain intact. Similar argument can be given when we focus on $n=1$ plane and Fig.~\ref{fig:boat4}(B). The internal fixed points can also be shown to render similar conclusions which is refrain from discussion to avoid repetition of similar arguments.

\subsection{\emph{Cooperation-level-driven resource feedback}} \label{Coop_l_res_feed}
In this case,
\begin{equation}
	f(p,q)=\frac{1}{1-r}\left[p+q\theta-1\right],
\end{equation}
and the corresponding resource dynamics, Eq.~(\ref{eq:resource_gen}), is
\begin{equation}
	\dot n=\frac{\epsilon}{1-r} n(1-n)\left[p+q\theta-1\right].
	\label{eq:resource_gen1}
\end{equation}
The condition $\dot n=0$ consequently yields the two boundary resource states, $n=0$ and $n=1$, together with the internal resource nullcline
\begin{equation}
	p+q\theta=1.
	\label{eq:resource_nullcline_rg}
\end{equation}
Combining these conditions with the evolutionary equilibrium condition $F(p,q,n)=0$ obtained from Eqs.~(\ref{eq:AD_p}) and (\ref{eq:AD_q}), the system admits three classes of non-isolated fixed points:
\begin{enumerate}[label=(\alph*)]
	\item non-isolated fixed points at the intersection of surfaces---$n=0$  and $F(p,q,n)=0$,
	\item non-isolated fixed points at the intersection of surfaces---$n=1$ and $F(p,q,n)=0$,
	\item non-isolated fixed points at the intersection of surfaces---$p+q\theta=1$ and $F(p,q,n)=0$, with $0<n<1$.
\end{enumerate}
We next examine the local stability of each of these fixed-point continua from the eigenvalues of the corresponding Jacobian.

\subsubsection{Non-isolated fixed points at the intersection of surfaces---$n=0$  and $F(p,q,n)=0$}
Evaluating the Jacobian along the depleted-resource $(n=0)$ fixed-point continuum yields the three eigenvalues
\begin{enumerate}[label=(\roman*)]\label{eq:eigen_dpeleted_theta_coop}
	\item $\lambda_1=\epsilon(p+q\theta-1)$,
	\item $\lambda_2=qA_0+(1-p)B_0$,
	\item $\lambda_3=0$.
\end{enumerate}
\subsubsection{Non-isolated fixed points at the intersection of surfaces---$n=1$  and $F(p,q,n)=0$}
The eigenvalues of the Jacobian matrix at the replete state ($n=1$) are given by:
\begin{enumerate}[label=(\roman*)]
	\item $\lambda_1=\epsilon(1-(p+\theta q))$,
	\item $\lambda_2=qA_1+(1-p)B_1$.
	\item $\lambda_3=0$.
\end{enumerate}
\subsubsection{Non-isolated fixed points at the intersection of surfaces---$p+\theta q=1$ and $F(p,q,n)=0$}
The eigenvalues of the Jacobian matrix at the intermediate state of resource $n\in(0,1)$ for $p+\theta q=1$ are given by: 
\begin{enumerate}[label=(\roman*)]	
	\item $\displaystyle
	\begin{aligned}[t]
		\lambda_1+\lambda_2
		={}&
		\frac{q}{(1-r)^2(1-r^2)}\times\\[-2pt]
		&\hspace{-1.5cm}{\scriptsize\left[{(\theta-1)G +(1+r)((1-n)(\beta_0-\gamma_0)+n(\beta_1-\gamma_1))}\right]},
	\end{aligned}
	$
	\item  $\lambda_1\lambda_2=\frac{-\epsilon n(1-n)q(1+\theta^2)}{(1-r)^2 (1-r^2)}H,$
	\item $\lambda_3=0$.
\end{enumerate}

\begin{figure}[ht!]
	\centering		
	\includegraphics[width=0.4\textwidth]{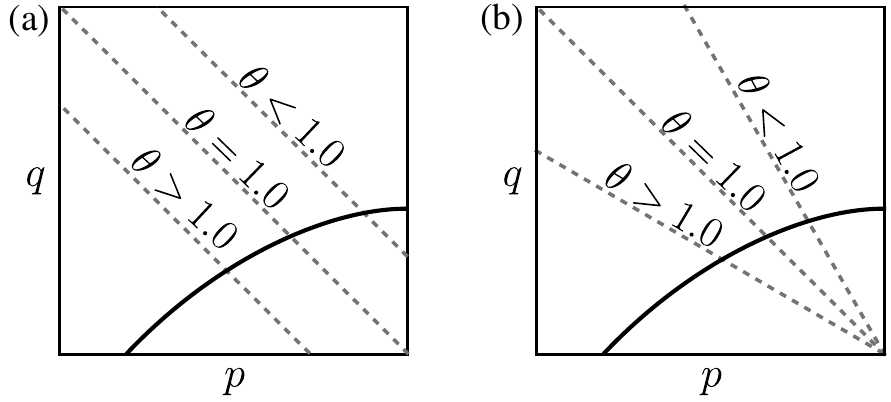}
	\caption{Schematic to understand the sensitivity of the fixed-point stability regions to the environmental feedback parameter $\theta$. The black solid curve represents the continuum of fixed points. The grey dashed lines correspond to the condition $\lambda_1=0$.  Subplot~(a) corresponds to the {reciprocity--generosity-driven resource feedback} and Subplot~(b) corresponds to the {cooperation-level-driven resource feedback.}}
	\label{fig:theta_sensitivity}
\end{figure}

We can comprehend the sensitivity of the results on $\theta$ and conclude about their qualitative robustness, exactly as we have done in the previous case. When one inspects the relevant schematic---Fig.~\ref{fig:theta_sensitivity}(b)---one notes that the only minor change in the details is that the lines corresponding to different $\theta$-values are not parallel but uniquely intersecting at $(p,q)=(1,0)$.

	\bibliographystyle{apsrev4-2} 
	\bibliography{mandal_etal_bibliography}

\begin{thebibliography}{132}%
\makeatletter
\providecommand \@ifxundefined [1]{%
 \@ifx{#1\undefined}
}%
\providecommand \@ifnum [1]{%
 \ifnum #1\expandafter \@firstoftwo
 \else \expandafter \@secondoftwo
 \fi
}%
\providecommand \@ifx [1]{%
 \ifx #1\expandafter \@firstoftwo
 \else \expandafter \@secondoftwo
 \fi
}%
\providecommand \natexlab [1]{#1}%
\providecommand \enquote  [1]{``#1''}%
\providecommand \bibnamefont  [1]{#1}%
\providecommand \bibfnamefont [1]{#1}%
\providecommand \citenamefont [1]{#1}%
\providecommand \href@noop [0]{\@secondoftwo}%
\providecommand \href [0]{\begingroup \@sanitize@url \@href}%
\providecommand \@href[1]{\@@startlink{#1}\@@href}%
\providecommand \@@href[1]{\endgroup#1\@@endlink}%
\providecommand \@sanitize@url [0]{\catcode `\\12\catcode `\$12\catcode
  `\&12\catcode `\#12\catcode `\^12\catcode `\_12\catcode `\%12\relax}%
\providecommand \@@startlink[1]{}%
\providecommand \@@endlink[0]{}%
\providecommand \url  [0]{\begingroup\@sanitize@url \@url }%
\providecommand \@url [1]{\endgroup\@href {#1}{\urlprefix }}%
\providecommand \urlprefix  [0]{URL }%
\providecommand \Eprint [0]{\href }%
\providecommand \doibase [0]{https://doi.org/}%
\providecommand \selectlanguage [0]{\@gobble}%
\providecommand \bibinfo  [0]{\@secondoftwo}%
\providecommand \bibfield  [0]{\@secondoftwo}%
\providecommand \translation [1]{[#1]}%
\providecommand \BibitemOpen [0]{}%
\providecommand \bibitemStop [0]{}%
\providecommand \bibitemNoStop [0]{.\EOS\space}%
\providecommand \EOS [0]{\spacefactor3000\relax}%
\providecommand \BibitemShut  [1]{\csname bibitem#1\endcsname}%
\let\auto@bib@innerbib\@empty
\bibitem [{\citenamefont {Hardin}(1968)}]{hardin1968}%
  \BibitemOpen
  \bibfield  {author} {\bibinfo {author} {\bibfnamefont {G.}~\bibnamefont
  {Hardin}},\ }\href {https://doi.org/10.1126/science.162.3859.1243} {\bibfield
   {journal} {\bibinfo  {journal} {Science}\ }\textbf {\bibinfo {volume}
  {162}},\ \bibinfo {pages} {1243–1248} (\bibinfo {year} {1968})}\BibitemShut
  {NoStop}%
\bibitem [{\citenamefont {Rankin}\ \emph {et~al.}(2007)\citenamefont {Rankin},
  \citenamefont {Bargum},\ and\ \citenamefont {Kokko}}]{Rankin2007}%
  \BibitemOpen
  \bibfield  {author} {\bibinfo {author} {\bibfnamefont {D.~J.}\ \bibnamefont
  {Rankin}}, \bibinfo {author} {\bibfnamefont {K.}~\bibnamefont {Bargum}},\
  and\ \bibinfo {author} {\bibfnamefont {H.}~\bibnamefont {Kokko}},\ }\href
  {https://doi.org/10.1016/j.tree.2007.07.009} {\bibfield  {journal} {\bibinfo
  {journal} {Trends in Ecology \& Evolution}\ }\textbf {\bibinfo {volume}
  {22}},\ \bibinfo {pages} {643–651} (\bibinfo {year} {2007})}\BibitemShut
  {NoStop}%
\bibitem [{\citenamefont {Frischmann}\ \emph {et~al.}(2019)\citenamefont
  {Frischmann}, \citenamefont {Marciano},\ and\ \citenamefont
  {Ramello}}]{Frischmann2019}%
  \BibitemOpen
  \bibfield  {author} {\bibinfo {author} {\bibfnamefont {B.~M.}\ \bibnamefont
  {Frischmann}}, \bibinfo {author} {\bibfnamefont {A.}~\bibnamefont
  {Marciano}},\ and\ \bibinfo {author} {\bibfnamefont {G.~B.}\ \bibnamefont
  {Ramello}},\ }\href {https://doi.org/10.1257/jep.33.4.211} {\bibfield
  {journal} {\bibinfo  {journal} {Journal of Economic Perspectives}\ }\textbf
  {\bibinfo {volume} {33}},\ \bibinfo {pages} {211–228} (\bibinfo {year}
  {2019})}\BibitemShut {NoStop}%
\bibitem [{\citenamefont {Janssen}\ \emph {et~al.}(2019)\citenamefont
  {Janssen}, \citenamefont {Smith-Heisters}, \citenamefont {Aggarwal},\ and\
  \citenamefont {Schoon}}]{Janssen2019}%
  \BibitemOpen
  \bibfield  {author} {\bibinfo {author} {\bibfnamefont {M.~A.}\ \bibnamefont
  {Janssen}}, \bibinfo {author} {\bibfnamefont {S.}~\bibnamefont
  {Smith-Heisters}}, \bibinfo {author} {\bibfnamefont {R.}~\bibnamefont
  {Aggarwal}},\ and\ \bibinfo {author} {\bibfnamefont {M.~L.}\ \bibnamefont
  {Schoon}},\ }\href {https://doi.org/10.1080/13504622.2019.1632266} {\bibfield
   {journal} {\bibinfo  {journal} {Environmental Education Research}\ }\textbf
  {\bibinfo {volume} {25}},\ \bibinfo {pages} {1587–1604} (\bibinfo {year}
  {2019})}\BibitemShut {NoStop}%
\bibitem [{\citenamefont {Ostrom}(2008)}]{Ostrom2008}%
  \BibitemOpen
  \bibfield  {author} {\bibinfo {author} {\bibfnamefont {E.}~\bibnamefont
  {Ostrom}},\ }\bibinfo {title} {Tragedy of the commons},\ in\ \href
  {https://doi.org/10.1057/978-1-349-95121-5_2047-1} {\emph {\bibinfo
  {booktitle} {The New Palgrave Dictionary of Economics}}}\ (\bibinfo
  {publisher} {Palgrave Macmillan UK},\ \bibinfo {year} {2008})\ p.\ \bibinfo
  {pages} {1–5}\BibitemShut {NoStop}%
\bibitem [{\citenamefont {Bromley}\ and\ \citenamefont
  {Cernea}()}]{bromley1989wbp}%
  \BibitemOpen
  \bibfield  {author} {\bibinfo {author} {\bibfnamefont {D.~W.}\ \bibnamefont
  {Bromley}}\ and\ \bibinfo {author} {\bibfnamefont {M.~M.}\ \bibnamefont
  {Cernea}},\ }\href
  {http://documents.worldbank.org/curated/en/548811468740174575} {\emph
  {\bibinfo {title} {The management of common property natural resources: Some
  conceptual and operational fallacies}}}\BibitemShut {NoStop}%
\bibitem [{\citenamefont {Zea‐Cabrera}\ \emph {et~al.}(2006)\citenamefont
  {Zea‐Cabrera}, \citenamefont {Iwasa}, \citenamefont {Levin},\ and\
  \citenamefont {Rodríguez‐Iturbe}}]{ZeaCabrera2006}%
  \BibitemOpen
  \bibfield  {author} {\bibinfo {author} {\bibfnamefont {E.}~\bibnamefont
  {Zea‐Cabrera}}, \bibinfo {author} {\bibfnamefont {Y.}~\bibnamefont
  {Iwasa}}, \bibinfo {author} {\bibfnamefont {S.}~\bibnamefont {Levin}},\ and\
  \bibinfo {author} {\bibfnamefont {I.}~\bibnamefont {Rodríguez‐Iturbe}},\
  }\href {http://dx.doi.org/10.1029/2005WR004514} {\bibfield  {journal}
  {\bibinfo  {journal} {Water Resources Research}\ }\textbf {\bibinfo {volume}
  {42}} (\bibinfo {year} {2006})}\BibitemShut {NoStop}%
\bibitem [{\citenamefont {Lal}(2009)}]{Lal2009}%
  \BibitemOpen
  \bibfield  {author} {\bibinfo {author} {\bibfnamefont {R.}~\bibnamefont
  {Lal}},\ }\bibinfo {title} {Tragedy of the global commons: Soil, water and
  air},\ in\ \href {https://doi.org/10.1007/978-90-481-2716-0_2} {\emph
  {\bibinfo {booktitle} {Climate Change, Intercropping, Pest Control and
  Beneficial Microorganisms}}}\ (\bibinfo  {publisher} {Springer Netherlands},\
  \bibinfo {year} {2009})\ p.\ \bibinfo {pages} {9–11}\BibitemShut {NoStop}%
\bibitem [{\citenamefont {Shiklomanov}(2000)}]{Shiklomanov2000}%
  \BibitemOpen
  \bibfield  {author} {\bibinfo {author} {\bibfnamefont {I.~A.}\ \bibnamefont
  {Shiklomanov}},\ }\href {https://doi.org/10.1080/02508060008686794}
  {\bibfield  {journal} {\bibinfo  {journal} {Water International}\ }\textbf
  {\bibinfo {volume} {25}},\ \bibinfo {pages} {11–32} (\bibinfo {year}
  {2000})}\BibitemShut {NoStop}%
\bibitem [{\citenamefont {Fraser}(2003)}]{Fraser2003}%
  \BibitemOpen
  \bibfield  {author} {\bibinfo {author} {\bibfnamefont {M.}~\bibnamefont
  {Fraser}},\ }\href {https://doi.org/10.1029/2003eo340009} {\bibfield
  {journal} {\bibinfo  {journal} {Eos, Transactions American Geophysical
  Union}\ }\textbf {\bibinfo {volume} {84}},\ \bibinfo {pages} {332–332}
  (\bibinfo {year} {2003})}\BibitemShut {NoStop}%
\bibitem [{\citenamefont {Ostrom}(1999)}]{Ostrom1999ARPS}%
  \BibitemOpen
  \bibfield  {author} {\bibinfo {author} {\bibfnamefont {E.}~\bibnamefont
  {Ostrom}},\ }\href {https://doi.org/10.1146/annurev.polisci.2.1.493}
  {\bibfield  {journal} {\bibinfo  {journal} {Annual Review of Political
  Science}\ }\textbf {\bibinfo {volume} {2}},\ \bibinfo {pages} {493–535}
  (\bibinfo {year} {1999})}\BibitemShut {NoStop}%
\bibitem [{\citenamefont {Pires}\ and\ \citenamefont
  {Moreto}(2011)}]{Pires2011EJCPR}%
  \BibitemOpen
  \bibfield  {author} {\bibinfo {author} {\bibfnamefont {S.~F.}\ \bibnamefont
  {Pires}}\ and\ \bibinfo {author} {\bibfnamefont {W.~D.}\ \bibnamefont
  {Moreto}},\ }\href {https://doi.org/10.1007/s10610-011-9141-3} {\bibfield
  {journal} {\bibinfo  {journal} {European Journal on Criminal Policy and
  Research}\ }\textbf {\bibinfo {volume} {17}},\ \bibinfo {pages} {101}
  (\bibinfo {year} {2011})}\BibitemShut {NoStop}%
\bibitem [{\citenamefont {Pandit}\ and\ \citenamefont
  {Thapa}(2003)}]{pandit2003ec}%
  \BibitemOpen
  \bibfield  {author} {\bibinfo {author} {\bibfnamefont {B.~H.}\ \bibnamefont
  {Pandit}}\ and\ \bibinfo {author} {\bibfnamefont {G.~B.}\ \bibnamefont
  {Thapa}},\ }\href {http://www.jstor.org/stable/44520692} {\bibfield
  {journal} {\bibinfo  {journal} {Environmental conservation}\ }\textbf
  {\bibinfo {volume} {30}},\ \bibinfo {pages} {283} (\bibinfo {year}
  {2003})}\BibitemShut {NoStop}%
\bibitem [{\citenamefont {Falster}\ and\ \citenamefont
  {Westoby}(2003)}]{Falster2003}%
  \BibitemOpen
  \bibfield  {author} {\bibinfo {author} {\bibfnamefont {D.~S.}\ \bibnamefont
  {Falster}}\ and\ \bibinfo {author} {\bibfnamefont {M.}~\bibnamefont
  {Westoby}},\ }\href {https://doi.org/10.1016/s0169-5347(03)00061-2}
  {\bibfield  {journal} {\bibinfo  {journal} {Trends in Ecology $\&$;
  Evolution}\ }\textbf {\bibinfo {volume} {18}},\ \bibinfo {pages} {337–343}
  (\bibinfo {year} {2003})}\BibitemShut {NoStop}%
\bibitem [{\citenamefont {Nowak}\ and\ \citenamefont
  {Sigmund}(1998)}]{Nowak1998}%
  \BibitemOpen
  \bibfield  {author} {\bibinfo {author} {\bibfnamefont {M.~A.}\ \bibnamefont
  {Nowak}}\ and\ \bibinfo {author} {\bibfnamefont {K.}~\bibnamefont
  {Sigmund}},\ }\href {https://doi.org/10.1038/31225} {\bibfield  {journal}
  {\bibinfo  {journal} {Nature}\ }\textbf {\bibinfo {volume} {393}},\ \bibinfo
  {pages} {573–577} (\bibinfo {year} {1998})}\BibitemShut {NoStop}%
\bibitem [{\citenamefont {LaPorte}\ \emph {et~al.}(2023)\citenamefont
  {LaPorte}, \citenamefont {Hilbe},\ and\ \citenamefont
  {Nowak}}]{LHN_2023_PLOS}%
  \BibitemOpen
  \bibfield  {author} {\bibinfo {author} {\bibfnamefont {P.}~\bibnamefont
  {LaPorte}}, \bibinfo {author} {\bibfnamefont {C.}~\bibnamefont {Hilbe}},\
  and\ \bibinfo {author} {\bibfnamefont {M.~A.}\ \bibnamefont {Nowak}},\ }\href
  {https://doi.org/10.1371/journal.pcbi.1010987} {\bibfield  {journal}
  {\bibinfo  {journal} {PLOS Computational Biology}\ }\textbf {\bibinfo
  {volume} {19}},\ \bibinfo {pages} {e1010987} (\bibinfo {year}
  {2023})}\BibitemShut {NoStop}%
\bibitem [{\citenamefont {Nowak}(2006{\natexlab{a}})}]{nowak2006science}%
  \BibitemOpen
  \bibfield  {author} {\bibinfo {author} {\bibfnamefont {M.~A.}\ \bibnamefont
  {Nowak}},\ }\href {https://doi.org/10.1126/science.1133755} {\bibfield
  {journal} {\bibinfo  {journal} {Science}\ }\textbf {\bibinfo {volume}
  {314}},\ \bibinfo {pages} {1560} (\bibinfo {year}
  {2006}{\natexlab{a}})}\BibitemShut {NoStop}%
\bibitem [{\citenamefont {Trivers}(1971)}]{Trivers1971}%
  \BibitemOpen
  \bibfield  {author} {\bibinfo {author} {\bibfnamefont {R.~L.}\ \bibnamefont
  {Trivers}},\ }\href {https://doi.org/10.1086/406755} {\bibfield  {journal}
  {\bibinfo  {journal} {The Quarterly Review of Biology}\ }\textbf {\bibinfo
  {volume} {46}},\ \bibinfo {pages} {35–57} (\bibinfo {year}
  {1971})}\BibitemShut {NoStop}%
\bibitem [{\citenamefont {Axelrod}\ and\ \citenamefont
  {Hamilton}(1981)}]{Axelrod1981}%
  \BibitemOpen
  \bibfield  {author} {\bibinfo {author} {\bibfnamefont {R.}~\bibnamefont
  {Axelrod}}\ and\ \bibinfo {author} {\bibfnamefont {W.~D.}\ \bibnamefont
  {Hamilton}},\ }\href {https://doi.org/10.1126/science.7466396} {\bibfield
  {journal} {\bibinfo  {journal} {Science}\ }\textbf {\bibinfo {volume}
  {211}},\ \bibinfo {pages} {1390–1396} (\bibinfo {year} {1981})}\BibitemShut
  {NoStop}%
\bibitem [{\citenamefont {Hilbe}\ \emph
  {et~al.}(2018{\natexlab{a}})\citenamefont {Hilbe}, \citenamefont
  {Chatterjee},\ and\ \citenamefont {Nowak}}]{Hilbe2018}%
  \BibitemOpen
  \bibfield  {author} {\bibinfo {author} {\bibfnamefont {C.}~\bibnamefont
  {Hilbe}}, \bibinfo {author} {\bibfnamefont {K.}~\bibnamefont {Chatterjee}},\
  and\ \bibinfo {author} {\bibfnamefont {M.~A.}\ \bibnamefont {Nowak}},\ }\href
  {https://doi.org/10.1038/s41562-018-0342-3} {\bibfield  {journal} {\bibinfo
  {journal} {Nature Human Behaviour}\ }\textbf {\bibinfo {volume} {2}},\
  \bibinfo {pages} {523–523} (\bibinfo {year}
  {2018}{\natexlab{a}})}\BibitemShut {NoStop}%
\bibitem [{\citenamefont {García}\ and\ \citenamefont {van
  Veelen}(2018)}]{Garca2018}%
  \BibitemOpen
  \bibfield  {author} {\bibinfo {author} {\bibfnamefont {J.}~\bibnamefont
  {García}}\ and\ \bibinfo {author} {\bibfnamefont {M.}~\bibnamefont {van
  Veelen}},\ }\href {http://dx.doi.org/10.3389/frobt.2018.00102} {\bibfield
  {journal} {\bibinfo  {journal} {Frontiers in Robotics and AI}\ }\textbf
  {\bibinfo {volume} {5}} (\bibinfo {year} {2018})}\BibitemShut {NoStop}%
\bibitem [{\citenamefont {Maynard~Smith}(1982)}]{Smith1982}%
  \BibitemOpen
  \bibfield  {author} {\bibinfo {author} {\bibfnamefont {J.}~\bibnamefont
  {Maynard~Smith}},\ }\href {http://dx.doi.org/10.1017/CBO9780511806292} {\emph
  {\bibinfo {title} {Evolution and the Theory of Games}}}\ (\bibinfo
  {publisher} {Cambridge University Press},\ \bibinfo {year}
  {1982})\BibitemShut {NoStop}%
\bibitem [{\citenamefont {Nowak}(2006{\natexlab{b}})}]{NOWAK2006}%
  \BibitemOpen
  \bibfield  {author} {\bibinfo {author} {\bibfnamefont {M.~A.}\ \bibnamefont
  {Nowak}},\ }\href {https://doi.org/10.2307/j.ctvjghw98} {\emph {\bibinfo
  {title} {Evolutionary Dynamics: Exploring the Equations of Life}}}\ (\bibinfo
   {publisher} {Harvard University Press},\ \bibinfo {year} {2006})\BibitemShut
  {NoStop}%
\bibitem [{\citenamefont {Maynard~Smith}\ and\ \citenamefont
  {Price}(1973)}]{SMITH1973}%
  \BibitemOpen
  \bibfield  {author} {\bibinfo {author} {\bibfnamefont {J.}~\bibnamefont
  {Maynard~Smith}}\ and\ \bibinfo {author} {\bibfnamefont {G.~R.}\ \bibnamefont
  {Price}},\ }\href {https://doi.org/10.1038/246015a0} {\bibfield  {journal}
  {\bibinfo  {journal} {Nature}\ }\textbf {\bibinfo {volume} {246}},\ \bibinfo
  {pages} {15–18} (\bibinfo {year} {1973})}\BibitemShut {NoStop}%
\bibitem [{\citenamefont {Hofbauer}\ and\ \citenamefont
  {Sigmund}(1998)}]{Hofbauer1998}%
  \BibitemOpen
  \bibfield  {author} {\bibinfo {author} {\bibfnamefont {J.}~\bibnamefont
  {Hofbauer}}\ and\ \bibinfo {author} {\bibfnamefont {K.}~\bibnamefont
  {Sigmund}},\ }\href {https://doi.org/10.1017/cbo9781139173179} {\emph
  {\bibinfo {title} {Evolutionary Games and Population Dynamics}}}\ (\bibinfo
  {publisher} {Cambridge University Press},\ \bibinfo {year}
  {1998})\BibitemShut {NoStop}%
\bibitem [{\citenamefont {Melbinger}\ and\ \citenamefont
  {Vergassola}(2015)}]{Melbinger2015}%
  \BibitemOpen
  \bibfield  {author} {\bibinfo {author} {\bibfnamefont {A.}~\bibnamefont
  {Melbinger}}\ and\ \bibinfo {author} {\bibfnamefont {M.}~\bibnamefont
  {Vergassola}},\ }\href {http://dx.doi.org/10.1038/srep15211} {\bibfield
  {journal} {\bibinfo  {journal} {Scientific Reports}\ }\textbf {\bibinfo
  {volume} {5}} (\bibinfo {year} {2015})}\BibitemShut {NoStop}%
\bibitem [{\citenamefont {Sih}\ \emph {et~al.}(2011)\citenamefont {Sih},
  \citenamefont {Ferrari},\ and\ \citenamefont {Harris}}]{Sih2011}%
  \BibitemOpen
  \bibfield  {author} {\bibinfo {author} {\bibfnamefont {A.}~\bibnamefont
  {Sih}}, \bibinfo {author} {\bibfnamefont {M.~C.~O.}\ \bibnamefont
  {Ferrari}},\ and\ \bibinfo {author} {\bibfnamefont {D.~J.}\ \bibnamefont
  {Harris}},\ }\href {https://doi.org/10.1111/j.1752-4571.2010.00166.x}
  {\bibfield  {journal} {\bibinfo  {journal} {Evolutionary Applications}\
  }\textbf {\bibinfo {volume} {4}},\ \bibinfo {pages} {367–387} (\bibinfo
  {year} {2011})}\BibitemShut {NoStop}%
\bibitem [{\citenamefont {Levins}(1968)}]{Levins1968}%
  \BibitemOpen
  \bibfield  {author} {\bibinfo {author} {\bibfnamefont {R.}~\bibnamefont
  {Levins}},\ }\href {https://doi.org/doi:10.1515/9780691209418} {\emph
  {\bibinfo {title} {Evolution in Changing Environments}}}\ (\bibinfo
  {publisher} {Princeton University Press},\ \bibinfo {address} {Princeton},\
  \bibinfo {year} {1968})\BibitemShut {NoStop}%
\bibitem [{\citenamefont {Kettlewell}(1955)}]{Kettlewell1955}%
  \BibitemOpen
  \bibfield  {author} {\bibinfo {author} {\bibfnamefont {H.~B.~D.}\
  \bibnamefont {Kettlewell}},\ }\href {https://doi.org/10.1038/hdy.1955.36}
  {\bibfield  {journal} {\bibinfo  {journal} {Heredity}\ }\textbf {\bibinfo
  {volume} {9}},\ \bibinfo {pages} {323–342} (\bibinfo {year}
  {1955})}\BibitemShut {NoStop}%
\bibitem [{\citenamefont {Brown}\ and\ \citenamefont
  {Kotler}(2004)}]{Brown2004}%
  \BibitemOpen
  \bibfield  {author} {\bibinfo {author} {\bibfnamefont {J.~S.}\ \bibnamefont
  {Brown}}\ and\ \bibinfo {author} {\bibfnamefont {B.~P.}\ \bibnamefont
  {Kotler}},\ }\href {https://doi.org/10.1111/j.1461-0248.2004.00661.x}
  {\bibfield  {journal} {\bibinfo  {journal} {Ecology Letters}\ }\textbf
  {\bibinfo {volume} {7}},\ \bibinfo {pages} {999–1014} (\bibinfo {year}
  {2004})}\BibitemShut {NoStop}%
\bibitem [{\citenamefont {Muscatine}\ and\ \citenamefont
  {Porter}(1977)}]{Muscatine1977}%
  \BibitemOpen
  \bibfield  {author} {\bibinfo {author} {\bibfnamefont {L.}~\bibnamefont
  {Muscatine}}\ and\ \bibinfo {author} {\bibfnamefont {J.~W.}\ \bibnamefont
  {Porter}},\ }\href {https://doi.org/10.2307/1297526} {\bibfield  {journal}
  {\bibinfo  {journal} {BioScience}\ }\textbf {\bibinfo {volume} {27}},\
  \bibinfo {pages} {454–460} (\bibinfo {year} {1977})}\BibitemShut {NoStop}%
\bibitem [{\citenamefont {Avila}\ and\ \citenamefont
  {Mullon}(2023)}]{Avila2023}%
  \BibitemOpen
  \bibfield  {author} {\bibinfo {author} {\bibfnamefont {P.}~\bibnamefont
  {Avila}}\ and\ \bibinfo {author} {\bibfnamefont {C.}~\bibnamefont {Mullon}},\
  }\href {http://dx.doi.org/10.1098/rstb.2021.0502} {\bibfield  {journal}
  {\bibinfo  {journal} {Philosophical Transactions of the Royal Society B}\
  }\textbf {\bibinfo {volume} {378}} (\bibinfo {year} {2023})}\BibitemShut
  {NoStop}%
\bibitem [{\citenamefont {Doebeli}(2011)}]{Doebeli2011}%
  \BibitemOpen
  \bibfield  {author} {\bibinfo {author} {\bibfnamefont {M.}~\bibnamefont
  {Doebeli}},\ }\href {https://doi.org/10.2307/j.ctt7rgw4} {\emph {\bibinfo
  {title} {Adaptive Diversification (MPB-48)}}}\ (\bibinfo  {publisher}
  {Princeton University Press},\ \bibinfo {year} {2011})\BibitemShut {NoStop}%
\bibitem [{\citenamefont {Lenski}(2017)}]{Lenski2017}%
  \BibitemOpen
  \bibfield  {author} {\bibinfo {author} {\bibfnamefont {R.~E.}\ \bibnamefont
  {Lenski}},\ }\href {https://doi.org/10.1038/ismej.2017.69} {\bibfield
  {journal} {\bibinfo  {journal} {The ISME Journal}\ }\textbf {\bibinfo
  {volume} {11}},\ \bibinfo {pages} {2181–2194} (\bibinfo {year}
  {2017})}\BibitemShut {NoStop}%
\bibitem [{\citenamefont {Meyer}\ \emph {et~al.}(2016)\citenamefont {Meyer},
  \citenamefont {Dobias}, \citenamefont {Medina}, \citenamefont {Servilio},
  \citenamefont {Gupta},\ and\ \citenamefont {Lenski}}]{Meyer2016}%
  \BibitemOpen
  \bibfield  {author} {\bibinfo {author} {\bibfnamefont {J.~R.}\ \bibnamefont
  {Meyer}}, \bibinfo {author} {\bibfnamefont {D.~T.}\ \bibnamefont {Dobias}},
  \bibinfo {author} {\bibfnamefont {S.~J.}\ \bibnamefont {Medina}}, \bibinfo
  {author} {\bibfnamefont {L.}~\bibnamefont {Servilio}}, \bibinfo {author}
  {\bibfnamefont {A.}~\bibnamefont {Gupta}},\ and\ \bibinfo {author}
  {\bibfnamefont {R.~E.}\ \bibnamefont {Lenski}},\ }\href
  {https://doi.org/10.1126/science.aai8446} {\bibfield  {journal} {\bibinfo
  {journal} {Science}\ }\textbf {\bibinfo {volume} {354}},\ \bibinfo {pages}
  {1301–1304} (\bibinfo {year} {2016})}\BibitemShut {NoStop}%
\bibitem [{\citenamefont {Robertson}(1991)}]{Robertson1991}%
  \BibitemOpen
  \bibfield  {author} {\bibinfo {author} {\bibfnamefont {D.~S.}\ \bibnamefont
  {Robertson}},\ }\href {https://doi.org/10.1016/s0022-5193(05)80393-5}
  {\bibfield  {journal} {\bibinfo  {journal} {Journal of Theoretical Biology}\
  }\textbf {\bibinfo {volume} {152}},\ \bibinfo {pages} {469–484} (\bibinfo
  {year} {1991})}\BibitemShut {NoStop}%
\bibitem [{\citenamefont {Post}\ and\ \citenamefont
  {Palkovacs}(2009)}]{Post2009}%
  \BibitemOpen
  \bibfield  {author} {\bibinfo {author} {\bibfnamefont {D.~M.}\ \bibnamefont
  {Post}}\ and\ \bibinfo {author} {\bibfnamefont {E.~P.}\ \bibnamefont
  {Palkovacs}},\ }\href {https://doi.org/10.1098/rstb.2009.0012} {\bibfield
  {journal} {\bibinfo  {journal} {Philosophical Transactions of the Royal
  Society B: Biological Sciences}\ }\textbf {\bibinfo {volume} {364}},\
  \bibinfo {pages} {1629–1640} (\bibinfo {year} {2009})}\BibitemShut
  {NoStop}%
\bibitem [{\citenamefont {Killingback}\ \emph {et~al.}(2010)\citenamefont
  {Killingback}, \citenamefont {Doebeli},\ and\ \citenamefont
  {Hauert}}]{Killingback2010}%
  \BibitemOpen
  \bibfield  {author} {\bibinfo {author} {\bibfnamefont {T.}~\bibnamefont
  {Killingback}}, \bibinfo {author} {\bibfnamefont {M.}~\bibnamefont
  {Doebeli}},\ and\ \bibinfo {author} {\bibfnamefont {C.}~\bibnamefont
  {Hauert}},\ }\href {https://doi.org/10.1162/biot_a_00019} {\bibfield
  {journal} {\bibinfo  {journal} {Biological Theory}\ }\textbf {\bibinfo
  {volume} {5}},\ \bibinfo {pages} {3–6} (\bibinfo {year}
  {2010})}\BibitemShut {NoStop}%
\bibitem [{\citenamefont {Passagem-Santos}\ and\ \citenamefont
  {Perfeito}(2018)}]{PassagemSantos2018}%
  \BibitemOpen
  \bibfield  {author} {\bibinfo {author} {\bibfnamefont {D.}~\bibnamefont
  {Passagem-Santos}}\ and\ \bibinfo {author} {\bibfnamefont {L.}~\bibnamefont
  {Perfeito}},\ }\href {http://dx.doi.org/10.1101/464362} {\  (\bibinfo {year}
  {2018})}\BibitemShut {NoStop}%
\bibitem [{\citenamefont {McKinnon}\ and\ \citenamefont
  {Rundle}(2002)}]{McKinnon2002}%
  \BibitemOpen
  \bibfield  {author} {\bibinfo {author} {\bibfnamefont {J.~S.}\ \bibnamefont
  {McKinnon}}\ and\ \bibinfo {author} {\bibfnamefont {H.~D.}\ \bibnamefont
  {Rundle}},\ }\href {https://doi.org/10.1016/s0169-5347(02)02579-x} {\bibfield
   {journal} {\bibinfo  {journal} {Trends in Ecology $\&$; Evolution}\ }\textbf
  {\bibinfo {volume} {17}},\ \bibinfo {pages} {480–488} (\bibinfo {year}
  {2002})}\BibitemShut {NoStop}%
\bibitem [{\citenamefont {Yoshida}\ \emph {et~al.}(2003)\citenamefont
  {Yoshida}, \citenamefont {Jones}, \citenamefont {Ellner}, \citenamefont
  {Fussmann},\ and\ \citenamefont {Hairston}}]{Yoshida2003}%
  \BibitemOpen
  \bibfield  {author} {\bibinfo {author} {\bibfnamefont {T.}~\bibnamefont
  {Yoshida}}, \bibinfo {author} {\bibfnamefont {L.~E.}\ \bibnamefont {Jones}},
  \bibinfo {author} {\bibfnamefont {S.~P.}\ \bibnamefont {Ellner}}, \bibinfo
  {author} {\bibfnamefont {G.~F.}\ \bibnamefont {Fussmann}},\ and\ \bibinfo
  {author} {\bibfnamefont {N.~G.}\ \bibnamefont {Hairston}},\ }\href
  {https://doi.org/10.1038/nature01767} {\bibfield  {journal} {\bibinfo
  {journal} {Nature}\ }\textbf {\bibinfo {volume} {424}},\ \bibinfo {pages}
  {303–306} (\bibinfo {year} {2003})}\BibitemShut {NoStop}%
\bibitem [{\citenamefont {Brooks}\ and\ \citenamefont
  {Dodson}(1965)}]{Brooks1965}%
  \BibitemOpen
  \bibfield  {author} {\bibinfo {author} {\bibfnamefont {J.~L.}\ \bibnamefont
  {Brooks}}\ and\ \bibinfo {author} {\bibfnamefont {S.~I.}\ \bibnamefont
  {Dodson}},\ }\href {https://doi.org/10.1126/science.150.3692.28} {\bibfield
  {journal} {\bibinfo  {journal} {Science}\ }\textbf {\bibinfo {volume}
  {150}},\ \bibinfo {pages} {28–35} (\bibinfo {year} {1965})}\BibitemShut
  {NoStop}%
\bibitem [{\citenamefont {Palkovacs}\ and\ \citenamefont
  {Post}(2008)}]{Palkovacs2008EcoevolutionaryIB}%
  \BibitemOpen
  \bibfield  {author} {\bibinfo {author} {\bibfnamefont {E.~P.}\ \bibnamefont
  {Palkovacs}}\ and\ \bibinfo {author} {\bibfnamefont {D.~M.}\ \bibnamefont
  {Post}},\ }\href {https://api.semanticscholar.org/CorpusID:31258860}
  {\bibfield  {journal} {\bibinfo  {journal} {Evolutionary Ecology Research}\
  }\textbf {\bibinfo {volume} {10}},\ \bibinfo {pages} {699} (\bibinfo {year}
  {2008})}\BibitemShut {NoStop}%
\bibitem [{\citenamefont {Post}\ \emph {et~al.}(2008)\citenamefont {Post},
  \citenamefont {Palkovacs}, \citenamefont {Schielke},\ and\ \citenamefont
  {Dodson}}]{Post2008}%
  \BibitemOpen
  \bibfield  {author} {\bibinfo {author} {\bibfnamefont {D.~M.}\ \bibnamefont
  {Post}}, \bibinfo {author} {\bibfnamefont {E.~P.}\ \bibnamefont {Palkovacs}},
  \bibinfo {author} {\bibfnamefont {E.~G.}\ \bibnamefont {Schielke}},\ and\
  \bibinfo {author} {\bibfnamefont {S.~I.}\ \bibnamefont {Dodson}},\ }\href
  {https://doi.org/10.1890/07-1216.1} {\bibfield  {journal} {\bibinfo
  {journal} {Ecology}\ }\textbf {\bibinfo {volume} {89}},\ \bibinfo {pages}
  {2019–2032} (\bibinfo {year} {2008})}\BibitemShut {NoStop}%
\bibitem [{\citenamefont {Reznick}\ \emph {et~al.}(1997)\citenamefont
  {Reznick}, \citenamefont {Shaw}, \citenamefont {Rodd},\ and\ \citenamefont
  {Shaw}}]{Reznick1997}%
  \BibitemOpen
  \bibfield  {author} {\bibinfo {author} {\bibfnamefont {D.~N.}\ \bibnamefont
  {Reznick}}, \bibinfo {author} {\bibfnamefont {F.~H.}\ \bibnamefont {Shaw}},
  \bibinfo {author} {\bibfnamefont {F.~H.}\ \bibnamefont {Rodd}},\ and\
  \bibinfo {author} {\bibfnamefont {R.~G.}\ \bibnamefont {Shaw}},\ }\href
  {https://doi.org/10.1126/science.275.5308.1934} {\bibfield  {journal}
  {\bibinfo  {journal} {Science}\ }\textbf {\bibinfo {volume} {275}},\ \bibinfo
  {pages} {1934–1937} (\bibinfo {year} {1997})}\BibitemShut {NoStop}%
\bibitem [{\citenamefont {Palkovacs}\ and\ \citenamefont
  {Post}(2009)}]{Palkovacs2009}%
  \BibitemOpen
  \bibfield  {author} {\bibinfo {author} {\bibfnamefont {E.~P.}\ \bibnamefont
  {Palkovacs}}\ and\ \bibinfo {author} {\bibfnamefont {D.~M.}\ \bibnamefont
  {Post}},\ }\href {https://doi.org/10.1890/08-1673.1} {\bibfield  {journal}
  {\bibinfo  {journal} {Ecology}\ }\textbf {\bibinfo {volume} {90}},\ \bibinfo
  {pages} {300–305} (\bibinfo {year} {2009})}\BibitemShut {NoStop}%
\bibitem [{\citenamefont {Grant}(1986)}]{Grant1986}%
  \BibitemOpen
  \bibfield  {author} {\bibinfo {author} {\bibfnamefont {P.~R.}\ \bibnamefont
  {Grant}},\ }\href {https://api.semanticscholar.org/CorpusID:267404090} {\emph
  {\bibinfo {title} {Ecology and Evolution of Darwin's Finches (Princeton
  Science Library Edition)}}}\ (\bibinfo  {publisher} {Princeton University
  Press},\ \bibinfo {year} {1986})\BibitemShut {NoStop}%
\bibitem [{\citenamefont {Hairston}\ \emph {et~al.}(2005)\citenamefont
  {Hairston}, \citenamefont {Ellner}, \citenamefont {Geber}, \citenamefont
  {Yoshida},\ and\ \citenamefont {Fox}}]{Hairston2005}%
  \BibitemOpen
  \bibfield  {author} {\bibinfo {author} {\bibfnamefont {N.~G.}\ \bibnamefont
  {Hairston}}, \bibinfo {author} {\bibfnamefont {S.~P.}\ \bibnamefont
  {Ellner}}, \bibinfo {author} {\bibfnamefont {M.~A.}\ \bibnamefont {Geber}},
  \bibinfo {author} {\bibfnamefont {T.}~\bibnamefont {Yoshida}},\ and\ \bibinfo
  {author} {\bibfnamefont {J.~A.}\ \bibnamefont {Fox}},\ }\href
  {https://doi.org/10.1111/j.1461-0248.2005.00812.x} {\bibfield  {journal}
  {\bibinfo  {journal} {Ecology Letters}\ }\textbf {\bibinfo {volume} {8}},\
  \bibinfo {pages} {1114–1127} (\bibinfo {year} {2005})}\BibitemShut
  {NoStop}%
\bibitem [{\citenamefont {Grant}\ and\ \citenamefont
  {Grant}(2006)}]{Grant2006}%
  \BibitemOpen
  \bibfield  {author} {\bibinfo {author} {\bibfnamefont {P.~R.}\ \bibnamefont
  {Grant}}\ and\ \bibinfo {author} {\bibfnamefont {B.~R.}\ \bibnamefont
  {Grant}},\ }\href {https://doi.org/10.1126/science.1128374} {\bibfield
  {journal} {\bibinfo  {journal} {Science}\ }\textbf {\bibinfo {volume}
  {313}},\ \bibinfo {pages} {224–226} (\bibinfo {year} {2006})}\BibitemShut
  {NoStop}%
\bibitem [{\citenamefont {Whitham}\ \emph {et~al.}(2006)\citenamefont
  {Whitham}, \citenamefont {Bailey}, \citenamefont {Schweitzer}, \citenamefont
  {Shuster}, \citenamefont {Bangert}, \citenamefont {LeRoy}, \citenamefont
  {Lonsdorf}, \citenamefont {Allan}, \citenamefont {DiFazio}, \citenamefont
  {Potts}, \citenamefont {Fischer}, \citenamefont {Gehring}, \citenamefont
  {Lindroth}, \citenamefont {Marks}, \citenamefont {Hart}, \citenamefont
  {Wimp},\ and\ \citenamefont {Wooley}}]{Whitham2006}%
  \BibitemOpen
  \bibfield  {author} {\bibinfo {author} {\bibfnamefont {T.~G.}\ \bibnamefont
  {Whitham}}, \bibinfo {author} {\bibfnamefont {J.~K.}\ \bibnamefont {Bailey}},
  \bibinfo {author} {\bibfnamefont {J.~A.}\ \bibnamefont {Schweitzer}},
  \bibinfo {author} {\bibfnamefont {S.~M.}\ \bibnamefont {Shuster}}, \bibinfo
  {author} {\bibfnamefont {R.~K.}\ \bibnamefont {Bangert}}, \bibinfo {author}
  {\bibfnamefont {C.~J.}\ \bibnamefont {LeRoy}}, \bibinfo {author}
  {\bibfnamefont {E.~V.}\ \bibnamefont {Lonsdorf}}, \bibinfo {author}
  {\bibfnamefont {G.~J.}\ \bibnamefont {Allan}}, \bibinfo {author}
  {\bibfnamefont {S.~P.}\ \bibnamefont {DiFazio}}, \bibinfo {author}
  {\bibfnamefont {B.~M.}\ \bibnamefont {Potts}}, \bibinfo {author}
  {\bibfnamefont {D.~G.}\ \bibnamefont {Fischer}}, \bibinfo {author}
  {\bibfnamefont {C.~A.}\ \bibnamefont {Gehring}}, \bibinfo {author}
  {\bibfnamefont {R.~L.}\ \bibnamefont {Lindroth}}, \bibinfo {author}
  {\bibfnamefont {J.~C.}\ \bibnamefont {Marks}}, \bibinfo {author}
  {\bibfnamefont {S.~C.}\ \bibnamefont {Hart}}, \bibinfo {author}
  {\bibfnamefont {G.~M.}\ \bibnamefont {Wimp}},\ and\ \bibinfo {author}
  {\bibfnamefont {S.~C.}\ \bibnamefont {Wooley}},\ }\href
  {https://doi.org/10.1038/nrg1877} {\bibfield  {journal} {\bibinfo  {journal}
  {Nature Reviews Genetics}\ }\textbf {\bibinfo {volume} {7}},\ \bibinfo
  {pages} {510–523} (\bibinfo {year} {2006})}\BibitemShut {NoStop}%
\bibitem [{\citenamefont {Weitz}\ \emph {et~al.}(2016)\citenamefont {Weitz},
  \citenamefont {Eksin}, \citenamefont {Paarporn}, \citenamefont {Brown},\ and\
  \citenamefont {Ratcliff}}]{weitz2016pnas}%
  \BibitemOpen
  \bibfield  {author} {\bibinfo {author} {\bibfnamefont {J.~S.}\ \bibnamefont
  {Weitz}}, \bibinfo {author} {\bibfnamefont {C.}~\bibnamefont {Eksin}},
  \bibinfo {author} {\bibfnamefont {K.}~\bibnamefont {Paarporn}}, \bibinfo
  {author} {\bibfnamefont {S.~P.}\ \bibnamefont {Brown}},\ and\ \bibinfo
  {author} {\bibfnamefont {W.~C.}\ \bibnamefont {Ratcliff}},\ }\href
  {https://doi.org/10.1073/pnas.1604096113} {\bibfield  {journal} {\bibinfo
  {journal} {Proc. Natl. Acad. Sci. U.S.A.}\ }\textbf {\bibinfo {volume}
  {113}},\ \bibinfo {pages} {E7518} (\bibinfo {year} {2016})}\BibitemShut
  {NoStop}%
\bibitem [{\citenamefont {Tilman}\ \emph {et~al.}(2020)\citenamefont {Tilman},
  \citenamefont {Plotkin},\ and\ \citenamefont {Ak\c{c}ay}}]{TPA_2020_NAT_COM}%
  \BibitemOpen
  \bibfield  {author} {\bibinfo {author} {\bibfnamefont {A.~R.}\ \bibnamefont
  {Tilman}}, \bibinfo {author} {\bibfnamefont {J.~B.}\ \bibnamefont
  {Plotkin}},\ and\ \bibinfo {author} {\bibfnamefont {E.}~\bibnamefont
  {Ak\c{c}ay}},\ }\href {http://dx.doi.org/10.1038/s41467-020-14531-6}
  {\bibfield  {journal} {\bibinfo  {journal} {Nature Communications}\ }\textbf
  {\bibinfo {volume} {11}} (\bibinfo {year} {2020})}\BibitemShut {NoStop}%
\bibitem [{\citenamefont {Du}\ \emph {et~al.}(2024)\citenamefont {Du},
  \citenamefont {Lu}, \citenamefont {Zhang}, \citenamefont {Shen},
  \citenamefont {Shi},\ and\ \citenamefont {Guo}}]{DLZ_2024_CHAOS}%
  \BibitemOpen
  \bibfield  {author} {\bibinfo {author} {\bibfnamefont {C.}~\bibnamefont
  {Du}}, \bibinfo {author} {\bibfnamefont {Y.}~\bibnamefont {Lu}}, \bibinfo
  {author} {\bibfnamefont {Y.}~\bibnamefont {Zhang}}, \bibinfo {author}
  {\bibfnamefont {C.}~\bibnamefont {Shen}}, \bibinfo {author} {\bibfnamefont
  {L.}~\bibnamefont {Shi}},\ and\ \bibinfo {author} {\bibfnamefont
  {H.}~\bibnamefont {Guo}},\ }\href {http://dx.doi.org/10.1063/5.0200761}
  {\bibfield  {journal} {\bibinfo  {journal} {Chaos: An Interdisciplinary
  Journal of Nonlinear Science}\ }\textbf {\bibinfo {volume} {34}} (\bibinfo
  {year} {2024})}\BibitemShut {NoStop}%
\bibitem [{\citenamefont {Chen}\ \emph {et~al.}(2025)\citenamefont {Chen},
  \citenamefont {Guan},\ and\ \citenamefont {Wu}}]{CGW_2025_PRE}%
  \BibitemOpen
  \bibfield  {author} {\bibinfo {author} {\bibfnamefont {Y.-D.}\ \bibnamefont
  {Chen}}, \bibinfo {author} {\bibfnamefont {J.-Y.}\ \bibnamefont {Guan}},\
  and\ \bibinfo {author} {\bibfnamefont {Z.-X.}\ \bibnamefont {Wu}},\ }\href
  {http://dx.doi.org/10.1103/PhysRevE.111.024305} {\bibfield  {journal}
  {\bibinfo  {journal} {Physical Review E}\ }\textbf {\bibinfo {volume} {111}}
  (\bibinfo {year} {2025})}\BibitemShut {NoStop}%
\bibitem [{\citenamefont {Lin}\ and\ \citenamefont {Weitz}(2019)}]{Lin2019}%
  \BibitemOpen
  \bibfield  {author} {\bibinfo {author} {\bibfnamefont {Y.-H.}\ \bibnamefont
  {Lin}}\ and\ \bibinfo {author} {\bibfnamefont {J.~S.}\ \bibnamefont
  {Weitz}},\ }\href {http://dx.doi.org/10.1103/PhysRevLett.122.148102}
  {\bibfield  {journal} {\bibinfo  {journal} {Physical Review Letters}\
  }\textbf {\bibinfo {volume} {122}} (\bibinfo {year} {2019})}\BibitemShut
  {NoStop}%
\bibitem [{\citenamefont {Bairagya}\ \emph {et~al.}(2021)\citenamefont
  {Bairagya}, \citenamefont {Mondal}, \citenamefont {Chowdhury},\ and\
  \citenamefont {Chakraborty}}]{BMC_2021_PRE}%
  \BibitemOpen
  \bibfield  {author} {\bibinfo {author} {\bibfnamefont {J.~D.}\ \bibnamefont
  {Bairagya}}, \bibinfo {author} {\bibfnamefont {S.~S.}\ \bibnamefont
  {Mondal}}, \bibinfo {author} {\bibfnamefont {D.}~\bibnamefont {Chowdhury}},\
  and\ \bibinfo {author} {\bibfnamefont {S.}~\bibnamefont {Chakraborty}},\
  }\href {https://doi.org/10.1103/physreve.104.044407} {\bibfield  {journal}
  {\bibinfo  {journal} {Physical Review E}\ }\textbf {\bibinfo {volume} {104}}
  (\bibinfo {year} {2021})}\BibitemShut {NoStop}%
\bibitem [{\citenamefont {Mondal}\ \emph {et~al.}(2022)\citenamefont {Mondal},
  \citenamefont {Pathak},\ and\ \citenamefont {Chakraborty}}]{Mondal2022JoPC}%
  \BibitemOpen
  \bibfield  {author} {\bibinfo {author} {\bibfnamefont {S.~S.}\ \bibnamefont
  {Mondal}}, \bibinfo {author} {\bibfnamefont {M.}~\bibnamefont {Pathak}},\
  and\ \bibinfo {author} {\bibfnamefont {S.}~\bibnamefont {Chakraborty}},\
  }\href {https://doi.org/10.1088/2632-072x/ac6c6e} {\bibfield  {journal}
  {\bibinfo  {journal} {Journal of Physics: Complexity}\ }\textbf {\bibinfo
  {volume} {3}},\ \bibinfo {pages} {025005} (\bibinfo {year}
  {2022})}\BibitemShut {NoStop}%
\bibitem [{\citenamefont {Bairagya}\ \emph {et~al.}(2023)\citenamefont
  {Bairagya}, \citenamefont {Mondal}, \citenamefont {Chowdhury},\ and\
  \citenamefont {Chakraborty}}]{BMC_2023_JOPC}%
  \BibitemOpen
  \bibfield  {author} {\bibinfo {author} {\bibfnamefont {J.~D.}\ \bibnamefont
  {Bairagya}}, \bibinfo {author} {\bibfnamefont {S.~S.}\ \bibnamefont
  {Mondal}}, \bibinfo {author} {\bibfnamefont {D.}~\bibnamefont {Chowdhury}},\
  and\ \bibinfo {author} {\bibfnamefont {S.}~\bibnamefont {Chakraborty}},\
  }\href {https://doi.org/10.1088/2632-072x/acc5cb} {\bibfield  {journal}
  {\bibinfo  {journal} {Journal of Physics: Complexity}\ }\textbf {\bibinfo
  {volume} {4}},\ \bibinfo {pages} {025002} (\bibinfo {year}
  {2023})}\BibitemShut {NoStop}%
\bibitem [{\citenamefont {Sohel~Mondal}\ \emph {et~al.}(2024)\citenamefont
  {Sohel~Mondal}, \citenamefont {Ray},\ and\ \citenamefont
  {Chakraborty}}]{MRC_2024_CHAOS}%
  \BibitemOpen
  \bibfield  {author} {\bibinfo {author} {\bibfnamefont {S.}~\bibnamefont
  {Sohel~Mondal}}, \bibinfo {author} {\bibfnamefont {A.}~\bibnamefont {Ray}},\
  and\ \bibinfo {author} {\bibfnamefont {S.}~\bibnamefont {Chakraborty}},\
  }\href {http://dx.doi.org/10.1063/5.0190800} {\bibfield  {journal} {\bibinfo
  {journal} {Chaos: An Interdisciplinary Journal of Nonlinear Science}\
  }\textbf {\bibinfo {volume} {34}} (\bibinfo {year} {2024})}\BibitemShut
  {NoStop}%
\bibitem [{\citenamefont {Mandal}\ \emph {et~al.}(2025)\citenamefont {Mandal},
  \citenamefont {Sarkar}, \citenamefont {Chakraborty},\ and\ \citenamefont
  {Dutta}}]{MSC_2025_PRSA}%
  \BibitemOpen
  \bibfield  {author} {\bibinfo {author} {\bibfnamefont {A.}~\bibnamefont
  {Mandal}}, \bibinfo {author} {\bibfnamefont {S.}~\bibnamefont {Sarkar}},
  \bibinfo {author} {\bibfnamefont {S.}~\bibnamefont {Chakraborty}},\ and\
  \bibinfo {author} {\bibfnamefont {P.~S.}\ \bibnamefont {Dutta}},\ }\href
  {http://dx.doi.org/10.1098/rspa.2024.0915} {\bibfield  {journal} {\bibinfo
  {journal} {Proceedings of the Royal Society A: Mathematical, Physical and
  Engineering Sciences}\ }\textbf {\bibinfo {volume} {481}} (\bibinfo {year}
  {2025})}\BibitemShut {NoStop}%
\bibitem [{\citenamefont {Hilbe}\ \emph
  {et~al.}(2018{\natexlab{b}})\citenamefont {Hilbe}, \citenamefont
  {{\v{S}}imsa}, \citenamefont {Chatterjee},\ and\ \citenamefont
  {Nowak}}]{hilbe2018nature}%
  \BibitemOpen
  \bibfield  {author} {\bibinfo {author} {\bibfnamefont {C.}~\bibnamefont
  {Hilbe}}, \bibinfo {author} {\bibfnamefont {{\v{S}}.}~\bibnamefont
  {{\v{S}}imsa}}, \bibinfo {author} {\bibfnamefont {K.}~\bibnamefont
  {Chatterjee}},\ and\ \bibinfo {author} {\bibfnamefont {M.~A.}\ \bibnamefont
  {Nowak}},\ }\href {https://doi.org/10.1038/s41586-018-0277-x} {\bibfield
  {journal} {\bibinfo  {journal} {Nature}\ }\textbf {\bibinfo {volume} {559}},\
  \bibinfo {pages} {246} (\bibinfo {year} {2018}{\natexlab{b}})}\BibitemShut
  {NoStop}%
\bibitem [{\citenamefont {Nowak}\ and\ \citenamefont
  {Sigmund}(1993)}]{Nowak1993}%
  \BibitemOpen
  \bibfield  {author} {\bibinfo {author} {\bibfnamefont {M.}~\bibnamefont
  {Nowak}}\ and\ \bibinfo {author} {\bibfnamefont {K.}~\bibnamefont
  {Sigmund}},\ }\href {https://doi.org/10.1038/364056a0} {\bibfield  {journal}
  {\bibinfo  {journal} {Nature}\ }\textbf {\bibinfo {volume} {364}},\ \bibinfo
  {pages} {56–58} (\bibinfo {year} {1993})}\BibitemShut {NoStop}%
\bibitem [{\citenamefont {Sigmund}(2010)}]{Sigmund_2010_PUP}%
  \BibitemOpen
  \bibfield  {author} {\bibinfo {author} {\bibfnamefont {K.}~\bibnamefont
  {Sigmund}},\ }\href {https://doi.org/doi:10.1515/9781400832255} {\emph
  {\bibinfo {title} {The Calculus of Selfishness}}}\ (\bibinfo  {publisher}
  {Princeton University Press},\ \bibinfo {address} {Princeton},\ \bibinfo
  {year} {2010})\BibitemShut {NoStop}%
\bibitem [{\citenamefont {Hauert}\ and\ \citenamefont
  {Schuster}(1997)}]{Hauert1997}%
  \BibitemOpen
  \bibfield  {author} {\bibinfo {author} {\bibfnamefont {C.}~\bibnamefont
  {Hauert}}\ and\ \bibinfo {author} {\bibfnamefont {H.~G.}\ \bibnamefont
  {Schuster}},\ }\href {https://doi.org/10.1098/rspb.1997.0073} {\bibfield
  {journal} {\bibinfo  {journal} {Proceedings of the Royal Society of London.
  Series B: Biological Sciences}\ }\textbf {\bibinfo {volume} {264}},\ \bibinfo
  {pages} {513–519} (\bibinfo {year} {1997})}\BibitemShut {NoStop}%
\bibitem [{\citenamefont {Taylor}\ and\ \citenamefont
  {Jonker}(1978)}]{Taylor1978}%
  \BibitemOpen
  \bibfield  {author} {\bibinfo {author} {\bibfnamefont {P.~D.}\ \bibnamefont
  {Taylor}}\ and\ \bibinfo {author} {\bibfnamefont {L.~B.}\ \bibnamefont
  {Jonker}},\ }\href {https://doi.org/10.1016/0025-5564(78)90077-9} {\bibfield
  {journal} {\bibinfo  {journal} {Mathematical Biosciences}\ }\textbf {\bibinfo
  {volume} {40}},\ \bibinfo {pages} {145–156} (\bibinfo {year}
  {1978})}\BibitemShut {NoStop}%
\bibitem [{\citenamefont {Schuster}\ and\ \citenamefont
  {Sigmund}(1983)}]{Schuster1983}%
  \BibitemOpen
  \bibfield  {author} {\bibinfo {author} {\bibfnamefont {P.}~\bibnamefont
  {Schuster}}\ and\ \bibinfo {author} {\bibfnamefont {K.}~\bibnamefont
  {Sigmund}},\ }\href {https://doi.org/10.1016/0022-5193(83)90445-9} {\bibfield
   {journal} {\bibinfo  {journal} {Journal of Theoretical Biology}\ }\textbf
  {\bibinfo {volume} {100}},\ \bibinfo {pages} {533–538} (\bibinfo {year}
  {1983})}\BibitemShut {NoStop}%
\bibitem [{\citenamefont {Schuster}\ and\ \citenamefont
  {Sigmund}(1985)}]{Schuster1985}%
  \BibitemOpen
  \bibfield  {author} {\bibinfo {author} {\bibfnamefont {P.}~\bibnamefont
  {Schuster}}\ and\ \bibinfo {author} {\bibfnamefont {K.}~\bibnamefont
  {Sigmund}},\ }\href {https://doi.org/10.1002/bbpc.19850890620} {\bibfield
  {journal} {\bibinfo  {journal} {Berichte der Bunsengesellschaft f\"{u}r
  physikalische Chemie}\ }\textbf {\bibinfo {volume} {89}},\ \bibinfo {pages}
  {668–682} (\bibinfo {year} {1985})}\BibitemShut {NoStop}%
\bibitem [{\citenamefont {Page}\ and\ \citenamefont {Nowak}(2002)}]{PAGE2002}%
  \BibitemOpen
  \bibfield  {author} {\bibinfo {author} {\bibfnamefont {K.~M.}\ \bibnamefont
  {Page}}\ and\ \bibinfo {author} {\bibfnamefont {M.~A.}\ \bibnamefont
  {Nowak}},\ }\href {https://doi.org/10.1006/jtbi.2002.3112} {\bibfield
  {journal} {\bibinfo  {journal} {Journal of Theoretical Biology}\ }\textbf
  {\bibinfo {volume} {219}},\ \bibinfo {pages} {93–98} (\bibinfo {year}
  {2002})}\BibitemShut {NoStop}%
\bibitem [{\citenamefont {Cressman}(2003)}]{cressman2003book}%
  \BibitemOpen
  \bibfield  {author} {\bibinfo {author} {\bibfnamefont {R.}~\bibnamefont
  {Cressman}},\ }\href {https://books.google.co.in/books?id=187BOmBvyvQC}
  {\emph {\bibinfo {title} {Evolutionary Dynamics and Extensive Form Games}}},\
  Economic learning and social evolution\ (\bibinfo  {publisher} {MIT Press},\
  \bibinfo {year} {2003})\BibitemShut {NoStop}%
\bibitem [{\citenamefont {Martinez-Vaquero}\ \emph {et~al.}(2012)\citenamefont
  {Martinez-Vaquero}, \citenamefont {Cuesta},\ and\ \citenamefont
  {Sánchez}}]{VCS_2012_PLOS}%
  \BibitemOpen
  \bibfield  {author} {\bibinfo {author} {\bibfnamefont {L.~A.}\ \bibnamefont
  {Martinez-Vaquero}}, \bibinfo {author} {\bibfnamefont {J.~A.}\ \bibnamefont
  {Cuesta}},\ and\ \bibinfo {author} {\bibfnamefont {A.}~\bibnamefont
  {Sánchez}},\ }\href {http://dx.doi.org/10.1371/journal.pone.0035135}
  {\bibfield  {journal} {\bibinfo  {journal} {PLoS ONE}\ }\textbf {\bibinfo
  {volume} {7}},\ \bibinfo {pages} {e35135} (\bibinfo {year}
  {2012})}\BibitemShut {NoStop}%
\bibitem [{\citenamefont {Fudenberg}\ and\ \citenamefont
  {Maskin}(1990)}]{Fundenberg1990}%
  \BibitemOpen
  \bibfield  {author} {\bibinfo {author} {\bibfnamefont {D.}~\bibnamefont
  {Fudenberg}}\ and\ \bibinfo {author} {\bibfnamefont {E.}~\bibnamefont
  {Maskin}},\ }\href {http://www.jstor.org/stable/2006583} {\bibfield
  {journal} {\bibinfo  {journal} {The American Economic Review}\ }\textbf
  {\bibinfo {volume} {80}},\ \bibinfo {pages} {274} (\bibinfo {year}
  {1990})}\BibitemShut {NoStop}%
\bibitem [{\citenamefont {Nowak}\ and\ \citenamefont
  {Sigmund}(2004)}]{Nowak2004}%
  \BibitemOpen
  \bibfield  {author} {\bibinfo {author} {\bibfnamefont {M.~A.}\ \bibnamefont
  {Nowak}}\ and\ \bibinfo {author} {\bibfnamefont {K.}~\bibnamefont
  {Sigmund}},\ }\href {https://doi.org/10.1126/science.1093411} {\bibfield
  {journal} {\bibinfo  {journal} {Science}\ }\textbf {\bibinfo {volume}
  {303}},\ \bibinfo {pages} {793–799} (\bibinfo {year} {2004})}\BibitemShut
  {NoStop}%
\bibitem [{\citenamefont {Geritz}\ \emph {et~al.}(1997)\citenamefont {Geritz},
  \citenamefont {Metz}, \citenamefont {Kisdi},\ and\ \citenamefont
  {Meszena}}]{Geritz1997}%
  \BibitemOpen
  \bibfield  {author} {\bibinfo {author} {\bibfnamefont {S.~A.~H.}\
  \bibnamefont {Geritz}}, \bibinfo {author} {\bibfnamefont {J.~A.~J.}\
  \bibnamefont {Metz}}, \bibinfo {author} {\bibfnamefont {E.}~\bibnamefont
  {Kisdi}},\ and\ \bibinfo {author} {\bibfnamefont {G.}~\bibnamefont
  {Meszena}},\ }\href {https://doi.org/10.1103/physrevlett.78.2024} {\bibfield
  {journal} {\bibinfo  {journal} {Physical Review Letters}\ }\textbf {\bibinfo
  {volume} {78}},\ \bibinfo {pages} {2024–2027} (\bibinfo {year}
  {1997})}\BibitemShut {NoStop}%
\bibitem [{\citenamefont {Geritz}\ \emph {et~al.}(1998)\citenamefont {Geritz},
  \citenamefont {Kisdi}, \citenamefont {Mesze´NA},\ and\ \citenamefont
  {Metz}}]{Geritz1998}%
  \BibitemOpen
  \bibfield  {author} {\bibinfo {author} {\bibfnamefont {S.}~\bibnamefont
  {Geritz}}, \bibinfo {author} {\bibfnamefont {E.}~\bibnamefont {Kisdi}},
  \bibinfo {author} {\bibfnamefont {G.}~\bibnamefont {Mesze´NA}},\ and\
  \bibinfo {author} {\bibfnamefont {J.}~\bibnamefont {Metz}},\ }\href
  {https://doi.org/10.1023/a:1006554906681} {\bibfield  {journal} {\bibinfo
  {journal} {Evolutionary Ecology}\ }\textbf {\bibinfo {volume} {12}},\
  \bibinfo {pages} {35–57} (\bibinfo {year} {1998})}\BibitemShut {NoStop}%
\bibitem [{\citenamefont {Dercole}\ and\ \citenamefont
  {Rinaldi}(2008)}]{Dercole_2008}%
  \BibitemOpen
  \bibfield  {author} {\bibinfo {author} {\bibfnamefont {F.}~\bibnamefont
  {Dercole}}\ and\ \bibinfo {author} {\bibfnamefont {S.}~\bibnamefont
  {Rinaldi}},\ }\href {https://doi.org/10.1515/9781400828340} {\emph {\bibinfo
  {title} {Analysis of Evolutionary Processes: The Adaptive Dynamics Approach
  and Its Applications: The Adaptive Dynamics Approach and Its Applications}}}\
  (\bibinfo  {publisher} {Princeton University Press},\ \bibinfo {year}
  {2008})\BibitemShut {NoStop}%
\bibitem [{\citenamefont {Kisdi}\ and\ \citenamefont
  {Geritz}(2009)}]{Kisdi2009}%
  \BibitemOpen
  \bibfield  {author} {\bibinfo {author} {\bibfnamefont {E.}~\bibnamefont
  {Kisdi}}\ and\ \bibinfo {author} {\bibfnamefont {S.~A.~H.}\ \bibnamefont
  {Geritz}},\ }\href {https://doi.org/10.1007/s00285-009-0300-9} {\bibfield
  {journal} {\bibinfo  {journal} {Journal of Mathematical Biology}\ }\textbf
  {\bibinfo {volume} {61}},\ \bibinfo {pages} {165–169} (\bibinfo {year}
  {2009})}\BibitemShut {NoStop}%
\bibitem [{\citenamefont {Bshary}\ and\ \citenamefont
  {Grutter}(2005)}]{Bshary2005}%
  \BibitemOpen
  \bibfield  {author} {\bibinfo {author} {\bibfnamefont {R.}~\bibnamefont
  {Bshary}}\ and\ \bibinfo {author} {\bibfnamefont {A.~S.}\ \bibnamefont
  {Grutter}},\ }\href {https://doi.org/10.1098/rsbl.2005.0344} {\bibfield
  {journal} {\bibinfo  {journal} {Biology Letters}\ }\textbf {\bibinfo {volume}
  {1}},\ \bibinfo {pages} {396–399} (\bibinfo {year} {2005})}\BibitemShut
  {NoStop}%
\bibitem [{\citenamefont {Kiers}\ \emph {et~al.}(2003)\citenamefont {Kiers},
  \citenamefont {Rousseau}, \citenamefont {West},\ and\ \citenamefont
  {Denison}}]{Kiers2003}%
  \BibitemOpen
  \bibfield  {author} {\bibinfo {author} {\bibfnamefont {E.~T.}\ \bibnamefont
  {Kiers}}, \bibinfo {author} {\bibfnamefont {R.~A.}\ \bibnamefont {Rousseau}},
  \bibinfo {author} {\bibfnamefont {S.~A.}\ \bibnamefont {West}},\ and\
  \bibinfo {author} {\bibfnamefont {R.~F.}\ \bibnamefont {Denison}},\ }\href
  {https://doi.org/10.1038/nature01931} {\bibfield  {journal} {\bibinfo
  {journal} {Nature}\ }\textbf {\bibinfo {volume} {425}},\ \bibinfo {pages}
  {78–81} (\bibinfo {year} {2003})}\BibitemShut {NoStop}%
\bibitem [{\citenamefont {Kiers}\ \emph {et~al.}(2011)\citenamefont {Kiers},
  \citenamefont {Duhamel}, \citenamefont {Beesetty}, \citenamefont {Mensah},
  \citenamefont {Franken}, \citenamefont {Verbruggen}, \citenamefont
  {Fellbaum}, \citenamefont {Kowalchuk}, \citenamefont {Hart}, \citenamefont
  {Bago}, \citenamefont {Palmer}, \citenamefont {West}, \citenamefont
  {Vandenkoornhuyse}, \citenamefont {Jansa},\ and\ \citenamefont
  {B\"{u}cking}}]{Kiers2011}%
  \BibitemOpen
  \bibfield  {author} {\bibinfo {author} {\bibfnamefont {E.~T.}\ \bibnamefont
  {Kiers}}, \bibinfo {author} {\bibfnamefont {M.}~\bibnamefont {Duhamel}},
  \bibinfo {author} {\bibfnamefont {Y.}~\bibnamefont {Beesetty}}, \bibinfo
  {author} {\bibfnamefont {J.~A.}\ \bibnamefont {Mensah}}, \bibinfo {author}
  {\bibfnamefont {O.}~\bibnamefont {Franken}}, \bibinfo {author} {\bibfnamefont
  {E.}~\bibnamefont {Verbruggen}}, \bibinfo {author} {\bibfnamefont {C.~R.}\
  \bibnamefont {Fellbaum}}, \bibinfo {author} {\bibfnamefont {G.~A.}\
  \bibnamefont {Kowalchuk}}, \bibinfo {author} {\bibfnamefont {M.~M.}\
  \bibnamefont {Hart}}, \bibinfo {author} {\bibfnamefont {A.}~\bibnamefont
  {Bago}}, \bibinfo {author} {\bibfnamefont {T.~M.}\ \bibnamefont {Palmer}},
  \bibinfo {author} {\bibfnamefont {S.~A.}\ \bibnamefont {West}}, \bibinfo
  {author} {\bibfnamefont {P.}~\bibnamefont {Vandenkoornhuyse}}, \bibinfo
  {author} {\bibfnamefont {J.}~\bibnamefont {Jansa}},\ and\ \bibinfo {author}
  {\bibfnamefont {H.}~\bibnamefont {B\"{u}cking}},\ }\href
  {https://doi.org/10.1126/science.1208473} {\bibfield  {journal} {\bibinfo
  {journal} {Science}\ }\textbf {\bibinfo {volume} {333}},\ \bibinfo {pages}
  {880–882} (\bibinfo {year} {2011})}\BibitemShut {NoStop}%
\bibitem [{\citenamefont {Fehr}\ and\ \citenamefont
  {G\"{a}chter}(2002)}]{Fehr2002}%
  \BibitemOpen
  \bibfield  {author} {\bibinfo {author} {\bibfnamefont {E.}~\bibnamefont
  {Fehr}}\ and\ \bibinfo {author} {\bibfnamefont {S.}~\bibnamefont
  {G\"{a}chter}},\ }\href {https://doi.org/10.1038/415137a} {\bibfield
  {journal} {\bibinfo  {journal} {Nature}\ }\textbf {\bibinfo {volume} {415}},\
  \bibinfo {pages} {137–140} (\bibinfo {year} {2002})}\BibitemShut {NoStop}%
\bibitem [{\citenamefont {Boyd}\ \emph {et~al.}(2003)\citenamefont {Boyd},
  \citenamefont {Gintis}, \citenamefont {Bowles},\ and\ \citenamefont
  {Richerson}}]{Boyd2003}%
  \BibitemOpen
  \bibfield  {author} {\bibinfo {author} {\bibfnamefont {R.}~\bibnamefont
  {Boyd}}, \bibinfo {author} {\bibfnamefont {H.}~\bibnamefont {Gintis}},
  \bibinfo {author} {\bibfnamefont {S.}~\bibnamefont {Bowles}},\ and\ \bibinfo
  {author} {\bibfnamefont {P.~J.}\ \bibnamefont {Richerson}},\ }\href
  {https://doi.org/10.1073/pnas.0630443100} {\bibfield  {journal} {\bibinfo
  {journal} {Proceedings of the National Academy of Sciences}\ }\textbf
  {\bibinfo {volume} {100}},\ \bibinfo {pages} {3531–3535} (\bibinfo {year}
  {2003})}\BibitemShut {NoStop}%
\bibitem [{\citenamefont {Bowles}\ and\ \citenamefont
  {Gintis}(2011)}]{bowles2011cooperative}%
  \BibitemOpen
  \bibfield  {author} {\bibinfo {author} {\bibfnamefont {S.}~\bibnamefont
  {Bowles}}\ and\ \bibinfo {author} {\bibfnamefont {H.}~\bibnamefont
  {Gintis}},\ }\href {https://books.google.co.in/books?id=dezaI9XMp0UC} {\emph
  {\bibinfo {title} {A Cooperative Species: Human Reciprocity and Its
  Evolution}}}\ (\bibinfo  {publisher} {Princeton University Press},\ \bibinfo
  {year} {2011})\BibitemShut {NoStop}%
\bibitem [{\citenamefont {Kiyonari}\ and\ \citenamefont
  {Barclay}(2008)}]{Kiyonari2008}%
  \BibitemOpen
  \bibfield  {author} {\bibinfo {author} {\bibfnamefont {T.}~\bibnamefont
  {Kiyonari}}\ and\ \bibinfo {author} {\bibfnamefont {P.}~\bibnamefont
  {Barclay}},\ }\href {https://doi.org/10.1037/a0011381} {\bibfield  {journal}
  {\bibinfo  {journal} {Journal of Personality and Social Psychology}\ }\textbf
  {\bibinfo {volume} {95}},\ \bibinfo {pages} {826–842} (\bibinfo {year}
  {2008})}\BibitemShut {NoStop}%
\bibitem [{\citenamefont {Hauert}(2010)}]{Hauert2010}%
  \BibitemOpen
  \bibfield  {author} {\bibinfo {author} {\bibfnamefont {C.}~\bibnamefont
  {Hauert}},\ }\href {https://doi.org/10.1016/j.jtbi.2010.08.009} {\bibfield
  {journal} {\bibinfo  {journal} {Journal of Theoretical Biology}\ }\textbf
  {\bibinfo {volume} {267}},\ \bibinfo {pages} {22–28} (\bibinfo {year}
  {2010})}\BibitemShut {NoStop}%
\bibitem [{\citenamefont {Stoop}\ \emph {et~al.}(2018)\citenamefont {Stoop},
  \citenamefont {van Soest},\ and\ \citenamefont {Vyrastekova}}]{Stoop2018}%
  \BibitemOpen
  \bibfield  {author} {\bibinfo {author} {\bibfnamefont {J.}~\bibnamefont
  {Stoop}}, \bibinfo {author} {\bibfnamefont {D.}~\bibnamefont {van Soest}},\
  and\ \bibinfo {author} {\bibfnamefont {J.}~\bibnamefont {Vyrastekova}},\
  }\href {https://doi.org/10.1016/j.jeem.2017.12.007} {\bibfield  {journal}
  {\bibinfo  {journal} {Journal of Environmental Economics and Management}\
  }\textbf {\bibinfo {volume} {88}},\ \bibinfo {pages} {300–310} (\bibinfo
  {year} {2018})}\BibitemShut {NoStop}%
\bibitem [{\citenamefont {Banerjee}\ \emph {et~al.}(2019)\citenamefont
  {Banerjee}, \citenamefont {Layek}, \citenamefont {Sasmal},\ and\
  \citenamefont {Ghosh}}]{Banerjee2019}%
  \BibitemOpen
  \bibfield  {author} {\bibinfo {author} {\bibfnamefont {J.}~\bibnamefont
  {Banerjee}}, \bibinfo {author} {\bibfnamefont {R.~K.}\ \bibnamefont {Layek}},
  \bibinfo {author} {\bibfnamefont {S.~K.}\ \bibnamefont {Sasmal}},\ and\
  \bibinfo {author} {\bibfnamefont {D.}~\bibnamefont {Ghosh}},\ }\href
  {https://doi.org/10.1209/0295-5075/126/18002} {\bibfield  {journal} {\bibinfo
   {journal} {EPL (Europhysics Letters)}\ }\textbf {\bibinfo {volume} {126}},\
  \bibinfo {pages} {18002} (\bibinfo {year} {2019})}\BibitemShut {NoStop}%
\bibitem [{\citenamefont {Clutton-Brock}\ and\ \citenamefont
  {Parker}(1995)}]{CluttonBrock1995}%
  \BibitemOpen
  \bibfield  {author} {\bibinfo {author} {\bibfnamefont {T.~H.}\ \bibnamefont
  {Clutton-Brock}}\ and\ \bibinfo {author} {\bibfnamefont {G.~A.}\ \bibnamefont
  {Parker}},\ }\href {https://doi.org/10.1038/373209a0} {\bibfield  {journal}
  {\bibinfo  {journal} {Nature}\ }\textbf {\bibinfo {volume} {373}},\ \bibinfo
  {pages} {209–216} (\bibinfo {year} {1995})}\BibitemShut {NoStop}%
\bibitem [{\citenamefont {Sasaki}\ and\ \citenamefont
  {Uchida}(2013)}]{Sasaki2013}%
  \BibitemOpen
  \bibfield  {author} {\bibinfo {author} {\bibfnamefont {T.}~\bibnamefont
  {Sasaki}}\ and\ \bibinfo {author} {\bibfnamefont {S.}~\bibnamefont
  {Uchida}},\ }\href {https://doi.org/10.1098/rspb.2012.2498} {\bibfield
  {journal} {\bibinfo  {journal} {Proceedings of the Royal Society B:
  Biological Sciences}\ }\textbf {\bibinfo {volume} {280}},\ \bibinfo {pages}
  {20122498} (\bibinfo {year} {2013})}\BibitemShut {NoStop}%
\bibitem [{\citenamefont {Nag~Chowdhury}\ \emph {et~al.}(2021)\citenamefont
  {Nag~Chowdhury}, \citenamefont {Kundu}, \citenamefont {Banerjee},
  \citenamefont {Perc},\ and\ \citenamefont {Ghosh}}]{NagChowdhury2021}%
  \BibitemOpen
  \bibfield  {author} {\bibinfo {author} {\bibfnamefont {S.}~\bibnamefont
  {Nag~Chowdhury}}, \bibinfo {author} {\bibfnamefont {S.}~\bibnamefont
  {Kundu}}, \bibinfo {author} {\bibfnamefont {J.}~\bibnamefont {Banerjee}},
  \bibinfo {author} {\bibfnamefont {M.}~\bibnamefont {Perc}},\ and\ \bibinfo
  {author} {\bibfnamefont {D.}~\bibnamefont {Ghosh}},\ }\href
  {https://doi.org/10.1016/j.jtbi.2021.110606} {\bibfield  {journal} {\bibinfo
  {journal} {Journal of Theoretical Biology}\ }\textbf {\bibinfo {volume}
  {518}},\ \bibinfo {pages} {110606} (\bibinfo {year} {2021})}\BibitemShut
  {NoStop}%
\bibitem [{\citenamefont {Chowdhury}\ \emph {et~al.}(2021)\citenamefont
  {Chowdhury}, \citenamefont {Kundu}, \citenamefont {Perc},\ and\ \citenamefont
  {Ghosh}}]{Chowdhury2021}%
  \BibitemOpen
  \bibfield  {author} {\bibinfo {author} {\bibfnamefont {S.~N.}\ \bibnamefont
  {Chowdhury}}, \bibinfo {author} {\bibfnamefont {S.}~\bibnamefont {Kundu}},
  \bibinfo {author} {\bibfnamefont {M.}~\bibnamefont {Perc}},\ and\ \bibinfo
  {author} {\bibfnamefont {D.}~\bibnamefont {Ghosh}},\ }\href
  {http://dx.doi.org/10.1098/rspa.2021.0397} {\bibfield  {journal} {\bibinfo
  {journal} {Proceedings of the Royal Society A: Mathematical, Physical and
  Engineering Sciences}\ }\textbf {\bibinfo {volume} {477}} (\bibinfo {year}
  {2021})}\BibitemShut {NoStop}%
\bibitem [{\citenamefont {Johnson}(2015)}]{Johnson2015}%
  \BibitemOpen
  \bibfield  {author} {\bibinfo {author} {\bibfnamefont {S.}~\bibnamefont
  {Johnson}},\ }\href {https://doi.org/10.1098/rsos.150223} {\bibfield
  {journal} {\bibinfo  {journal} {Royal Society Open Science}\ }\textbf
  {\bibinfo {volume} {2}},\ \bibinfo {pages} {150223} (\bibinfo {year}
  {2015})}\BibitemShut {NoStop}%
\bibitem [{\citenamefont {Greenwood}(2016)}]{Greenwood2016}%
  \BibitemOpen
  \bibfield  {author} {\bibinfo {author} {\bibfnamefont {G.~W.}\ \bibnamefont
  {Greenwood}},\ }in\ \href {https://doi.org/10.1109/cig.2016.7860402} {\emph
  {\bibinfo {booktitle} {2016 IEEE Conference on Computational Intelligence and
  Games (CIG)}}}\ (\bibinfo  {publisher} {IEEE},\ \bibinfo {year} {2016})\ p.\
  \bibinfo {pages} {1–7}\BibitemShut {NoStop}%
\bibitem [{\citenamefont {Chen}\ and\ \citenamefont
  {Szolnoki}(2018)}]{Chen2018}%
  \BibitemOpen
  \bibfield  {author} {\bibinfo {author} {\bibfnamefont {X.}~\bibnamefont
  {Chen}}\ and\ \bibinfo {author} {\bibfnamefont {A.}~\bibnamefont
  {Szolnoki}},\ }\href {http://dx.doi.org/10.1371/journal.pcbi.1006347}
  {\bibfield  {journal} {\bibinfo  {journal} {PLOS Computational Biology}\
  }\textbf {\bibinfo {volume} {14}},\ \bibinfo {pages} {e1006347} (\bibinfo
  {year} {2018})}\BibitemShut {NoStop}%
\bibitem [{\citenamefont {Sigmund}\ \emph {et~al.}(2001)\citenamefont
  {Sigmund}, \citenamefont {Hauert},\ and\ \citenamefont
  {Nowak}}]{Sigmund2001}%
  \BibitemOpen
  \bibfield  {author} {\bibinfo {author} {\bibfnamefont {K.}~\bibnamefont
  {Sigmund}}, \bibinfo {author} {\bibfnamefont {C.}~\bibnamefont {Hauert}},\
  and\ \bibinfo {author} {\bibfnamefont {M.~A.}\ \bibnamefont {Nowak}},\ }\href
  {https://doi.org/10.1073/pnas.161155698} {\bibfield  {journal} {\bibinfo
  {journal} {Proceedings of the National Academy of Sciences}\ }\textbf
  {\bibinfo {volume} {98}},\ \bibinfo {pages} {10757–10762} (\bibinfo {year}
  {2001})}\BibitemShut {NoStop}%
\bibitem [{\citenamefont {Van~Lange}\ \emph {et~al.}(2014)\citenamefont
  {Van~Lange}, \citenamefont {Rockenbach},\ and\ \citenamefont
  {Yamagishi}}]{Lange_2014_BOOK_OUP}%
  \BibitemOpen
  \bibinfo {editor} {\bibfnamefont {P.~A.}\ \bibnamefont {Van~Lange}}, \bibinfo
  {editor} {\bibfnamefont {B.}~\bibnamefont {Rockenbach}},\ and\ \bibinfo
  {editor} {\bibfnamefont {T.}~\bibnamefont {Yamagishi}},\ eds.,\ \href
  {https://doi.org/10.1093/acprof:oso/9780199300730.001.0001} {\emph {\bibinfo
  {title} {Reward and Punishment in Social Dilemmas}}}\ (\bibinfo  {publisher}
  {Oxford University Press},\ \bibinfo {year} {2014})\BibitemShut {NoStop}%
\bibitem [{\citenamefont {Góis}\ \emph {et~al.}(2019)\citenamefont {Góis},
  \citenamefont {Santos}, \citenamefont {Pacheco},\ and\ \citenamefont
  {Santos}}]{Gis2019}%
  \BibitemOpen
  \bibfield  {author} {\bibinfo {author} {\bibfnamefont {A.~R.}\ \bibnamefont
  {Góis}}, \bibinfo {author} {\bibfnamefont {F.~P.}\ \bibnamefont {Santos}},
  \bibinfo {author} {\bibfnamefont {J.~M.}\ \bibnamefont {Pacheco}},\ and\
  \bibinfo {author} {\bibfnamefont {F.~C.}\ \bibnamefont {Santos}},\ }\href
  {http://dx.doi.org/10.1038/s41598-019-52524-8} {\bibfield  {journal}
  {\bibinfo  {journal} {Scientific Reports}\ }\textbf {\bibinfo {volume} {9}}
  (\bibinfo {year} {2019})}\BibitemShut {NoStop}%
\bibitem [{\citenamefont {Liu}\ \emph {et~al.}(2019)\citenamefont {Liu},
  \citenamefont {Li}, \citenamefont {Alam}, \citenamefont {Chen},\ and\
  \citenamefont {Wu}}]{Liu2019}%
  \BibitemOpen
  \bibfield  {author} {\bibinfo {author} {\bibfnamefont {J.}~\bibnamefont
  {Liu}}, \bibinfo {author} {\bibfnamefont {M.}~\bibnamefont {Li}}, \bibinfo
  {author} {\bibfnamefont {M.}~\bibnamefont {Alam}}, \bibinfo {author}
  {\bibfnamefont {Y.}~\bibnamefont {Chen}},\ and\ \bibinfo {author}
  {\bibfnamefont {T.}~\bibnamefont {Wu}},\ }\href
  {https://doi.org/10.1007/s11036-018-1166-0} {\bibfield  {journal} {\bibinfo
  {journal} {Mobile Networks and Applications}\ }\textbf {\bibinfo {volume}
  {24}},\ \bibinfo {pages} {1279–1294} (\bibinfo {year} {2019})}\BibitemShut
  {NoStop}%
\bibitem [{\citenamefont {Fang}\ \emph {et~al.}(2019)\citenamefont {Fang},
  \citenamefont {Benko}, \citenamefont {Perc}, \citenamefont {Xu},\ and\
  \citenamefont {Tan}}]{Fang2019}%
  \BibitemOpen
  \bibfield  {author} {\bibinfo {author} {\bibfnamefont {Y.}~\bibnamefont
  {Fang}}, \bibinfo {author} {\bibfnamefont {T.~P.}\ \bibnamefont {Benko}},
  \bibinfo {author} {\bibfnamefont {M.}~\bibnamefont {Perc}}, \bibinfo {author}
  {\bibfnamefont {H.}~\bibnamefont {Xu}},\ and\ \bibinfo {author}
  {\bibfnamefont {Q.}~\bibnamefont {Tan}},\ }\href
  {https://doi.org/10.1098/rspa.2019.0349} {\bibfield  {journal} {\bibinfo
  {journal} {Proceedings of the Royal Society A: Mathematical, Physical and
  Engineering Sciences}\ }\textbf {\bibinfo {volume} {475}},\ \bibinfo {pages}
  {20190349} (\bibinfo {year} {2019})}\BibitemShut {NoStop}%
\bibitem [{\citenamefont {Ozono}\ \emph {et~al.}(2020)\citenamefont {Ozono},
  \citenamefont {Kamijo},\ and\ \citenamefont {Shimizu}}]{Ozono2020}%
  \BibitemOpen
  \bibfield  {author} {\bibinfo {author} {\bibfnamefont {H.}~\bibnamefont
  {Ozono}}, \bibinfo {author} {\bibfnamefont {Y.}~\bibnamefont {Kamijo}},\ and\
  \bibinfo {author} {\bibfnamefont {K.}~\bibnamefont {Shimizu}},\ }\href
  {http://dx.doi.org/10.1038/s41598-020-64930-4} {\bibfield  {journal}
  {\bibinfo  {journal} {Scientific Reports}\ }\textbf {\bibinfo {volume} {10}}
  (\bibinfo {year} {2020})}\BibitemShut {NoStop}%
\bibitem [{\citenamefont {Nowak}\ and\ \citenamefont
  {Sigmund}(1990)}]{Nowak1990}%
  \BibitemOpen
  \bibfield  {author} {\bibinfo {author} {\bibfnamefont {M.}~\bibnamefont
  {Nowak}}\ and\ \bibinfo {author} {\bibfnamefont {K.}~\bibnamefont
  {Sigmund}},\ }\href {https://doi.org/10.1007/bf00049570} {\bibfield
  {journal} {\bibinfo  {journal} {Acta Applicandae Mathematicae}\ }\textbf
  {\bibinfo {volume} {20}},\ \bibinfo {pages} {247–265} (\bibinfo {year}
  {1990})}\BibitemShut {NoStop}%
\bibitem [{\citenamefont {Nowak}\ and\ \citenamefont
  {Sigmund}(1989)}]{Nowak1989}%
  \BibitemOpen
  \bibfield  {author} {\bibinfo {author} {\bibfnamefont {M.}~\bibnamefont
  {Nowak}}\ and\ \bibinfo {author} {\bibfnamefont {K.}~\bibnamefont
  {Sigmund}},\ }\href {https://doi.org/10.1016/s0022-5193(89)80146-8}
  {\bibfield  {journal} {\bibinfo  {journal} {Journal of Theoretical Biology}\
  }\textbf {\bibinfo {volume} {137}},\ \bibinfo {pages} {21–26} (\bibinfo
  {year} {1989})}\BibitemShut {NoStop}%
\bibitem [{\citenamefont {Hofbauer}\ and\ \citenamefont
  {Sigmund}(1990)}]{Hofbauer1990}%
  \BibitemOpen
  \bibfield  {author} {\bibinfo {author} {\bibfnamefont {J.}~\bibnamefont
  {Hofbauer}}\ and\ \bibinfo {author} {\bibfnamefont {K.}~\bibnamefont
  {Sigmund}},\ }\href {https://doi.org/10.1016/0893-9659(90)90051-c} {\bibfield
   {journal} {\bibinfo  {journal} {Applied Mathematics Letters}\ }\textbf
  {\bibinfo {volume} {3}},\ \bibinfo {pages} {75–79} (\bibinfo {year}
  {1990})}\BibitemShut {NoStop}%
\bibitem [{\citenamefont {Dieckmann}\ and\ \citenamefont
  {Law}(1996)}]{Dieckmann1996JMB}%
  \BibitemOpen
  \bibfield  {author} {\bibinfo {author} {\bibfnamefont {U.}~\bibnamefont
  {Dieckmann}}\ and\ \bibinfo {author} {\bibfnamefont {R.}~\bibnamefont
  {Law}},\ }\href {https://doi.org/10.1007/bf02409751} {\bibfield  {journal}
  {\bibinfo  {journal} {Journal of Mathematical Biology}\ }\textbf {\bibinfo
  {volume} {34}},\ \bibinfo {pages} {579} (\bibinfo {year} {1996})}\BibitemShut
  {NoStop}%
\bibitem [{\citenamefont {Metz}\ \emph {et~al.}(1992)\citenamefont {Metz},
  \citenamefont {Nisbet},\ and\ \citenamefont {Geritz}}]{Metz1992tee}%
  \BibitemOpen
  \bibfield  {author} {\bibinfo {author} {\bibfnamefont {J.}~\bibnamefont
  {Metz}}, \bibinfo {author} {\bibfnamefont {R.}~\bibnamefont {Nisbet}},\ and\
  \bibinfo {author} {\bibfnamefont {S.}~\bibnamefont {Geritz}},\ }\href
  {https://doi.org/10.1016/0169-5347(92)90073-k} {\bibfield  {journal}
  {\bibinfo  {journal} {Trends Ecol. Evol.}\ }\textbf {\bibinfo {volume} {7}},\
  \bibinfo {pages} {198} (\bibinfo {year} {1992})}\BibitemShut {NoStop}%
\bibitem [{\citenamefont {Fudenberg}\ and\ \citenamefont
  {Levine}(1998)}]{fudenberg1998theory}%
  \BibitemOpen
  \bibfield  {author} {\bibinfo {author} {\bibfnamefont {D.}~\bibnamefont
  {Fudenberg}}\ and\ \bibinfo {author} {\bibfnamefont {D.}~\bibnamefont
  {Levine}},\ }\href {https://books.google.co.in/books?id=G6vTQFluxuEC} {\emph
  {\bibinfo {title} {The Theory of Learning in Games}}},\ Economics Learning
  and Social Evolution Series\ (\bibinfo  {publisher} {MIT Press},\ \bibinfo
  {year} {1998})\BibitemShut {NoStop}%
\bibitem [{\citenamefont {Friedman}(1996)}]{F96EJ}%
  \BibitemOpen
  \bibfield  {author} {\bibinfo {author} {\bibfnamefont {D.}~\bibnamefont
  {Friedman}},\ }\href {https://doi.org/10.2307/2234928} {\bibfield  {journal}
  {\bibinfo  {journal} {The Economic Journal}\ }\textbf {\bibinfo {volume}
  {106}},\ \bibinfo {pages} {1} (\bibinfo {year} {1996})}\BibitemShut {NoStop}%
\bibitem [{\citenamefont {Sadhukhan}\ \emph {et~al.}(2022)\citenamefont
  {Sadhukhan}, \citenamefont {Shukla},\ and\ \citenamefont
  {Chakraborty}}]{Sadhukhan2022jop}%
  \BibitemOpen
  \bibfield  {author} {\bibinfo {author} {\bibfnamefont {S.}~\bibnamefont
  {Sadhukhan}}, \bibinfo {author} {\bibfnamefont {A.}~\bibnamefont {Shukla}},\
  and\ \bibinfo {author} {\bibfnamefont {S.}~\bibnamefont {Chakraborty}},\
  }\href {https://doi.org/10.1088/2632-072x/ac4d10} {\bibfield  {journal}
  {\bibinfo  {journal} {Journal of Physics: Complexity}\ }\textbf {\bibinfo
  {volume} {3}},\ \bibinfo {pages} {015009} (\bibinfo {year}
  {2022})}\BibitemShut {NoStop}%
\bibitem [{\citenamefont {Kurokawa}(2019)}]{Kurokawa2019}%
  \BibitemOpen
  \bibfield  {author} {\bibinfo {author} {\bibfnamefont {S.}~\bibnamefont
  {Kurokawa}},\ }\href {https://doi.org/10.1016/j.ecocom.2019.100778}
  {\bibfield  {journal} {\bibinfo  {journal} {Ecological Complexity}\ }\textbf
  {\bibinfo {volume} {40}},\ \bibinfo {pages} {100778} (\bibinfo {year}
  {2019})}\BibitemShut {NoStop}%
\bibitem [{\citenamefont {Kurokawa}\ \emph {et~al.}(2010)\citenamefont
  {Kurokawa}, \citenamefont {Wakano},\ and\ \citenamefont
  {Ihara}}]{Kurokawa2010}%
  \BibitemOpen
  \bibfield  {author} {\bibinfo {author} {\bibfnamefont {S.}~\bibnamefont
  {Kurokawa}}, \bibinfo {author} {\bibfnamefont {J.~Y.}\ \bibnamefont
  {Wakano}},\ and\ \bibinfo {author} {\bibfnamefont {Y.}~\bibnamefont
  {Ihara}},\ }\href {https://doi.org/10.1016/j.tpb.2010.03.002} {\bibfield
  {journal} {\bibinfo  {journal} {Theoretical Population Biology}\ }\textbf
  {\bibinfo {volume} {77}},\ \bibinfo {pages} {257–262} (\bibinfo {year}
  {2010})}\BibitemShut {NoStop}%
\bibitem [{\citenamefont {Park}\ \emph {et~al.}(2017)\citenamefont {Park},
  \citenamefont {Kim},\ and\ \citenamefont {Jeong}}]{Park2017}%
  \BibitemOpen
  \bibfield  {author} {\bibinfo {author} {\bibfnamefont {H.~J.}\ \bibnamefont
  {Park}}, \bibinfo {author} {\bibfnamefont {B.~J.}\ \bibnamefont {Kim}},\ and\
  \bibinfo {author} {\bibfnamefont {H.-C.}\ \bibnamefont {Jeong}},\ }\href
  {http://dx.doi.org/10.1103/PhysRevE.95.042314} {\bibfield  {journal}
  {\bibinfo  {journal} {Physical Review E}\ }\textbf {\bibinfo {volume} {95}}
  (\bibinfo {year} {2017})}\BibitemShut {NoStop}%
\bibitem [{\citenamefont {Marsden}\ and\ \citenamefont
  {McCracken}(1976)}]{Marsden1976}%
  \BibitemOpen
  \bibfield  {author} {\bibinfo {author} {\bibfnamefont {J.~E.}\ \bibnamefont
  {Marsden}}\ and\ \bibinfo {author} {\bibfnamefont {M.}~\bibnamefont
  {McCracken}},\ }\href {https://doi.org/10.1007/978-1-4612-6374-6} {\emph
  {\bibinfo {title} {The Hopf Bifurcation and Its Applications}}}\ (\bibinfo
  {publisher} {Springer New York},\ \bibinfo {year} {1976})\BibitemShut
  {NoStop}%
\bibitem [{\citenamefont {Nowak}\ and\ \citenamefont {Sigmund}(1992)}]{NS1993}%
  \BibitemOpen
  \bibfield  {author} {\bibinfo {author} {\bibfnamefont {M.~A.}\ \bibnamefont
  {Nowak}}\ and\ \bibinfo {author} {\bibfnamefont {K.}~\bibnamefont
  {Sigmund}},\ }\href {https://doi.org/10.1038/355250a0} {\bibfield  {journal}
  {\bibinfo  {journal} {Nature}\ }\textbf {\bibinfo {volume} {355}},\ \bibinfo
  {pages} {250} (\bibinfo {year} {1992})}\BibitemShut {NoStop}%
\bibitem [{\citenamefont {Nowak}\ and\ \citenamefont
  {Sigmund}(2005)}]{Nowak2005}%
  \BibitemOpen
  \bibfield  {author} {\bibinfo {author} {\bibfnamefont {M.~A.}\ \bibnamefont
  {Nowak}}\ and\ \bibinfo {author} {\bibfnamefont {K.}~\bibnamefont
  {Sigmund}},\ }\href {https://doi.org/10.1038/nature04131} {\bibfield
  {journal} {\bibinfo  {journal} {Nature}\ }\textbf {\bibinfo {volume} {437}},\
  \bibinfo {pages} {1291–1298} (\bibinfo {year} {2005})}\BibitemShut
  {NoStop}%
\bibitem [{\citenamefont {McAvoy}\ and\ \citenamefont
  {Hauert}(2017)}]{McAvoy2017}%
  \BibitemOpen
  \bibfield  {author} {\bibinfo {author} {\bibfnamefont {A.}~\bibnamefont
  {McAvoy}}\ and\ \bibinfo {author} {\bibfnamefont {C.}~\bibnamefont
  {Hauert}},\ }\href {https://doi.org/10.1016/j.tpb.2016.09.004} {\bibfield
  {journal} {\bibinfo  {journal} {Theoretical Population Biology}\ }\textbf
  {\bibinfo {volume} {113}},\ \bibinfo {pages} {13–22} (\bibinfo {year}
  {2017})}\BibitemShut {NoStop}%
\bibitem [{\citenamefont {Ostrom}(2015)}]{Ostrom2015}%
  \BibitemOpen
  \bibfield  {author} {\bibinfo {author} {\bibfnamefont {E.}~\bibnamefont
  {Ostrom}},\ }\href {https://doi.org/10.1017/cbo9781316423936} {\emph
  {\bibinfo {title} {Governing the Commons: The Evolution of Institutions for
  Collective Action}}}\ (\bibinfo  {publisher} {Cambridge University Press},\
  \bibinfo {year} {2015})\BibitemShut {NoStop}%
\bibitem [{\citenamefont {Cole}\ and\ \citenamefont {Ostrom}(2010)}]{Cole2010}%
  \BibitemOpen
  \bibfield  {author} {\bibinfo {author} {\bibfnamefont {D.~H.}\ \bibnamefont
  {Cole}}\ and\ \bibinfo {author} {\bibfnamefont {E.}~\bibnamefont {Ostrom}},\
  }\href {http://dx.doi.org/10.2139/ssrn.1656418} {\bibfield  {journal}
  {\bibinfo  {journal} {SSRN Electronic Journal}\ } (\bibinfo {year}
  {2010})}\BibitemShut {NoStop}%
\bibitem [{\citenamefont {Sigmund}\ \emph {et~al.}(2010)\citenamefont
  {Sigmund}, \citenamefont {De~Silva}, \citenamefont {Traulsen},\ and\
  \citenamefont {Hauert}}]{SST_2010_NAT}%
  \BibitemOpen
  \bibfield  {author} {\bibinfo {author} {\bibfnamefont {K.}~\bibnamefont
  {Sigmund}}, \bibinfo {author} {\bibfnamefont {H.}~\bibnamefont {De~Silva}},
  \bibinfo {author} {\bibfnamefont {A.}~\bibnamefont {Traulsen}},\ and\
  \bibinfo {author} {\bibfnamefont {C.}~\bibnamefont {Hauert}},\ }\href
  {https://doi.org/10.1038/nature09203} {\bibfield  {journal} {\bibinfo
  {journal} {Nature}\ }\textbf {\bibinfo {volume} {466}},\ \bibinfo {pages}
  {861–863} (\bibinfo {year} {2010})}\BibitemShut {NoStop}%
\bibitem [{\citenamefont {Perc}(2012)}]{P_2012_Sc_R}%
  \BibitemOpen
  \bibfield  {author} {\bibinfo {author} {\bibfnamefont {M.}~\bibnamefont
  {Perc}},\ }\href {http://dx.doi.org/10.1038/srep00344} {\bibfield  {journal}
  {\bibinfo  {journal} {Scientific Reports}\ }\textbf {\bibinfo {volume} {2}}
  (\bibinfo {year} {2012})}\BibitemShut {NoStop}%
\bibitem [{\citenamefont {Kareva}\ \emph {et~al.}(2013)\citenamefont {Kareva},
  \citenamefont {Morin},\ and\ \citenamefont {Karev}}]{kareva2013bmb}%
  \BibitemOpen
  \bibfield  {author} {\bibinfo {author} {\bibfnamefont {I.}~\bibnamefont
  {Kareva}}, \bibinfo {author} {\bibfnamefont {B.}~\bibnamefont {Morin}},\ and\
  \bibinfo {author} {\bibfnamefont {G.}~\bibnamefont {Karev}},\ }\href
  {https://doi.org/10.1007/s11538-012-9804-3} {\bibfield  {journal} {\bibinfo
  {journal} {Bulletin of Mathematical Biology}\ }\textbf {\bibinfo {volume}
  {75}},\ \bibinfo {pages} {565–588} (\bibinfo {year} {2013})}\BibitemShut
  {NoStop}%
\bibitem [{\citenamefont {West}\ \emph {et~al.}(2006)\citenamefont {West},
  \citenamefont {Griffin}, \citenamefont {Gardner},\ and\ \citenamefont
  {Diggle}}]{west2006social}%
  \BibitemOpen
  \bibfield  {author} {\bibinfo {author} {\bibfnamefont {S.~A.}\ \bibnamefont
  {West}}, \bibinfo {author} {\bibfnamefont {A.~S.}\ \bibnamefont {Griffin}},
  \bibinfo {author} {\bibfnamefont {A.}~\bibnamefont {Gardner}},\ and\ \bibinfo
  {author} {\bibfnamefont {S.~P.}\ \bibnamefont {Diggle}},\ }\href
  {https://doi.org/10.1038/nrmicro1461} {\bibfield  {journal} {\bibinfo
  {journal} {Nature Reviews Microbiology}\ }\textbf {\bibinfo {volume} {4}},\
  \bibinfo {pages} {597–607} (\bibinfo {year} {2006})}\BibitemShut {NoStop}%
\bibitem [{\citenamefont {Nadell}\ \emph {et~al.}(2016)\citenamefont {Nadell},
  \citenamefont {Drescher},\ and\ \citenamefont {Foster}}]{nadell2016spatial}%
  \BibitemOpen
  \bibfield  {author} {\bibinfo {author} {\bibfnamefont {C.~D.}\ \bibnamefont
  {Nadell}}, \bibinfo {author} {\bibfnamefont {K.}~\bibnamefont {Drescher}},\
  and\ \bibinfo {author} {\bibfnamefont {K.~R.}\ \bibnamefont {Foster}},\
  }\href {https://doi.org/10.1038/nrmicro.2016.84} {\bibfield  {journal}
  {\bibinfo  {journal} {Nature Reviews Microbiology}\ }\textbf {\bibinfo
  {volume} {14}},\ \bibinfo {pages} {589–600} (\bibinfo {year}
  {2016})}\BibitemShut {NoStop}%
\bibitem [{\citenamefont {Raina}\ \emph {et~al.}(2018)\citenamefont {Raina},
  \citenamefont {Eme}, \citenamefont {Pollock}, \citenamefont {Spang},
  \citenamefont {Archibald},\ and\ \citenamefont
  {Williams}}]{raina2018microbial}%
  \BibitemOpen
  \bibfield  {author} {\bibinfo {author} {\bibfnamefont {J.-B.}\ \bibnamefont
  {Raina}}, \bibinfo {author} {\bibfnamefont {L.}~\bibnamefont {Eme}}, \bibinfo
  {author} {\bibfnamefont {F.~J.}\ \bibnamefont {Pollock}}, \bibinfo {author}
  {\bibfnamefont {A.}~\bibnamefont {Spang}}, \bibinfo {author} {\bibfnamefont
  {J.~M.}\ \bibnamefont {Archibald}},\ and\ \bibinfo {author} {\bibfnamefont
  {T.~A.}\ \bibnamefont {Williams}},\ }\href
  {http://dx.doi.org/10.1242/bio.032524} {\bibfield  {journal} {\bibinfo
  {journal} {Biology Open}\ }\textbf {\bibinfo {volume} {7}} (\bibinfo {year}
  {2018})}\BibitemShut {NoStop}%
\bibitem [{\citenamefont {Smith}\ \emph {et~al.}(2019)\citenamefont {Smith},
  \citenamefont {Shorten}, \citenamefont {Altermann}, \citenamefont {Roy},\
  and\ \citenamefont {McNabb}}]{smith2019evolution}%
  \BibitemOpen
  \bibfield  {author} {\bibinfo {author} {\bibfnamefont {N.~W.}\ \bibnamefont
  {Smith}}, \bibinfo {author} {\bibfnamefont {P.~R.}\ \bibnamefont {Shorten}},
  \bibinfo {author} {\bibfnamefont {E.}~\bibnamefont {Altermann}}, \bibinfo
  {author} {\bibfnamefont {N.~C.}\ \bibnamefont {Roy}},\ and\ \bibinfo {author}
  {\bibfnamefont {W.~C.}\ \bibnamefont {McNabb}},\ }\href
  {http://dx.doi.org/10.3389/fevo.2019.00153} {\bibfield  {journal} {\bibinfo
  {journal} {Frontiers in Ecology and Evolution}\ }\textbf {\bibinfo {volume}
  {7}} (\bibinfo {year} {2019})}\BibitemShut {NoStop}%
\bibitem [{\citenamefont {Harrison}\ \emph {et~al.}(2007)\citenamefont
  {Harrison}, \citenamefont {Paul}, \citenamefont {Massey},\ and\ \citenamefont
  {Buckling}}]{harrison2008microbial}%
  \BibitemOpen
  \bibfield  {author} {\bibinfo {author} {\bibfnamefont {F.}~\bibnamefont
  {Harrison}}, \bibinfo {author} {\bibfnamefont {J.}~\bibnamefont {Paul}},
  \bibinfo {author} {\bibfnamefont {R.~C.}\ \bibnamefont {Massey}},\ and\
  \bibinfo {author} {\bibfnamefont {A.}~\bibnamefont {Buckling}},\ }\href
  {https://doi.org/10.1038/ismej.2007.96} {\bibfield  {journal} {\bibinfo
  {journal} {The ISME Journal}\ }\textbf {\bibinfo {volume} {2}},\ \bibinfo
  {pages} {49–55} (\bibinfo {year} {2007})}\BibitemShut {NoStop}%
\bibitem [{\citenamefont {Drescher}\ \emph {et~al.}(2014)\citenamefont
  {Drescher}, \citenamefont {Nadell}, \citenamefont {Stone}, \citenamefont
  {Wingreen},\ and\ \citenamefont {Bassler}}]{drescher2014biofilm}%
  \BibitemOpen
  \bibfield  {author} {\bibinfo {author} {\bibfnamefont {K.}~\bibnamefont
  {Drescher}}, \bibinfo {author} {\bibfnamefont {C.~D.}\ \bibnamefont
  {Nadell}}, \bibinfo {author} {\bibfnamefont {H.~A.}\ \bibnamefont {Stone}},
  \bibinfo {author} {\bibfnamefont {N.~S.}\ \bibnamefont {Wingreen}},\ and\
  \bibinfo {author} {\bibfnamefont {B.~L.}\ \bibnamefont {Bassler}},\ }\href
  {https://doi.org/10.1016/j.cub.2013.10.030} {\bibfield  {journal} {\bibinfo
  {journal} {Current Biology}\ }\textbf {\bibinfo {volume} {24}},\ \bibinfo
  {pages} {50–55} (\bibinfo {year} {2014})}\BibitemShut {NoStop}%
\bibitem [{\citenamefont {Luo}\ \emph {et~al.}(2025)\citenamefont {Luo},
  \citenamefont {Lin}, \citenamefont {Chen},\ and\ \citenamefont
  {Szolnoki}}]{Luo2025}%
  \BibitemOpen
  \bibfield  {author} {\bibinfo {author} {\bibfnamefont {J.}~\bibnamefont
  {Luo}}, \bibinfo {author} {\bibfnamefont {D.}~\bibnamefont {Lin}}, \bibinfo
  {author} {\bibfnamefont {X.}~\bibnamefont {Chen}},\ and\ \bibinfo {author}
  {\bibfnamefont {A.}~\bibnamefont {Szolnoki}},\ }\href
  {http://dx.doi.org/10.1063/5.0262821} {\bibfield  {journal} {\bibinfo
  {journal} {Chaos: An Interdisciplinary Journal of Nonlinear Science}\
  }\textbf {\bibinfo {volume} {35}} (\bibinfo {year} {2025})}\BibitemShut
  {NoStop}%
\bibitem [{\citenamefont {Wakano}(2007)}]{Wakano2007}%
  \BibitemOpen
  \bibfield  {author} {\bibinfo {author} {\bibfnamefont {J.~Y.}\ \bibnamefont
  {Wakano}},\ }\href {https://doi.org/10.1016/j.jtbi.2007.04.008} {\bibfield
  {journal} {\bibinfo  {journal} {Journal of Theoretical Biology}\ }\textbf
  {\bibinfo {volume} {247}},\ \bibinfo {pages} {616–622} (\bibinfo {year}
  {2007})}\BibitemShut {NoStop}%
\bibitem [{\citenamefont {Strogatz}(2018)}]{Strogatz2018}%
  \BibitemOpen
  \bibfield  {author} {\bibinfo {author} {\bibfnamefont {S.~H.}\ \bibnamefont
  {Strogatz}},\ }\href {https://doi.org/10.1201/9780429492563} {\emph {\bibinfo
  {title} {Nonlinear Dynamics and Chaos}}}\ (\bibinfo  {publisher} {CRC
  Press},\ \bibinfo {year} {2018})\BibitemShut {NoStop}%
\bibitem [{\citenamefont {Pandit}\ \emph {et~al.}(2018)\citenamefont {Pandit},
  \citenamefont {Mukhopadhyay},\ and\ \citenamefont
  {Chakraborty}}]{Pandit2018}%
  \BibitemOpen
  \bibfield  {author} {\bibinfo {author} {\bibfnamefont {V.}~\bibnamefont
  {Pandit}}, \bibinfo {author} {\bibfnamefont {A.}~\bibnamefont
  {Mukhopadhyay}},\ and\ \bibinfo {author} {\bibfnamefont {S.}~\bibnamefont
  {Chakraborty}},\ }\href {http://dx.doi.org/10.1063/1.5011955} {\bibfield
  {journal} {\bibinfo  {journal} {Chaos: An Interdisciplinary Journal of
  Nonlinear Science}\ }\textbf {\bibinfo {volume} {28}} (\bibinfo {year}
  {2018})}\BibitemShut {NoStop}%
\bibitem [{\citenamefont {Hummert}\ \emph {et~al.}(2014)\citenamefont
  {Hummert}, \citenamefont {Bohl}, \citenamefont {Basanta}, \citenamefont
  {Deutsch}, \citenamefont {Werner}, \citenamefont {Theißen}, \citenamefont
  {Schroeter},\ and\ \citenamefont {Schuster}}]{HBB_2014_MOLB}%
  \BibitemOpen
  \bibfield  {author} {\bibinfo {author} {\bibfnamefont {S.}~\bibnamefont
  {Hummert}}, \bibinfo {author} {\bibfnamefont {K.}~\bibnamefont {Bohl}},
  \bibinfo {author} {\bibfnamefont {D.}~\bibnamefont {Basanta}}, \bibinfo
  {author} {\bibfnamefont {A.}~\bibnamefont {Deutsch}}, \bibinfo {author}
  {\bibfnamefont {S.}~\bibnamefont {Werner}}, \bibinfo {author} {\bibfnamefont
  {G.}~\bibnamefont {Theißen}}, \bibinfo {author} {\bibfnamefont
  {A.}~\bibnamefont {Schroeter}},\ and\ \bibinfo {author} {\bibfnamefont
  {S.}~\bibnamefont {Schuster}},\ }\href {https://doi.org/10.1039/c3mb70602h}
  {\bibfield  {journal} {\bibinfo  {journal} {Molecular BioSystems}\ }\textbf
  {\bibinfo {volume} {10}},\ \bibinfo {pages} {3044–3065} (\bibinfo {year}
  {2014})}\BibitemShut {NoStop}%
\bibitem [{\citenamefont {Sadhukhan}\ \emph {et~al.}(2021)\citenamefont
  {Sadhukhan}, \citenamefont {Chattopadhyay},\ and\ \citenamefont
  {Chakraborty}}]{SCR_2021_PRE}%
  \BibitemOpen
  \bibfield  {author} {\bibinfo {author} {\bibfnamefont {S.}~\bibnamefont
  {Sadhukhan}}, \bibinfo {author} {\bibfnamefont {R.}~\bibnamefont
  {Chattopadhyay}},\ and\ \bibinfo {author} {\bibfnamefont {S.}~\bibnamefont
  {Chakraborty}},\ }\href {https://doi.org/10.1103/PhysRevResearch.3.013009}
  {\bibfield  {journal} {\bibinfo  {journal} {Phys. Rev. Research}\ }\textbf
  {\bibinfo {volume} {3}},\ \bibinfo {pages} {013009} (\bibinfo {year}
  {2021})}\BibitemShut {NoStop}%
\bibitem [{\citenamefont {Hauert}(2002)}]{H_2002_IJBC}%
  \BibitemOpen
  \bibfield  {author} {\bibinfo {author} {\bibfnamefont {C.}~\bibnamefont
  {Hauert}},\ }\href {https://doi.org/10.1142/s0218127402005273} {\bibfield
  {journal} {\bibinfo  {journal} {International Journal of Bifurcation and
  Chaos}\ }\textbf {\bibinfo {volume} {12}},\ \bibinfo {pages} {1531–1548}
  (\bibinfo {year} {2002})}\BibitemShut {NoStop}%
\end{thebibliography}%
	
\end{document}